\documentclass[11pt,hyper,letterpaper]{JHEP3}
\usepackage[dvips]{epsfig}
\usepackage{graphicx}
\usepackage{dcolumn}
\usepackage{bm}
\usepackage{amsmath,amssymb,epsf,amsfonts}
\usepackage{bbm}
\usepackage{slashed}

\newcommand{\be}{\begin{equation}}
\newcommand{\ee}{\end{equation}}
\newcommand{\ba}{\begin{eqnarray}}
\newcommand{\ea}{\end{eqnarray}}
\newcommand{\bi}{\begin{itemize}}
\newcommand{\ei}{\end{itemize}}
\newcommand{\ben}{\begin{enumerate}}
\newcommand{\een}{\end{enumerate}}

\newcommand{\ie}{{\it i.e.}\ }

\def\Tr{{\rm Tr}}

\newcommand{\N}{{\mathcal N}}

\newcommand{\ra}{\rightarrow}
\newcommand{\cp}{\mathbbm{CP}^3}

\renewcommand{\a}{\alpha}
\renewcommand{\b}{\beta}

\newcommand{\e}{\epsilon}
\newcommand{\G}{\Gamma}
\newcommand{\g}{\gamma}
\renewcommand{\k}{\kappa}

\newcommand{\m}{\mu}

\newcommand{\om}{\omega}
\renewcommand{\r}{\rho}
\newcommand{\s}{\sigma}

\newcommand{\z}{\zeta}
\newcommand{\p}{\partial}

\newcommand{\sh}{\sharp}
\newcommand{\nn}{\nonumber}
\newcommand{\lp}{\left(}
\newcommand{\rp}{\right)}

\def\a{\alpha}
\def\b{\beta}

\def\G{\Gamma}

\def\g{\gamma}

\def\k{\kappa}             

\def\m{\mu}

\def\th{\theta}                   
\def\r{\rho}                                     
\def\s{\sigma}                                   

\def\z{\zeta}

\def\G{\Gamma}

\newcommand{\de}{\delta}

\title{ \LARGE Adding Flavor to $\mathbf{AdS_4/CFT_3}$}
\author{Martin Ammon, Johanna Erdmenger, Ren\'e Meyer, Andy O'Bannon,\!$^1$ and Timm Wrase\!$^{1,2}$\footnotemark[1]
\\
$^1$Max-Planck-Institut f\"{u}r Physik (Werner-Heisenberg-Institut) \\ F\"{o}hringer Ring 6, 80805 M\"{u}nchen, Germany
\\
$^2$Institut f\"{u}r Theoretische Physik, Leibniz Universit\"{a}t Hannover \\ Appelstra{\ss}e 2, 30167, Hannover, Germany}

\footnotetext[1]{E-mail addresses: \email{ammon, jke, meyer, ahob @mppmu.mpg.de, timm.wrase@itp.uni-hannover.de}}

\date{\today}

\abstract{Aharony, Bergman, Jafferis, and Maldacena have proposed that the low-energy description of multiple M2-branes at a $\mathbbm{C}^4/\mathbbm{Z}_k$ singularity is a (2+1)-dimensional $\N=6$ supersymmetric $U(N_c) \times U(N_c)$ Chern-Simons matter theory, the ABJM theory.  In the large-$N_c$ limit, its holographic dual is supergravity in $AdS_4 \times S^7/\mathbbm{Z}_k$. We study various ways to add fields that transform in the fundamental representation of the gauge groups, \textit{i.e.} flavor fields, to the ABJM theory. We work in a probe limit and perform analyses in both the supergravity and field theory descriptions. In the supergravity description we find a large class of supersymmetric embeddings of probe flavor branes. In the field theory description, we present a general method to determine the couplings of the flavor fields to the fields of the ABJM theory. We then study four examples in detail: codimension-zero $\N=3$ supersymmetric flavor, described in supergravity by Kaluza-Klein monopoles or D6-branes; codimension-one $\N=(0,6)$ supersymmetric chiral flavor, described by D8-branes; codimension-one $\N=(3,3)$ supersymmetric non-chiral flavor, described by M5/D4-branes; codimension-two $\N=4$ supersymmetric flavor, described by M2/D2-branes. Finally we discuss special physical equivalences between brane embeddings in M-theory, and their interpretation in the field theory description.}

\keywords{AdS/CFT correspondence, Gauge/gravity correspondence}
\preprint{MPP-2009-52\\ ITP-UH-17/09}

\begin{document}

\section{Introduction}
\label{introduction}

Recently the study of the low-energy description of multiple M2-branes has seen great progress. In particular, the authors of ref. \cite{Aharony:2008ug}, following the work of refs. \cite{Bagger:2006sk,Gustavsson:2007vu,Bagger:2007jr,Bagger:2007vi,VanRaamsdonk:2008ft,Lambert:2008et,Distler:2008mk}, proposed that $N_c$ coincident M2-branes probing a $\mathbbm{C}^4/\mathbbm{Z}_k$ singularity have a low-energy description as a particular $\N=6$ supersymmetric Chern-Simons-matter theory. We will refer to the theory of ref. \cite{Aharony:2008ug} as ``the ABJM theory.''

The ABJM theory provides new information about the anti-de Sitter/Conformal Field Theory (AdS/CFT) correspondence \cite{Maldacena:1997re,Witten:1998qj,Gubser:1998bc}. Given $N_c$ M2-branes at a $\mathbbm{C}^4/\mathbbm{Z}_k$ singularity, when we take $N_c \ra \infty$ we can replace the M2-branes with their near-horizon geometry, which is (3+1)-dimensional AdS space times a $\mathbbm{Z}_k$ orbifold of a seven-sphere, $AdS_4 \times S^7/\mathbbm{Z}_k$. The natural conjecture was that the low-energy theory of M2-branes, in the large-$N_c$ limit, is dual to eleven-dimensional supergravity on $AdS_4 \times S^7/\mathbbm{Z}_k$. The exact low-energy theory of multiple M2-branes was unknown before ref. \cite{Aharony:2008ug}, however.

Additionally, via AdS/CFT, the ABJM theory may also have practical applications as a solvable toy model for certain condensed matter systems, for example systems whose low-energy dynamics is dominated by a quantum critical point and/or is described by a strongly-coupled Chern-Simons-matter theory.

The authors of ref. \cite{Aharony:2008ug} derive the $\N=6$ supersymmetric Chern-Simons-matter theory, and its relation to M2-branes, from a particular brane construction in type IIB string theory. The initial configuration includes two stacks of $N_c$ D3-branes, an NS5-brane, and a $(1,k)$5-brane. At this stage we can identify the low-energy theory of the D3-branes as a (2+1)-dimensional $\N=3$ supersymmetric Chern-Simons-matter theory with product gauge group $U(N_c) \times U(N_c)$ and with equal but opposite Chern-Simons levels $k$ and $-k$ for the two gauge groups, which we denote as $U(N_c)_k \times U(N_c)_{-k}$. Performing a T-duality, an uplift to M-theory, and a certain kind of ``near-horizon'' limit\footnote{Throughout this paper we will use quotation marks to distinguish this ``near-horizon'' limit, which we explain in detail below, and the usual near-horizon limit of a stack of branes, for example the near-horizon limit of M2-branes, which produces $AdS_4$. We define the ``near-horizon'' limit precisely in section \ref{typeIIAandM}, and perform the limit explicitly in appendix \ref{app:IIBtoM}.}, we obtain M2-branes probing $\mathbbm{C}^4/\mathbbm{Z}_k$. A supersymmetry enhancement occurs in the ``near-horizon'' limit, from $\N=3$ to $\N=6$ supersymmetry. The low-energy theory of the M2-branes (the ABJM theory) is thus an $\N=6$ supersymmetric $U(N_c)_k \times U(N_c)_{-k}$ theory with adjoint and bifundamental fields. Upon taking $N_c \ra \infty$ and replacing the M2-branes with their near-horizon geometry, the appropriate description is eleven-dimensional supergravity on $AdS_4 \times S^7 / \mathbbm{Z}_k$. We may then take $k \ra \infty$, where the appropriate description becomes type IIA supergravity on $AdS_4 \times \cp$.

In this paper we deform the ABJM theory by introducing fields in the fundamental representation of the gauge groups, \textit{i.e} flavor fields. We study flavor fields both in the brane construction and also in the field theory. In particular, for a given brane construction, we present a general recipe to determine the couplings of the flavor fields to the fields of the ABJM theory. We then apply our general recipe to four examples, where our goals are to write the field theory Lagrangians and to compare the symmetries of the string/gravity description and the field theory.

To explain how we  add flavor to the ABJM theory, we first recall how to add flavor in the ``usual'' AdS/CFT correspondence, which arises from the study of D3-branes in flat space. Here the gravity theory is type IIB supergravity in the near-horizon geometry of D3-branes, which is $AdS_5 \times S^5$, and the dual strongly-coupled CFT is (3+1)-dimensional $\N=4$ supersymmetric $SU(N_c)$ Yang-Mills (SYM) theory in the 't Hooft limit $N_c \ra \infty$ and additionally with large 't Hooft coupling $\lambda \equiv g_{YM}^2 N_c \ra \infty$.

To introduce flavor fields, we return to D3-branes in flat space and introduce additional open string degrees of freedom, \textit{i.e.} additional D-branes. The standard example is D7-branes that intersect the D3-branes along their (3+1)-dimensional worldvolume \cite{Karch:2002sh}. The endpoints of 3-7 and 7-3 strings act as pointlike excitations in the $N_c$ or $\bar{N}_c$ of $SU(N_c)$ on the D3-branes' worldvolume. The additional branes are thus called ``flavor branes.'' By now a large literature exists on the physical properties of these gauge/gravity models with flavor, beginning with refs. \cite{Karch:2002sh,Kruczenski:2003be,Sakai:2003wu,Babington:2003vm} (for a review see ref. \cite{Erdmenger:2007cm}).

To obtain $AdS_5 \times S^5$ then requires taking $N_c \ra \infty$. If we keep the number $N_f$ of D7-branes fixed as $N_c \ra \infty$, so that $N_f \ll N_c$, then we may neglect the D7-branes' contribution to the stress-energy tensor, and hence we may ignore their effect on the metric\footnote{Analogous statements apply for the other fields of supergravity.}. This limit is called the ``probe limit'' because in this limit the D7-brane ``cleanly probes'' the geometry without deforming it. In the field theory the probe limit amounts to ignoring quantum effects due to the flavor, such as the running of the coupling, because such effects are suppressed by $N_f/N_c$. In the language of perturbation theory, we are discarding diagrams with quark (and/or squark) loops (sometimes called the ``quenched approximation'').

Sometimes the flavor fields will be confined to a lower-dimensional defect, for example, a supersymmetric D3/D5 intersection can give rise to probe D5-branes along $AdS_4 \times S^2$ inside $AdS_5 \times S^5$, which describes supersymmetric flavor fields confined to propagate only in (2+1)-dimensions of the (3+1)-dimensional theory, \textit{i.e.} along a codimension-one defect \cite{Karch:2000gx,DeWolfe:2001pq,Erdmenger:2002ex}. In the construction of the associated supersymmetric defect field theory, a convenient step was to write the ambient fields in terms of the lower-dimensional superspace appropriate for the defect, based on the previous results of refs. \cite{Hori:2000ic,Sethi:1997zza,Kapustin:1998pb,Mirabelli:1997aj}.  The codimension-two case of a D3-brane probe wrapping $AdS_3 \times S^1$ inside $AdS_5 \times S^5$ was studied in ref. \cite{Constable:2002xt}.

To add supersymmetric flavor to the ABJM theory, we will add supersymmetric flavor branes in the type IIB brane construction of the ABJM theory. We begin by listing all supersymmetric Dp-branes that are extended along the coordinate axes in the type IIB construction (see table \ref{table:IIB} in section \ref{gravityanalysis}). We then perform a T-duality, a lift to M-theory, and the ``near-horizon'' limit to determine where these branes end up in M-theory on $\mathbbm{C}^4/\mathbbm{Z}_k$, and we compute the amount of supersymmetry the object (brane or Kaluza-Klein (KK) monopole) preserves (table \ref{table:flat} in appendix \ref{app:probeflat} lists a few examples). Lastly we take $N_c \ra \infty$ and determine where the objects end up in M-theory on $AdS_4 \times S^7/\mathbbm{Z}_k$, and for a few examples we compute the amount of supersymmetry the objects preserve (see appendix \ref{app:probeads}).

All of the above analysis occurs on the gravity side of the AdS/CFT correspondence. How do we determine the dual field theory, including the flavor fields? As a systematic approach to construct the field theory, we start a few steps ``before'' ABJM's type IIB construction. We begin in type IIB with just a D3/Dp intersection, where the Dp-brane is the flavor brane. The actions for many D3/Dp systems are known (see for example refs. \cite{Karch:2002sh,DeWolfe:2001pq,Erdmenger:2002ex,Constable:2002xt,Gomis:2006sb,Harvey:2007ab,Buchbinder:2007ar,Harvey:2008zz}). We then follow what happens to the action, step-by-step, during ABJM's type IIB construction. Two steps are crucially important in this procedure. The first is the addition of the NS5-branes, which impose boundary conditions that set to zero some of the degrees of freedom, as explained in refs. \cite{Hanany:1996ie,Gaiotto:2008sa}. The second is when we take the ``near-horizon'' limit (after T-dualizing in $x^6$ and lifting to M-theory). In the field theory, this corresponds to taking a low-energy limit and writing an effective theory valid on scales below the Chern-Simons mass scale $g_{YM}^2 k /(4\pi)$ (with $g_{YM}$ the Yang-Mills coupling of the (2+1)-dimensional theory). Roughly speaking, the action will be the known D3/Dp action after 1.) imposing the NS5-brane boundary conditions and 2.) taking the low-energy limit. The resulting action is the answer for the field theory, and should have the correct symmetries. As we will see, however, this procedure is not always easy to implement in concrete examples.

We apply our general procedure to four examples. We will add four different flavor branes in type IIB, which become four objects (branes or KK monopoles) in M-theory on $\mathbbm{C}^4/\mathbbm{Z}_k$. The four branes, and the type of flavor fields they describe, are as follows:

\begin{itemize}
\item A D5-brane that becomes a D6-brane in type IIA and a KK monopole in M-theory, and which introduces codimension-zero $\N=3$ supersymmetric flavor fields.

\item A D7-brane that becomes a D8-brane in type IIA and an M9-brane\footnote{Here, and throughout the paper, ``M9-brane'' will refer to the still-mysterious M-theory description of D8-branes. The only part of the M-theory description that we really use is the name ``M9-brane,'' however. In most cases, thinking of this object as a D8-brane in type IIA suffices. For more on the conjectured M9-brane, see ref. \cite{Bergshoeff:1998bs} and references therein.} in M-theory, and which introduces codimension-one, chiral, $\N=(0,6)$ supersymmetric flavor fields.

\item A D3-brane that becomes a D4-brane in type IIA and an M5-brane in M-theory, and which introduces codimension-one, non-chiral, $\N=(3,3)$ supersymmetric flavor fields.

\item A D3-brane that becomes a D2-brane in type IIA and an M2-brane in M-theory, and which introduces codimension-two, $\N=4$ supersymmetric flavor fields.
\end{itemize}
In all four cases, we describe the location of the object in M-theory on $\mathbbm{C}^4/\mathbbm{Z}_k$, calculate the number of supercharges it preserves, and identify the isometries of the background that it preserves. For the first two cases, we write the field theory Lagrangian describing the coupling of the flavor fields to the fields of the ABJM theory and match the symmetries between the field theory and supergravity descriptions. In the third and fourth cases we take some first steps toward constructing the Lagrangians, commenting in particular on the symmetries.

In our analysis we find that many different Dp-branes in type IIB, for example Dp-branes with different embeddings or even Dp-branes of different dimensionality, become the same object in M-theory. Furthermore, the embeddings of many such M-theory objects may be mapped into one another via an $SU(4)$ isometry transformation (as first noted in ref. \cite{Hikida:2009tp}). When that occurs, we call the two objects ``$SU(4)$-equivalent.'' A natural question is what $SU(4)$ equivalence means in the field theory. In simple terms, $SU(4)$ equivalence occurs when two different theories flow to the same low-energy fixed point (corresponding to two type IIB Dp-branes becoming the same object after the ``near-horizon'' limit). We will discuss $SU(4)$ equivalence in more detail, and provide some examples, below.

Throughout this paper we work in the probe limit whenever applicable. In other words, whenever we have $N_f$ flavor branes and we take $N_c \ra \infty$, we will keep $N_f$ fixed, such that $N_f \ll N_c$. We will always consider $N_f$ coincident flavor branes; we will never separate the flavor branes from one another. We will also consider only massless flavor fields.

We will end our introduction by reviewing similar studies of probe branes in $AdS_4/CFT_3$, to compare and contrast with our study.

The authors of ref. \cite{Myers:2006qr} considered M2-branes on $\mathbbm{C}^4$, added probe M5-branes (codimension one) and M2-branes (codimension two), and computed the spectrum of geometric fluctuations of the probe branes. In type IIA, these probes become D4-branes and D2-branes, respectively. Our analysis extends that of ref. \cite{Myers:2006qr} in two ways. First, we consider branes probing $\mathbbm{C}^4/\mathbbm{Z}_k$ with $k\geq2$ rather than $\mathbbm{C}^4$. Second, we consider supersymmetric M2- and M5-branes, but we also consider a KK monopole as well as the M9/D8-brane. We do not study fluctuations of our probes, however.

In type IIA on $AdS_4 \times \cp$, the authors of refs. \cite{Fujita:2009kw,Chandrasekhar:2009ey} studied D4-branes extended along $AdS_3 \times \mathbbm{CP}^1$ and the authors of ref. \cite{Fujita:2009kw} studied D8-branes extended along $AdS_3 \times \cp$. We study these objects on the gravity side, but our analysis extends that of ref. \cite{Fujita:2009kw} primarily on the field theory side: for the D8-brane, we write the Lagrangian describing the coupling of the flavor fields to the fields of the ABJM theory.

The authors of refs. \cite{Hohenegger:2009as,Gaiotto:2009tk,Hikida:2009tp} studied the KK monopole (D5-brane in type IIB, D6-brane in type IIA), on both sides, gravity and field theory\footnote{An extension of the analysis of refs.  \cite{Hohenegger:2009as,Gaiotto:2009tk,Hikida:2009tp} to a more general system appears in ref. \cite{Fujita:2009xz}.}. Although we will have little to add to the physics in this case, it serves as an especially nice example of our general recipe before we consider more complicated cases.

This paper is organized as follows. In section \ref{reviewabjm} we review the ABJM theory and its brane construction. Section \ref{generalanalysis} explains in general terms how to add flavor probe branes to the ABJM theory and discusses in detail the complications that can arise in deriving the field theory. The next sections apply this general procedure to four examples. Section \ref{codzero} is dedicated to codimension-zero flavor, section \ref{codonechiral} is dedicated to codimension-one chiral flavor, section \ref{codonenonchiral} is dedicated to codimension-one non-chiral flavor, and section \ref{codtwo} is dedicated to codimension-two flavor. In section \ref{su4} we discuss $SU(4)$ equivalence. We conclude with some discussion in section \ref{conclusion}. We collect various technical results in four appendices.

\section{Review of the ABJM Theory}
\label{reviewabjm}

In this section we review the ABJM theory \cite{Aharony:2008ug}. In particular, we review its field content, Lagrangian, and symmetries. We also review the type IIB (and type IIA) brane construction of the theory, and its large-$N$ supergravity dual.

\subsection{The Gauge Theory}
\label{abjmtheory}

Let us begin by writing the Lagrangian and reviewing the symmetries of the ABJM theory. The theory is a $U(N_c) \times U(N_c)$ gauge theory with a Chern-Simons term for each gauge group factor. The two Chern-Simons terms have equal but opposite levels, $k$ and $-k$, which we denote by $U(N_c)_{k} \times U(N_c)_{-k}$.

The nicest way to write the Lagrangian is in $\N=2$ superspace. Our conventions are those of ref. \cite{Ivanov:1991fn}. We use a mostly-plus Minkowski metric. The (2+1)-dimensional $\N=2$ supersymmetry algebra includes two Majorana spinors, which we will combine into a single complex spinor\footnote{We can obtain the (2+1)-dimensional  $\N=2$ supersymmetry algebra via dimensional reduction of the (3+1)-dimensional $\N=1$ supersymmetry algebra. $\theta^{\alpha}$ is precisely the single complex spinor of the (3+1)-dimensional $\N=1$ supersymmetry algebra.} $\theta^{\alpha}$, and its complex conjugate $\bar \theta^{\alpha}$, with $\alpha = 1,2$ the spinor index. The superspace covariant derivatives are then
\be
D_{\alpha} = \frac{\partial}{\partial \theta^{\alpha}} + i \left(\gamma^{\mu} \bar{\theta} \right)_{\alpha} \frac{\partial}{\partial x^{\mu}}, \qquad \bar{D}_{\alpha} = - \frac{\partial}{\partial \bar{\theta}^{\alpha}} - i \left( \theta \gamma^{\mu} \right)_{\alpha} \frac{\partial}{\partial x^{\mu}},
\ee
where $\gamma^0 = i \sigma_2$, $\gamma^1 = \sigma_1$, and $\gamma^{2} = \sigma_3$, with $\sigma_1$, $\sigma_2$, and $\sigma_3$ the usual Pauli matrices. A chiral superfield $\phi$ obeys $\bar{D}_{\alpha} \phi = 0$.

The ABJM theory includes the following fields:
\begin{enumerate}
\item Two $\N=2$ vector superfields $V_i$, one for each gauge group, hence $i=1,2$ labels the $U(N_c)$ factor. An $\N=2$ vector superfield includes a vector potential $A_{\mu}$, a real scalar field $\sigma$, two real (Majorana) gauginos, and an auxiliary real scalar field $D$, all in the adjoint representation of the gauge group.
\item Two $\N=2$ chiral superfields $\Phi_i$, each of which is in the adjoint representation. An $\N=2$ chiral superfield includes two real (Majorana) fermions, two real scalars, and a complex auxiliary scalar $F$.
\item Four $\N=2$ chiral superfields, $A_1$, $A_2$, $B_1$ and $B_2$, where $A_1$ and $A_2$ are in the bifundamental $(N_c,\overline{N_c})$ representation and the $B_1$ and $B_2$ are in the anti-bifundamental $(\overline{N_c},N_c)$ representation.
\end{enumerate}
We will divide the action into three pieces,
\be
S_{\text{ABJM}} = S_{\text{CS}} + S_{\text{bifund}} + S_{\text{pot}},
\ee
where, in $\N=2$ superspace,
\ba
S_{\text{CS}} & = & - i \, \frac{k}{4 \pi} \int d^3x \, d^4\theta \int_0^1 dt \, \Tr \left( V_1 \bar{D}^{\alpha} \left( e^{t V_1} D_{\alpha} e^{-t V_1} \right) - V_2 \bar{D}^{\alpha} \left( e^{tV_2} D_{\alpha} e^{-tV_2} \right) \right), \\
S_{\text{bifund}} & = & - \int d^3x \, d^4\theta \, \Tr \left( \bar{A}_a e^{-V_1} A_a e^{V_2} + \bar{B}_a e^{-V_2} B_a e^{V_1} \right),\\
S_{\text{pot}} & = & \int d^3x \, d^2\theta \, W + c.c.,
\ea
with the superpotential
\be
\label{abjmsupo1}
W = - \frac{k}{8\pi} \Tr \left( \Phi_1^2 - \Phi_2^2 \right) + \Tr\left(B_a \Phi_1 A_a \right) + \Tr\left( A_a \Phi_2 B_a \right).
\ee
In $S_{\text{bifund}}$ and the superpotential, summation over $a=1,2$ is implicit. All traces are taken in the fundamental representation. Without the superpotential the action has $\N=2$ supersymmetry. The chiral superfields $\Phi_i$ combine with the corresponding $V_i$ to form $\N=4$ vector multiplets, although the Chern-Simons terms only preserve $\N=3$ supersymmetry. The form of the superpotential is completely fixed by $\N=3$ supersymmetry (see for example ref.  \cite{Gaiotto:2007qi}).

The fields $\Phi_i$ have no kinetic terms, hence at low energy they can be integrated out, yielding the superpotential
\be
\label{abjmsupo2}
W_{\text{ABJM}} = \frac{2 \pi}{k} \varepsilon^{ab} \, \varepsilon^{\dot{a} \dot{b}} \, \Tr \left( A_a B_{\dot{a}} A_b B_{\dot{b}} \right),
\ee
which clearly exhibits an $SU(2)$ symmetry acting on $A_a$ and a separate $SU(2)$ symmetry acting on $B_{\dot{a}}$. We denote this symmetry as $SU(2)_A \times SU(2)_B$. The R-symmetry of the theory, $SO(3)_R \equiv SU(2)_R$, does not commute with the $SU(2)_A \times SU(2)_B$: under the $SU(2)_R$ symmetry,  $(A_1,B_1^*)$ and $(A_2,B_2^*)$ are each a doublet. We thus conclude that the full symmetry is $SU(4)$, under which $(A_1,A_2,B_1^*,B_2^*)$ transforms as a \textbf{4}. As argued in ref. \cite{Aharony:2008ug}, the supercharges also transform under the $SU(4)$, hence the full R-symmetry is $SU(4)_R \equiv SO(6)_R$, and hence the theory is in fact $\N=6$ supersymmetric.

We emphasize that at low energy the supersymmetry is enhanced, where by ``low energy'' we mean energies lower than the mass, $g_{YM}^2 k / (4 \pi)$ (here we use a normalization for the kinetic terms of the vector multiplet with a $1/g_{YM}^2$ in front), of the fields in the $\N=4$ vector multiplet\footnote{When we integrate out the fermions in the vector multiplets, we may worry that the Chern-Simons level will change: the adjoint fermions have the same mass with the same sign within the $U(N_c)$ multiplet, but with the opposite sign of fermions in the other $U(N_c)$, so the Chern-Simons level should be shifted by $\pm N_c$. The massive gauge fields cancel that shift, however \cite{Kao:1995gf}.}. We will see this supersymmetry enhancement again shortly, in the brane construction of the theory.

The theory additionally has a $U(1)_b$ ``baryon number'' symmetry under which $A_i \rightarrow e^{i \alpha} A_i$ and $B_i \rightarrow e^{-i \alpha} B_i$. Remarkably, the theory also has a parity symmetry, which involves inverting one spatial coordinate (say $x^1 \rightarrow - x^1$), exchanging the two gauge groups, and performing charge conjugation on all of the fields.

Finally, as shown in ref. \cite{Aharony:2008ug}, the moduli space of the theory is $\mathbbm{C}^4 / \mathbbm{Z}_k$, where the $\mathbbm{Z}_k$ acts as $(A_1,A_2,B_1^*,B_2^*) \rightarrow e^{2 \pi i/k} (A_1,A_2,B_1^*,B_2^*)$, where here $A_a$ and $B_a$ denote only the scalar component of the corresponding superfields.

\subsection{Type IIB Construction}
\label{typeIIBconstruction}

In this section we review the type IIB brane construction (of ref. \cite{Aharony:2008ug}) leading to the $\N=6$ Chern-Simons-matter theory with gauge group $U(N_c)_k \times U(N_c)_{-k}$ described above. Consider the following brane setup in type IIB string theory
\begin{center}
        \begin{tabular}{|c|cccccccccc|}\hline
                &0&1&2&3&4&5&6&7&8&9\\ \hline
NS5&$\bullet$&$\bullet$&$\bullet$&$\bullet$&$\bullet$&$\bullet$&--&--&--&--\\
NS5$^{\prime}$&$\bullet$&$\bullet$&$\bullet$&$\bullet$&$\bullet$&$\bullet$&--&--&--&--\\
$N_c$ D3&$\bullet$&$\bullet$&$\bullet$&--&--&--&$\bullet$&--&--&--\\
$k$ D5&$\bullet$&$\bullet$&$\bullet$&$\bullet$&$\bullet$&--&--&--&--&$\bullet$\\ \hline
        \end{tabular}
\end{center}
where the $x^6$ direction is a circle. The NS5- and NS5$^{\prime}$-branes are separated in the $x^6$ direction. The $N_c$ D3-branes, which are extended in the $x^6$ direction, break on the NS5-branes. The $k$ D5-branes and the NS5$^{\prime}$-brane are coincident in $x^6$.

The D3-branes, together with the NS5- and NS5$^{\prime}$-branes, give rise to an $\N = 4$ supersymmetric (2+1)-dimensional $U(N_c) \times U(N_c)$ Yang-Mills theory \cite{Hanany:1996ie}. The bosonic part of the $\N=4$ vector multiplet in each $U(N_c)$ gauge group consists of the (2+1)-dimensional components of the D3-brane worldvolume gauge field together with the three real scalars describing each D3-brane's position in the $(x^3,x^4,x^5) \equiv (345)$ directions. Recall from the last subsection that each $\N=4$ vector multiplet consists of an $\N=2$ vector multiplet $V_i$ and an $\N=2$ chiral multiplet $\Phi_i$. The real scalars are the two real scalars in $\Phi_i$ plus the real scalar $\sigma_i$ in $V_i$, which thus form a vector representation of $SO(3)_R$. Similarly, the auxiliary fields $D$ and $F$ form a vector of the R-symmetry.

The theory also has (anti-)bifundamental $\N=2$ chiral multiplets, coming from strings stretched between the two stacks of D3-branes. These are the fields $A_a$ and $B_a$ of the last subsection, with $a=1,2$.

The $k$ D5-branes coincident with the NS5$^{\prime}$-branes introduce massless D3/D5 strings, and break the supersymmetry to $\N=2$. The field theory thus has $k$ massless $\N=2$ chiral multiplets in the fundamental and $k$ massless $\N=2$ chiral multiplets in the anti-fundamental of \textit{each} $U(N_c)$ factor.

What does any of this have to do with Chern-Simons theory? If we can give the fundamental and anti-fundamental fields the same mass, then via the parity anomaly these fields will produce Chern-Simons terms at low energy. More precisely, we need \textit{real} masses of \textit{equal} sign. As argued in ref. \cite{Bergman:1999na} (see also ref. \cite{Kitao:1998mf}), the deformation that produces such masses is to bind the $k$ D5-branes to the NS5$^{\prime}$-brane, producing a $(1,k)$5-brane. To preserve $\N=2$ supersymmetry, the $(1,k)$5-brane must be tilted at an angle $\theta$ in the $(59)$ plane, which we denote by $[5,9]_\theta$. The angle $\theta$ depends on the complex axion-dilaton $\tau = \frac i {g_s} + \chi$ as $\theta = \arg(\tau) - \arg(k+\tau).$ In what follows, we will always set $\tau=i$. Such a deformation actually gives the fundamental and anti-fundamental fields infinite mass. Integrating out these fields then produces Chern-Simons terms with levels $k$ and $-k$ for the two $U(N_c)$ gauge groups. Moreover, we can enhance the supersymmetry to $\N=3$ if we additionally rotate the $(1,k)$5-brane by the same angle $\theta$ in the $(37)$ and $(48)$ planes. We thus arrive at the brane construction
\begin{center}
        \begin{tabular}{|c|cccccccccc|}\hline
                &0&1&2&3&4&5&6&7&8&9\\ \hline
NS5&$\bullet$&$\bullet$&$\bullet$&$\bullet$&$\bullet$&$\bullet$&--&--&--&--\\
$(1,k)$5&$\bullet$&$\bullet$&$\bullet$&$[3,7]_\theta$&$[4,8]_\theta$&$[5,9]_\theta$&--&--&--&--\\
$N_c$ D3&$\bullet$&$\bullet$&$\bullet$&--&--&--&$\bullet$&--&--&--\\\hline
        \end{tabular}
\end{center}

We will henceforth refer to the final brane configuration above as ``the type IIB setup.'' The field theory associated with this setup is an $\N=3$ $U(N_c)_{k} \times U(N_c)_{-k}$ Yang-Mills theory with Chern-Simons terms and four massless bifundamental matter multiplets $(A_a,B_b)$. We saw above that this theory flows in the infrared (meaning energies below $g_{YM}^2k/(4\pi)$) to the $\N=6$ superconformal $U(N_c)_{k} \times U(N_c)_{-k}$ Chern-Simons theory with the same bifundamental matter content. The easiest way to see that happen in the brane setup is to T-dualize and then lift to M-theory.

\subsection{Type IIA and M-theory Descriptions}
\label{typeIIAandM}

If we perform a T-duality along $x^6$ then the type IIB brane setup above turns into the following type IIA configuration: the $N_c$ D3-branes become $N_c$ D2-branes in the $(012)$ directions. The NS5-brane along $(012345)$ becomes a KK monopole associated with the $x^6$ circle. The $(1,k)$5-brane becomes a KK monopole in the $(0123)\, [3,7]_\theta \, [4,8]_\theta \, [5,9]_\theta$ directions associated with the $x^6$ circle. Normally $k$ D5-branes would appear as $k$ D6-branes in type IIA string theory. Here the $k$ D5-branes bound into the $(1,k)$5-brane appear as D6-brane flux on the KK monopole. The configuration in type IIA string theory is thus
\begin{center}
        \begin{tabular}{|c|cccccccccc|}\hline
                &0&1&2&3&4&5&6&7&8&9\\ \hline
$N_c$ D2&$\bullet$&$\bullet$&$\bullet$&--&--&--&--&--&--&--\\
KK monopole&$\bullet$&$\bullet$&$\bullet$&$\bullet$&$\bullet$&$\bullet$&--&--&--&--\\
KK monopole with D6-brane flux&$\bullet$&$\bullet$&$\bullet$&$[3,7]_\theta$&$[4,8]_\theta$&$[5,9]_\theta$&--&--&--&--\\
\hline
        \end{tabular}
\end{center}

We can now lift the configuration to M-theory, introducing a second circle direction, which we will denote $x^{\sh}$. The D2-branes become M2-branes, whereas the KK monopole associated with the $x^6$ circle remains unchanged. Normally a D6-brane would lift to a KK monopole associated with the $x^{\sh}$ circle, hence the KK monopole with D6-brane flux becomes a KK monopole associated with a circle on the $(6,\sh)$ torus. Notice that the two 5-branes in the type IIB picture (\textit{i.e.} the NS5-brane and the $(1,k)$5-brane) lift to pure geometry in M-theory.

The spacetime is now $\mathbbm{R}^{1,2} \times X_8$, where the M2-branes are extended along $\mathbb{R}^{1,2}$ and $X_8$ is the spacetime generated by the KK monopoles. The space $X_8$ preserves $3/16$ of the 32 supersymmetries of M-theory. (Adding the M2-branes (with the right orientation) does not break any additional supersymmetries.) We thus expect the M2-branes' worldvolume theory to have $\N=3$ supersymmetry.

The enhancement of supersymmetry that we saw in the field theory occurs when we take a ``near-horizon'' limit, which we define as follows. At the intersection point of the two KK monopoles, the singularity of the space $X_8$ is locally $\mathbbm{C}^4 / \mathbbm{Z}_k$. Denoting the complex coordinates of $\mathbbm{C}^4$ by $z^i$, the action of the $\mathbbm{Z}_k$ is $z^j \rightarrow e^{2\pi i / k} z^j.$ The ``near-horizon'' limit means retaining only the $\mathbbm{C}^4 / \mathbbm{Z}_k$ singularity of the full $X_8$ space. We will often refer to this as ``zooming in'' on the singularity. $\mathbbm{C}^4 / \mathbbm{Z}_k$ preserves 12 supersymmetries, or $3/8$ of the 32 supersymmetries of M-theory. We write the metric of $X_8$, and take the ``near-horizon'' limit, explicitly in appendix \ref{app:IIBtoM}.

12 real supercharges is of course the correct amount for a (2+1)-dimensional $\N=6$ supersymmetric theory. Recall also that the moduli space of the $\N=6$ Chern-Simons-matter theory is precisely $\mathbbm{C}^4/\mathbbm{Z}_k$.  Furthermore, $\mathbbm{C}^4 \cong \mathbbm{R}^8$ has an $SO(8)$ isometry, of which only $SU(4) \times U(1)$ remains after the $\mathbbm{Z}_k$ orbifold.  These symmetries match the $SU(4)_R \times U(1)_b$ symmetry of the $\N=6$ Chern-Simons theory. The central conclusion of ref. \cite{Aharony:2008ug} was therefore that the $\N=6$ superconformal $U(N_c)_k \times U(N_c)_{-k}$ Chern-Simons matter theory of section \ref{abjmtheory} describes the low-energy dynamics of $N_c$ coincident M2-branes at the $\mathbbm{C}^4 / \mathbbm{Z}_k$ singularity.

Recalling that, in the field theory, the $\mathbbm{Z}_k$ acts on the bifundamentals as $(A_1,A_2,B_1^*,B_2^*) \rightarrow e^{2 \pi i / k } (A_1,A_2,B_1^*,B_2^*)$, and also that they transform as a \textbf{4} of $SU(4)_R$, we can (roughly) identify $(z^1,z^2,z^3,z^4)$ with $(A_1,A_2,B_1^*,B_2^*)$, where here $A_a$ and $B_a$ represent the bosonic components of the corresponding superfields. The $U(1)_b$ symmetry of the field theory thus appears as a phase shift $z^i \rightarrow e^{i \alpha} z^i$ (which is equivalent to shifts in the $x^{\sh}$ circle, as we show in appendix \ref{app:IIBtoM}).

\subsection{The Dual Gravity Theory}
\label{dualgravity}

Consider $N_c$ M2-branes at the $\mathbbm{C}^4/\mathbbm{Z}_k$ singularity. If we take $N_c \ra \infty$, we can replace the M2-branes with their near-horizon geometry, $AdS_4 \times S^7/\mathbbm{Z}_k$. The natural conjecture then is that eleven-dimensional supergravity on $AdS_4 \times S^7/\mathbbm{Z}_k$ is holographically dual to the $\N=6$ supersymmetric $U(N_c)_k \times U(N_c)_{-k}$ Chern-Simons-matter theory at large $N_c$. The $AdS_4$ radius of curvature $R$ is related to the 't Hooft coupling $\lambda = N_c / k$ and the Chern-Simons level $k$ as (with $\ell_p$ the eleven-dimensional Planck length),
\begin{equation}
\frac{R^3}{\ell_p^3} = 4 \pi \sqrt{2kN_c} = 4 \pi k \sqrt{2 \lambda}.
\end{equation}
We can thus trust the M-theory description in the strong 't Hooft coupling limit $\lambda \rightarrow \infty$. If we write the $S^7$ as a circle fibration over $\mathbbm{CP}^3$, then the $\mathbbm{Z}_k$ orbifold acts on the fiber direction. The radius of the circle in Planck units is on the order of $R/k\ell_p \propto (N_ck)^{1/6}/k$, so we can only trust the solution when $N_c \gg k^5$. In short, when $N_c \ra \infty$ such that $N_c \gg k^5$ (which implies $\lambda = N_c/k \ra \infty$), the $\N=6$ supersymmetric $U(N_c)_k \times U(N_c)_{-k}$ Chern-Simons-matter theory is dual holographically to eleven-dimensional supergravity on $AdS_4 \times S^7/\mathbbm{Z}_k$.

When $k^5 \gg N_c \gg k$, where again $\lambda \ra \infty$, the circle becomes small and the appropriate description is in terms of type IIA supergravity on the spacetime $AdS_4 \times \cp$.

\section{General Analysis of Probe Flavor}
\label{generalanalysis}

In this section we discuss how to add flavor to the ABJM theory in general terms. We first discuss the gravity analysis and then the field theory analysis. More specifically, we explain in this section exactly what we compute on the gravity side and what we want to compute on the field theory side.

Our general approach is to add flavor branes in the type IIB setup and follow what happens to them, in both the gravity and field theory descriptions, in the construction of the ABJM theory (T-duality, lift to M-theory, etc.). Why start with type IIB? The main reason is that the brane description in type IIB provides an easy starting point for constructing the field theory.

\subsection{Gravity Analysis}
\label{gravityanalysis}

On the gravity side, we introduce flavor branes in the type IIB setup. To limit our search for supersymmetric probes, we impose four constraints. First, we consider only D1-, D3-, D5- and D7-branes. D(-1)-branes do not introduce flavor degrees of freedom on the D3-brane worldvolume, and D9-branes are unstable without orientifold planes, so we will not consider these cases. Second, we do not separate any probes from the D3-branes in overall transverse directions. Third, when we consider multiple probes, \textit{i.e.} $N_f > 1$, we do not separate them from each other, so that they retain a $U(N_f)$ symmetry. Fourth, we consider only probes aligned along the coordinate axes. More generally the probe brane could be at an angle with respect to these axes. We studied a few special cases of probes at angles (see appendix \ref{app:IIBprobesusy}) and found that all such probes appeared to preserve as much as, or less, supersymmetry, as the probes listed below, \ie they never exhibit enhanced supersymmetry.

The counting of supercharges left unbroken by our probes in this background is a straightforward exercise, the details of which appear in appendix \ref{app:IIBprobesusy}. The main result of appendix \ref{app:IIBprobesusy} is table \ref{table:IIB}, which appears below. Table \ref{table:IIB} lists the flavor Dp-branes we study, exactly where they are located in the type IIB setup, and the number of real supercharges each Dp-brane preserves. Although for specific calculations we focused on the Dp-branes listed in table \ref{table:IIB}, most of our comments in this section will be applicable more generally.

A very important fact (mentioned in appendix \ref{app:IIBprobesusy}) is that when $k=0$, such that the type IIB setup includes just NS5-branes and no $(1,k)$5-brane, all of our flavor branes preserve 4 real supercharges, except for two cases that preserve 8 supercharges. The first case preserving 8 supercharges is D3-branes along (0126), which are of course coincident with the D3-branes whose low-energy dynamics we are interested in. The second case preserving 8 supercharges is D5-branes along (012789), which were first studied in ref. \cite{Hanany:1996ie}. For all cases, table \ref{table:IIB} indicates the number of supercharges that remain unbroken after forming the $(1,k)$5-brane.

The first column of table \ref{table:IIB} lists the type of brane in the type IIB construction while the second column lists the resulting type IIA description, obtained by T-dualizing in $x^6$, and the third column lists the M-theory description, obtained by lifting to eleven dimensions. A type IIB D-brane that becomes a D6-brane in type IIA will lift to a KK monopole associated with the M-theory circle, which we have indicated with ``KK.'' The fourth column lists the codimension of the defect to which the flavor fields will be confined in the (2+1)-dimensional Chern-Simons-matter theory. The fifth column indicates the directions in which the probe brane is extended in the IIB construction. The $SO(3)$ symmetry that acts simultaneously on the directions $(345)$ and $(789)$ gives rise to other supersymmetric branes, related to the ones in the table by $SO(3)$ transformations. We have indicated this by $()$. For example, the first brane could extend along $(07)$, $(08)$ or $(09)$. The last two columns of the table indicate the number of real supercharges preserved by the probe brane or anti-brane. Recall that for codimension-zero branes the number of preserved supercharges must be even, but for higher codimension the brane may preserve an odd number of real supercharges.

\begin{table}[h!]
\begin{center}
\begin{tabular}{|c|c|c|c|c|c|c|}
  \hline
  Type IIB & Type IIA & M theory & codim & wrapping & SUSY & SUSY (anti)\\ \hline \hline
  D1 & D2 & M2 & 2 & 0(7) & 2 & 2  \\\hline
  D3 & D2 & M2 & 0 & 0126 & 6 & 0 \\\hline
  D3 & D4 & M5 & 1 & 01(37) & 3 & 3 \\\hline
  D3 & D4 & M5 & 1 & 01(38) & 2 & 2\\\hline
  D3 & D2 & M2 & 2 & 0(34)6 & 2 & 2\\\hline
  D3 & D2 & M2 & 2 & 06(78) & 2 & 2\\\hline
  D5 & D6 & KK & 0 & 012(347) & 2 & 2 \\\hline
  D5 & D6 & KK & 0 & 012(349) & 4 & 2\\\hline
  D5 & D6 & KK & 0 & 012789 & 6 & 0\\\hline
  D5 & D4 & M5 & 1 & 013456 & 3 & 3\\\hline
  D5 & D4 & M5 & 1 & 01(378)6 & 2 & 2\\\hline
  D5 & D4 & M5 & 1 & 01(389)6 & 3 & 3\\\hline
  D5 & D6 & KK & 2 & 0(34)789 & 2 & 2\\\hline
  D7 & D6 & KK & 0 & 0126(3478) & 2 & 4\\\hline
  D7 & D6 & KK & 0 & 0126(3479) & 2 & 2\\\hline
  D7 & D8 & M9 & 1 & 01345789 & 3 & 3\\\hline
\end{tabular}
\end{center}
\caption{List of D-branes (extended along the coordinate axes) that we can add to the type IIB construction while still preserving some supersymmetry. For more details, see the accompanying paragraph.}\label{table:IIB}
\end{table}

As reviewed in section \ref{typeIIBconstruction}, to go from the type IIB setup to M2-branes on $\mathbbm{C}^4/\mathbbm{Z}_k$, we T-dualize in $x^6$, lift to M-theory, and take the ``near-horizon'' limit. We can easily determine what type of object the flavor Dp-branes become in M-theory: we obtain M2-, M5-, and M9-branes or KK monopoles. More difficult to determine is the exact position of the object on $\mathbbm{C}^4/\mathbbm{Z}_k$. To find that, we take the straightforward approach. We compute explicitly the coordinate transformations from the type IIB coordinates to the coordinates $(z^1,z^2,z^3,z^4)$ of $\mathbbm{C}^4/\mathbbm{Z}_k$. We present the details of the computation in appendix \ref{app:IIBtoM}. Given the embedding of a Dp-brane in type IIB, we can then immediately write the embedding of the corresponding object in M-theory on $\mathbbm{C}^4/\mathbbm{Z}_k$.

Once we know the location of the M-brane or KK monopole in $\mathbbm{C}^4/\mathbbm{Z}_k$, we can compute the amount of supersymmetry and the isometries that the object preserves. The details of those calculations, for a subset of our examples, appear in appendix \ref{app:probeflat}. Our results are summarized in table \ref{table:flat} in appendix \ref{app:probeflat}. The locations of our objects on $\mathbbm{C}^4/\mathbbm{Z}_k$ are more complicated to explain, however, so we will not reproduce table \ref{table:flat} here. We also studied a few examples of M-branes or KK monopoles in the near-horizon geometry of very many M2-branes, $AdS_4 \times S^7/\mathbbm{Z}_k$. The details of those calculations appear in appendix \ref{app:probeads}. Knowing what symmetries the object preserves is, of course, extremely helpful when constructing the dual field theory.

In our analysis of objects on $\mathbbm{C}^4/\mathbbm{Z}_k$, we make use of a helpful tool, originally used in ref. \cite{Hikida:2009tp}, which we call ``$SU(4)$ equivalence.'' The basic idea is that two different Dp-branes in type IIB can become the same object in M-theory on $\mathbbm{C}^4/\mathbbm{Z}_k$. More specifically, two Dp-branes of different dimensionality and/or located in different places in the type IIB setup (and hence possibly preserving different symmetries) can actually become the same object in M-theory. At work here is the ``near-horizon'' limit, which ``erases'' many of the details of the type IIB embedding.\footnote{That of course was an essential feature in the brane construction of the ABJM theory: the ``erasure'' produced the (super)symmetry enhancement.} To be still more precise, the two Dp-branes can become the same type of object, two M5-branes for example, but located in two different places, \textit{i.e.} with two different embeddings into $\mathbbm{C}^4/\mathbbm{Z}_k$. If we can rotate one object into the other via an $SU(4)$ isometry, however, then the two objects are physically equivalent. We may thus work with either one, and any physical results will be valid for both. On a technical level, some things may be easier to calculate for one embedding than for the other, for example the calculation of the number of preserved supercharges. We will present some explicit examples of $SU(4)$ equivalence, and discuss its field theory meaning, in section \ref{su4}.

\subsection{Field Theory Analysis}
\label{fieldtheoryanalysis}

Eleven-dimensional supergravity on $AdS_4 \times S^7/\mathbbm{Z}_k$ is dual to the ABJM theory with Chern-Simons level $k$, and $N_c$ large (such that $N_c \gg k^5$). What is the dual field theory when we add one of our flavor M-branes or KK monopoles, however? If the object preserves a large amount of symmetry, then that symmetry may be enough to determine the form of the field theory action. That will not always be the case, of course, so we want a more general method to determine the field theory. We will now describe a general ``recipe,'' one that is actually very straightforward and, in principle at least, is guaranteed to give the correct field theory for any flavor Dp-brane in the type IIB setup. Our recipe actually begins a few steps ``before'' the type IIB setup. We begin with D3-branes alone (so no NS5- or $(1,k)$5-branes) and flavor Dp-branes. The recipe then consists of four steps, as follows.

\bigskip

\textbf{Step 1: Construct the D3/Dp Theory}

In type IIB consider D3-branes alone in flat space, so let $x^6$ be non-compact and remove the NS5- and $(1,k)$5-branes. We then add supersymmetric flavor Dp-branes. In general, we next need to determine the low-energy theory ``living'' on the D3-branes, including the couplings to the (defect) flavor fields. We will generically call that theory ``the D3/Dp theory.'' Fortunately, for many examples the D3/Dp theory is already known. The following table lists various D3/Dp systems for which the field theory has been determined explicitly. The first column indicates the D3/Dp system. The second column indicates the number of Neumann-Dirichlet (ND) directions. The third column indicates the dimension of the intersection (the subspace of the D3-brane worldvolume in which the flavor fields propagate). The fourth column lists references in which the D3/Dp theory is written explicitly. All of the systems listed preserve 8 real supercharges. To our knowledge, as of this writing the table below represents a complete list of D3/Dp systems for which the field theories have been written explicitly in the published literature.

\begin{center}
        \begin{tabular}{|c|c|c|c|}\hline
        D3/Dp&$\#$ ND & Intersection&Reference(s)\\ \hline
D3/D7&4&(3+1)&\cite{Erdmenger:2007cm}\\ \hline
D3/D5&4&(2+1)&\cite{DeWolfe:2001pq,Erdmenger:2002ex}\\ \hline
D3/D3&4&(1+1)&\cite{Constable:2002xt}\\ \hline
D3/D7&8&(1+1)&\cite{Harvey:2007ab,Buchbinder:2007ar,Harvey:2008zz}\\ \hline
D3/D5&8&(0+1)&\cite{Gomis:2006sb}\\ \hline
        \end{tabular}
\end{center}

\noindent Every Dp-brane in table \ref{table:IIB} is described by one of the theories above, except for the D3/D1 system. Recall that if the D3/Dp intersection has 4 Neumann-Dirichlet (ND) directions then the corresponding flavor fields (from 3-p and p-3 strings) will produce non-chiral flavor, simply because the fields are arranged in hypermultiplets \cite{Johnson:2003gi}, whereas with 8 ND directions we can obtain chiral flavor, as occurs for the 8 ND D3/D7 intersection \cite{Harvey:2007ab,Buchbinder:2007ar,Harvey:2008zz}.

\bigskip

\textbf{Step 2: Add the NS5-branes}

Now we ask what happens when we construct the ABJM theory from the D3-branes. First we introduce the NS5- and NS5$^{\prime}$-branes along $(012345)$ and separated in $x^6$ (which for now is still non-compact), and let the D3-branes end on them in $x^6$. From a field theory point of view, adding the NS5-branes has two effects. The first effect is that on the D3-brane worldvolume the $x^6$ direction is now finite in extent, so the low-energy effective theory on the D3-brane worldvolume will be a (2+1)-dimensional field theory. In other words, we must perform a dimensional reduction in the $x^6$ direction. The (3+1)-dimensional $\N=4$ multiplet decomposes into two (2+1)-dimensional $\N=4$ multiplets, a vector multiplet and a hypermultiplet. The second effect of the NS5-branes is to impose boundary conditions that ``kill'' (\ie set to zero) the adjoint (2+1)-dimensional $\N=4$ hypermultiplet \cite{Hanany:1996ie,Gaiotto:2008sa}. We will call these ``the NS5-brane boundary conditions.'' We must thus take the D3/Dp action we wrote in Step 1 and perform a dimensional reduction in $x^6$ and then determine what couplings remain after we impose the NS5-brane boundary conditions.

For this procedure, a crucial distinction is whether the flavor Dp-brane is extended in $x^6$ or not. If not, then we need only dimensionally reduce and impose boundary conditions on the fields of the (3+1)-dimensional $\N=4$ supersymmetric Yang-Mills theory. If the flavor Dp-brane is extended in $x^6$, then we must also perform a dimensional reduction and impose boundary conditions on the flavor fields. In this paper we study examples in which we can avoid doing these operations explicitly.

We will also mention an alternative, but entirely equivalent, way to perform Step 2,\footnote{We thank Ingo Kirsch for mentioning this alternative approach to us.} namely to perform two T-dualities, one along $x^6$ and another along one of the directions 3, 4 and 5 (along the NS5-branes but transverse to the D3-branes). Strictly speaking, here we must assume that $x^6$ is compact, and that we have two stacks of D3-branes, giving rise to two $U(N_c)$ gauge groups, as in the type IIB construction of the ABJM theory. The NS5-branes ultimately become the orbifold space $\mathbbm{C}^2/\mathbbm{Z}_2 \times \mathbbm{C}$, the D3-branes become D3-branes located at the orbifold singularity, and the flavor branes become some Dq-branes (with $q=p$, $p+2$, or $p-2$), which may be wrapping some part of $\mathbbm{C}^2/\mathbbm{Z}_2 \times \mathbbm{C}$ \cite{Karch:1998yv,Karch:1998uy,Park:1999eb}. We can then use well-known machinery for studying D-branes on orbifolds (see refs. \cite{Karch:1998yv,Karch:1998uy,Park:1999eb,Gimon:1996rq} and references therein) to determine the field theory.

\bigskip

\textbf{Step 3: Compactify $x^6$, form the $(1,k)$5-brane, and lift to M-theory}

Now we compactify $x^6$ and add another stack of D3-branes, so that the gauge group of the D3-branes' worldvolumes is $U(N_c)\times U(N_c)$. If the flavor Dp-brane is localized in $x^6$, then in this paper we will always introduce two stacks of flavor Dp-branes, each with $N_f$ Dp-branes, located at opposite sides of the $x^6$ circle, away from the NS5-branes. We will thus obtain open strings stretched from each stack of Dp-branes to the corresponding stack of D3-branes, and hence we obtain massless fields in the fundamental representation of each gauge group factor. We then introduce the $k$ D5-branes along $(012349)$, bind them to the NS5$^{\prime}$-brane to form a $(1,k)$5-brane, and then rotate the $(1,k)$5-brane. None of these operations affect the form of the action in our flavor sector: they correspond to adding \textit{additional} flavor fields, which then acquire mass terms and are integrated out, producing the Chern-Simons terms. The action in \textit{our} flavor sector (\textit{i.e.} the coupling to adjoint fields, coming from 3-p and p-3 strings) is unchanged\footnote{We can make a more direct argument for why these operations do not affect our flavor action, for flavor Dp-branes not extended along $x^6$. We can start with the D3/Dp intersection and immediately add an NS5-brane and the $(1,k)$5-brane. Once again, we first do a dimensional reduction to (2+1) dimensions. We then impose a boundary condition for the NS5-brane and a separate boundary condition for the $(1,k)$5-brane. Together these set to zero the $\N=4$ hypermultiplet and introduce a Chern-Simons term \cite{Hanany:1996ie,Gaiotto:2008sa,Kitao:1998mf}. In these operations, the only changes in the flavor sector are the same that occur with just NS5-branes: some couplings are eliminated when the boundary conditions set adjoint fields to zero. Otherwise the action in the flavor sector does not change. For flavor Dp-branes extended along $x^6$, more work may be required to determine the effect of the $(1,k)$5-brane boundary condition on the flavor fields, along the lines of ref. \cite{Gaiotto:2008sa}.}. We then T-dualize to type IIA and lift to M-theory. The action in the flavor sector is unchanged in those two steps. In particular, notice that the symmetries will be unchanged. We thus arrive in M-theory on the manifold $X^8$ mentioned in section \ref{typeIIAandM}.

\bigskip

\textbf{Step 4: Take the low-energy limit}

In the supergravity description, the last step is to ``zoom in'' on the $\mathbbm{C}^4/\mathbbm{Z}_k$ singularity of $X^8$, which appears in the field theory description as a low-energy limit. More precisely, we are doing effective field theory: we want to write a theory valid on scales below the mass of the $\N=4$ vector multiplet, $g_{YM}^2 k / (4\pi)$. Following the rules of effective field theory, in the low-energy action we must write all terms consistent with the symmetries, which in particular means supersymmetry and R-symmetry. If we can determine the coefficients of these terms (using for example supersymmetry), then the action we obtain is the correct action of the theory. Furthermore, as in the ABJM theory, to determine whether a (super)symmetry enhancement occurred, a helpful step is to integrate out the fields $\Phi_i$. We emphasize that integrating out the $\Phi_i$ does not change the theory, however. The equations of motion for the $\Phi_i$'s are simply algebraic constraints: the theory already has whatever symmetry it has before we formally integrate out the $\Phi_i$'s.

\bigskip

Our recipe has advantages and disadvantages. Let us first consider the advantages. One advantage is the fact that, in principle at least, our recipe is guaranteed to produce the correct field theory. Another advantage is the fact that the input for our recipe is a known D3/Dp theory, that is, our recipe is a kind of ``machine'' that takes a known D3/Dp theory and outputs the field theory for flavor fields coupled to the ABJM theory. Notice also that the action we obtain in the flavor sector will generally be valid for all values of $N_c$ and $k$, although we will primarily be interested in the limits where gauge-gravity duality is under best control (such as $N_c \gg k^5$).

Now let us consider some disadvantages. Although in principle our recipe is guaranteed to work, in practice some of the steps can be difficult. Indeed, having studied many of the Dp-branes listed in table \ref{table:IIB}, we can say from experience that Steps 2 and 4 often present technical challenges, especially in cases where the flavor fields are confined to a defect.

In Step 2 for example, for defect flavor fields, na\"ively imposing the NS5-brane boundary conditions often leaves us with field content that does not easily fit into simple representations of the defect's supersymmetry group. (We know what supersymmetry the system should have from the gravity analysis.) In such cases, a more rigorous analysis of supersymmetry-preserving boundary conditions, along the lines of ref. \cite{Gaiotto:2008sa}, may be required. We will see an example of this in section \ref{codonenonchiral}.

As for Step 4, several special issues arise. Step 4 often requires careful analysis of supersymmetric non-renormalization theorems. In the ABJM construction (without flavor), we begin with an $\N=3$ supersymmetric Yang-Mills-Chern-Simons-matter theory. An important feature of $\N=3$ supersymmetry is that the action is fully determined by the symmetry \cite{Gaiotto:2007qi}. That means that the low-energy limit consists only of discarding the kinetic terms for the $\N=4$ vector multiplet (while leaving the Chern-Simons terms). The action cannot change otherwise, for example the superpotential cannot acquire new terms, and the K\"ahler potential cannot be renormalized. When we add defect flavor, however, the Lorentz symmetry of the ABJM theory is broken to the subgroup that leaves the defect invariant, and the amount of supersymmetry is also reduced. In such cases a prerequisite for Step 4 is to re-examine non-renormalization theorems for defect theories. For the defect field theories corresponding to the D3/D5 and D3/D3 intersections ((3+1)-dimensional $\N=4$ SYM with defect flavor), proofs of non-renormalization appear in refs. \cite{Erdmenger:2002ex,Constable:2002xt}. Our examples in sections \ref{codzero} and \ref{codonechiral} have enough symmetry to avoid this issue.

Step 4 also involves integrating out the fields $\Phi_i$. For codimension-zero flavor fields that is usually straightforward. If our flavor fields are codimension one or two, however, this procedure is more difficult. In particular, we would need to decompose the (2+1)-dimensional fields of the ABJM theory into lower-dimensional multiplets, and then integrate out the lower-dimensional fields corresponding to the $\Phi_i$.

A useful strategy for Step 4 is to ``work backwards,'' that is, to use symmetries of the \textit{gravity} description to guess the final result. In other words, given the symmetries on the gravity side, we can write all possible terms consistent with \textit{those} symmetries in the field theory. In cases where a symmetry enhancement occurs, we must demonstrate that these are all the terms allowed by the \textit{original} symmetry, so that we ``retroactively'' justify the result.

Lastly, let us explain the field theory meaning of $SU(4)$ equivalence. On the gravity side, $SU(4)$ equivalence was the statement that two Dp-branes in the type IIB setup, which may be located in different places or even have different dimensionality (but which must have the same codimension in (2+1) dimensions) become the same type of object in M-theory on $\mathbbm{C}^4/\mathbbm{Z}_k$, where the embeddings of the the two objects are related by an $SU(4)$ isometry transformation. In the field theory what is happening is simply that two different theories, with different symmetries for example (including possibly different amounts of supersymmetry), flow to the same low-energy fixed point in Step 4. We will discuss that further, and present an explicit example, in section \ref{su4}.

To illustrate various features of our recipe, we now turn to several examples.

\section{Codimension Zero $\N=3$ Supersymmetric Flavor}
\label{codzero}

In this section we study codimension-zero $\N=3$ supersymmetric flavor fields, which have already been studied in refs. \cite{Hohenegger:2009as,Gaiotto:2009tk,Hikida:2009tp}. For us this section serves as a particularly nice illustration of our general recipe. Compared to other examples, however, this example lacks many interesting features. For example, no supersymmetry enhancement occurs in the ``near-horizon'' limit, as we will review.

\subsection{Supergravity with KK-monopoles/D6-branes}
\label{codzerogravity}

To obtain codimension-zero $\N=3$ supersymmetric flavor, we follow refs. \cite{Hohenegger:2009as,Gaiotto:2009tk,Hikida:2009tp}, and add D5-branes extended along (012789) in the type IIB setup. We are free to choose their position on the $x^6$ circle. We will add two stacks of D5-branes, each with $N_f$ coincident D5-branes, on opposite sides of the circle, away from the NS5-brane and $(1,k)$5-brane. The strings from the D5-branes to the two stacks of D3-branes thus introduce massless flavor in both gauge groups. As shown in table \ref{table:IIB} in section \ref{gravityanalysis}, these D5-branes preserve 6 real supercharges in the type IIB setup.

After T-duality in $x^6$ the D5-branes become D6-branes. The $2 N_f$ D6-branes are coincident, and have a $U(2N_f)$ symmetry broken to $U(N_f) \times U(N_f)$ by a $\mathbbm{Z}_2$-valued Wilson line, as explained in ref. \cite{Hikida:2009tp}. (The Wilson line simply tells us where the D5-branes were in type IIB.) After uplift to M-theory and the ``near-horizon'' limit, the D6-branes become KK monopoles associated with the $x^{\sh}$ circle in M-theory on $\mathbbm{C}^4/\mathbbm{Z}_k$. The authors of ref. \cite{Hikida:2009tp} argue that the embedding of the KK monopole is described by the equations $z^1 =\bar{z}^3$, $z^2 =\bar{z}^4$ in $\mathbbm{C}^4/\mathbbm{Z}_k$. The authors of ref. \cite{Hikida:2009tp} then showed that, by using the $SU(4)$ symmetry of $\mathbbm{C}^4/\mathbbm{Z}_k$, we can map this embedding to Im$(z^i)=0, \forall i$. In other words, the two embeddings are $SU(4)$ equivalent. The symmetries preserved by the KK monopoles are easier to see in the latter embedding, however. In the latter embedding, the KK monopoles are extended along (012) and Re$(z^i), \forall i$.

The circle direction associated with the KK monopoles corresponds to the $U(1)_b$ symmetry of the background, so the KK monopoles preserve this symmetry. The KK monopoles break the $SU(4)$ symmetry to an $SO(4)$ under which $(z^1,z^2,z^3,z^4)$ transforms as a \textbf{4}. The total symmetry group that the KK monopoles preserve is $SO(4)\times U(1)_b = SU(2) \times SU(2) \times U(1)_b$ \cite{Hohenegger:2009as,Gaiotto:2009tk,Hikida:2009tp}. In appendix \ref{app:probeflat} we find that the KK monopoles preserve 6 real supercharges.

If we take $N_c \rightarrow \infty$, we can replace the M2-branes by their near-horizon geometry, which is $AdS_4 \times S^7/\mathbbm{Z}_k$. The KK monopoles are extended along $AdS_4$ and wrap a three cycle in $S^7/\mathbbm{Z}_k$. In appendix \ref{app:probeads} we analyze the $\kappa$-symmetry condition for these monopoles and find that after the near-horizon limit the number of preserved supercharges has doubled to 12.

For large $k$, such that $k^5 \gg N_c$ and the appropriate description is type IIA on $AdS_4 \times \cp$, the monopoles become D6-branes wrapping $AdS_4 \times \mathbbm{RP}^3$ \cite{Hohenegger:2009as,Gaiotto:2009tk,Hikida:2009tp}.

To summarize: as explained in refs. \cite{Hohenegger:2009as,Gaiotto:2009tk,Hikida:2009tp}, in the type IIB setup we can add D5-branes that produce fundamental matter for both gauge group factors. These become KK monopoles in M-theory on $\mathbbm{C}^4/\mathbbm{Z}_k$. These KK monopoles preserve 6 real supercharges, so we expect a dual field theory with $\mathcal{N}=3$ superconformal symmetry. The corresponding R-symmetry group has to be $SO(3)$, which fits into the symmetry found above. We now proceed to review the dual field theories constructed in refs. \cite{Hohenegger:2009as,Gaiotto:2009tk,Hikida:2009tp}, and check that it has the right symmetries and amount of supersymmetry.

\subsection{The Field Theory}
\label{codzerofieldtheory}

We will first review the theory described in refs. \cite{Hohenegger:2009as,Gaiotto:2009tk,Hikida:2009tp}, and then ``re-derive'' it using our recipe.

In the type IIB setup we introduce two stacks of $N_f$ coincident D5-branes along $(012789)$ on opposite sides of the $x^6$ circle. These D5-branes preserve the $\N=3$ supersymmetry of the type IIB setup (see appendix \ref{app:IIBprobesusy}). The strings stretched between each stack of D3-branes and each stack of D5-branes will produce $\N=2$ chiral superfields transforming in the $U(N_f)$ and $U(N_c)$ representations $(\bar N_f, N_c)$ and $(N_f, \bar N_c)$ of each $U(N_c)$. We will denote these as $Q_i$ and $\tilde{Q}_i$, respectively, where again $i$ labels the gauge group, $i=1,2$. The field $Q_1$, for example, transforms in the $\bar{N}_f$ representation of $U(N_f)$ and the $N_c$ representation of the ``first'' ($i=1$) $U(N_c)$ gauge group, while $\tilde{Q}_1$ transforms in the conjugate representations, $N_f$ and $\bar{N}_c$. For notational simplicity, we will suppress flavor indices.

$\N=3$ supersymmetry completely determines the action \cite{Gaiotto:2007qi}. The kinetic terms of the flavor fields are
\be
\label{eq:D7kineticterms}
S_{\text{fund}} = - \int d^3x \, d^4 \theta \left( \bar{Q}_i e^{-V_i} Q_i + \tilde{Q}_i e^{V_i} \bar{\tilde{Q}}_i \right).
\ee
Here we have left summation over $i$ implicit. The superpotential now has extra terms,
\be
\label{eq:D7superpotentialbeforeintegratingout}
W = - \frac{k}{8 \pi} \Tr ( \Phi_1^2 - \Phi_2^2 ) + \Tr (B_a \Phi_1 A_a) + \Tr (A_a \Phi_2 B_a) + \tilde{Q}_1 \Phi_1 Q_1  - \tilde{Q}_2 \Phi_2 Q_2.
\ee
At low energy we again integrate out $\Phi_1$ and $\Phi_2$, which gives
\begin{align}
W & =  \frac{2 \pi}{k} \Tr \left[ (A_aB_a + Q_1 \tilde Q_1)^2  - (B_aA_a - Q_2 \tilde Q_2)^2 \right] \, .\label{eq:D7superpotential}
\end{align}

Now let us derive the action above using our recipe.

\bigskip

\textbf{Step 1: Construct the D3/D5 Theory}

We return to type IIB and consider D3-branes alone in flat space, so for now let $x^6$ be non-compact and remove the NS5- and $(1,k)$5-brane. We then add $N_f$ flavor D5-branes, which intersect the D3-branes in (2+1) dimensions. This D3/D5 intersection has 4 ND directions and preserves 8 real supercharges.

The D3/D5 theory was constructed in refs. \cite{DeWolfe:2001pq,Erdmenger:2002ex}. In the flavor sector we have two $\N=2$ chiral superfields (which comprise an $\N=4$ hypermultiplet), which of course propagate only in (2+1) dimensions. In the adjoint sector, we start with the theory on the D3-branes, (3+1)-dimensional $\N=4$ SYM theory. The (3+1)-dimensional $\N=4$ multiplet decomposes into two (2+1)-dimensional $\N=4$ multiplets, a vector multiplet and a hypermultiplet. The (2+1)-dimensional $\N=4$ vector multiplet then further decomposes into an $\N=2$ vector multiplet and an $\N=2$ chiral multiplet. The kinetic term for the flavor fields is then precisely the one above, \ie the flavor fields have the standard coupling to the $\N=2$ vector superfield. The superpotential is also precisely the one above (at least, the terms  involving the flavor fields are the same), \ie a coupling to the adjoint $\N=2$ chiral superfield from the $\N=4$ vector multiplet. (See for example eq. (4.7) in ref. \cite{Erdmenger:2002ex}.) The entire action preserves $\N=4$ supersymmetry, that is, 8 real supercharges. We emphasize that the flavor fields do not couple to the (2+1)-dimensional $\N=4$ hypermultiplet at all. 

\bigskip

\textbf{Step 2: Add the NS5-branes}

We add the NS5-brane and NS5$^{\prime}$-brane along $(012345)$ and separated in $x^6$ (which for now is still non-compact), and let the D3-branes end on them in $x^6$. We first perform a dimensional reduction in $x^6$, which does not affect the flavor action in this case (since it is already (2+1)-dimensional). The NS5-brane boundary conditions set to zero the (2+1)-dimensional $\N=4$ hypermultiplet. As we mentioned, however, the flavor fields do not couple to the $\N=4$ hypermultiplet, so this step actually has no effect on the action in the flavor sector. The theory retains $\N=4$ supersymmetry.

\bigskip

\textbf{Step 3: Compactify $x^6$, form the $(1,k)$5-brane, and lift to M-theory}

Now we compactify $x^6$ and add another stack of D3-branes, so that the gauge group of the D3-branes' worldvolumes is $U(N_c)\times U(N_c)$, and another stack of $N_f$ D5-branes. We thus obtain two sets of flavor fields, the $Q_i$ and $\tilde{Q}_i$, with $i=1,2$, mentioned above, and correspondingly, two copies of the defect action. As we argued in section \ref{generalanalysis}, forming the $(1,k)$5-brane and lifting to M-theory does not change the defect action. The action acquires Chern-Simons terms, however, which break the supersymmetry to $\N=3$.

\bigskip

\textbf{Step 4: Take the low-energy limit}

Lastly, we must take the low-energy limit, which means writing all terms consistent with the symmetries of the field theory. Our theory has $\N=3$ supersymmetry. As mentioned above, $\N=3$ supersymmetry completely determines the form of the action \cite{Gaiotto:2007qi}. The flavor action thus remains the same, and hence we arrive at eqs. (\ref{eq:D7kineticterms}) and (\ref{eq:D7superpotentialbeforeintegratingout}). The very last step is to integrate out the $\Phi_i$, as we did above, the result being eq. (\ref{eq:D7superpotential}).

\bigskip

Once we have the result for the field theory action, we must ask whether any symmetry enhancement occurred in the low-energy limit, as happened in the ABJM theory without flavor.

Inspecting the superpotential above, we can see that the theory retains the $U(1)_b$ ``baryon number'' symmetry under which $A_a \rightarrow e^{i \alpha} A_a$ and $B_a \rightarrow e^{-i \alpha} B_a$. The theory additionally has a global $U(N_f) \times U(N_f)$ flavor symmetry, of which the overall, diagonal $U(1)$ (usually also called ``baryon number'') acts as $Q_i \rightarrow e^{-i \beta} Q_i$ and $\tilde{Q}_i \rightarrow e^{i \beta} \tilde{Q}_i$.

From the superpotential in eq. (\ref{eq:D7superpotential}) we can also see that codimension-zero flavor breaks the $SU(2)_A \times SU(2)_B$ symmetry to the diagonal subgroup that leaves invariant the product of fields $A_a B_a$. If we perform $SU(2)_A$ and $SU(2)_B$ transformations,
\be
\begin{pmatrix} A_1 \\ A_2 \end{pmatrix} \ra e^{i m_j \sigma_j} \begin{pmatrix} A_1 \\ A_2 \end{pmatrix}, \qquad \begin{pmatrix} B_1 \\ B_2 \end{pmatrix} \ra e^{i n_j \sigma_j} \begin{pmatrix} B_1 \\ B_2 \end{pmatrix} \nonumber
\ee
where $m_j$ and $n_j$ are the parameters of the transformation, and the $\sigma_j$ are the Pauli matrices ($j=1,2,3$), then  we have
\be
A_aB_a = \begin{pmatrix} A_1,A_2 \end{pmatrix} \begin{pmatrix} B_1 \\ B_2 \end{pmatrix} \ra \begin{pmatrix} A_1,A_2 \end{pmatrix} e^{i m_j \sigma_j^T} e^{i n_j \sigma_j} \begin{pmatrix} B_1 \\ B_2 \end{pmatrix} \nonumber
\ee
\ba
e^{i m_j \sigma_j^T} e^{i n_j \sigma_j} & = & \mathbbm{1}_2 + i m_j \sigma_j^T + i n_j \sigma_j  + \ldots \nonumber \\ & = & \mathbbm{1}_2 + i (m_1 + n_1) \sigma_1 + i (-m_2 + n_2) \sigma_2 + i (m_3 + n_3) \sigma_3 )+ \ldots \nonumber
\ea
where we have expanded the exponentials, $\mathbbm{1}_2$ stands for the $2\times2$ identity matrix, and $\ldots$ stands for terms of higher order in $m_j$ and $n_j$. By demanding that the terms linear in $m_j$ and $n_j$ vanish, we find that only the subspace of the $SU(2)_A\times SU(2)_B$ algebra where $n_1 = -m_1$, $n_2=m_2$ and $n_3 = -m_3$ leaves $A_a B_a$, and hence the superpotential, invariant. We will denote this diagonal subgroup $SU(2)_D$.

The theory also has $\N=3$ supersymmetry and hence retains the $SU(2)_R$ symmetry, under which $(A_1,B_1^*)$, $(A_2,B_2^*)$, and $(Q_i,\bar{\tilde{Q}}_i)$ transform as doublets. In the ABJM theory without flavor, the superpotential exhibited the symmetry $SU(2)_A \times SU(2)_B$, which does not commute with $SU(2)_R$. The conclusion was that in fact the full R-symmetry was $SU(4)$, and hence the theory had $\N=6$ supersymmetry, as we reviewed in section \ref{abjmtheory}.

The crucial question is thus whether or not $SU(2)_R$ and the $SU(2)_D$ subgroup of $SU(2)_A \times SU(2)_B$ commute. As mentioned in refs. \cite{Hohenegger:2009as,Gaiotto:2009tk,Hikida:2009tp}, they do commute, as we will now show explicitly\footnote{The fact that $SU(2)_R$ and $SU(2)_D$ commute is a familiar feature of $SU(4)$. The $SU(4)$ algebra has two obvious $SU(2) \times SU(2)$ sub-algebras, whose diagonal $SU(2)$'s commute with one another. In the ABJM theory these are $SU(2)_A \times SU(2)_B$, with diagonal $SU(2)_D$, and $SU(2)_1 \times SU(2)_2$, with diagonal $SU(2)_R$. Here $SU(1)_1$ acts on $(A_1,B_1^*)$ as a doublet and leaves $(B_2^*,A_2)$ invariant, while for $SU(2)_2$ $(A_1,B_1^*)$ is invariant and $(B_2^*,A_2)$ is a doublet.}. Let the $4 \times 4$ matrices $\de_j^R = i \sigma_j \otimes 1_2$ and
\be
\de_j^A = \left( \begin{matrix} i \sigma_j & 0_2 \\ 0_2 & 0_2 \end{matrix} \right) \,,\quad \de_j^B = \left( \begin{matrix} 0_2 & 0_2 \\ 0_2&-i \sigma_{j}^* \end{matrix} \right)\,, \nonumber
\ee
represent the generators of $SU(2)_R$, $SU(2)_A$ and $SU(2)_B$ that act on the vector $(A_1,A_2,B_1^*,B_2^*)$. Here $0_2$ represents the $2 \times 2$ null matrix. We then find
\ba
\left[\de_1^R,\de_j^A\right] &=& i \sigma_2 \otimes \sigma_j \,,\quad \left[\de_2^R,\de_j^A\right] = -i \sigma_1 \otimes \sigma_j\,,\quad \left[\de_3^R,\de_j^A\right] = 0\,, \nonumber \\
\left[\de_1^R,\de_j^B\right] &=& i \sigma_2 \otimes \sigma_{j}^* \,,\quad \left[\de_2^R,\de_j^B\right] = -i \sigma_1 \otimes \sigma_{j}^*\,,\quad \left[\de_3^R,\de_j^B\right] = 0\,. \nonumber
\ea
and hence we immediately find that the subgroup $SU(2)_D$ commutes with $SU(2)_R$:
\be
\left[\de_j^R,\de_1^A - \de_1^B \right] = \left[\de_j^R,\de_2^A+\de_2^B \right] = \left[\de_j^R,\de_3^A - \de_3^B \right] = 0. \nn
\ee
The $SU(2)_R$ is therefore not enhanced, so the system has only $\N=3$ supersymmetry.

To summarize: classically the theory has $\N=3$ superconformal symmetry, with bosonic subgroup $SO(3,2)$, and global symmetry $SU(2)_R \times SU(2)_D \times U(1)_b \times U(N_f) \times U(N_f)$ which matches perfectly with the symmetries in the supergravity description above.

As this case was a rather trivial example of our recipe, we now turn to slightly more involved examples, in particular, examples that exhibit supersymmetry enhancement.

\section{Codimension One $\N=(0,6)$ Supersymmetric Flavor}
\label{codonechiral}

In this section and the next we study two different probe branes that introduce codimension-one flavor fields, that is, flavor fields propagating in a (1+1)-dimensional subspace of the (2+1)-dimensional ABJM theory. The branes we study, a D7/D8/M9-brane and a D3/D4/M5-brane, were first studied in type IIA on $AdS_4 \times \cp$ in ref. \cite{Fujita:2009kw}. We review and extend the gravity results of ref. \cite{Fujita:2009kw}, and write the dual field theory Lagrangian explicitly for the D7/D8/M9-brane.

In (1+1) dimensions the supersymmetries divide into left- and right-handed sectors. We begin in this section with a chiral codimension-one theory, which preserves $\N=(0,6)$ supersymmetry. In the next section we study non-chiral flavor.

\subsection{Supergravity with M9/D8-brane probes}
\label{codoneM9}

We begin by adding D7-branes extended along (01345789) in the type IIB setup. We are free to choose their position on the $x^6$ circle. We will add two stacks of D7-branes, each with $N_f$ coincident D7-branes, on opposite sides of the circle, away from the NS5-brane and $(1,k)$5-brane. The strings from the D7-branes to the two stacks of D3-branes introduce massless flavor in both gauge groups. (In contrast, the authors of ref. \cite{Fujita:2009kw} considered matter fields that coupled only to a single gauge group.) Notice also that the D7-branes and D3-branes have 8 ND directions, hence the flavor fields will be chiral, as we mentioned in section \ref{fieldtheoryanalysis}. As shown in table \ref{table:IIB} in section \ref{gravityanalysis}, these D7-branes preserve 3 real supercharges in the type IIB setup.

After T-duality in $x^6$ the D7-branes become D8-branes. The $2 N_f$ D8-branes are coincident, and have a $U(2N_f)$ symmetry broken to $U(N_f) \times U(N_f)$ by a $\mathbbm{Z}_2$-valued Wilson line (similar to what happened in section \ref{codzerogravity}). After uplift to M-theory and the ``near-horizon'' limit, the D8-branes become M9-branes extended along (01) and along all of $\mathbbm{C}^4/\mathbbm{Z}_k$.  Obviously the branes preserve the full $SU(4) \times U(1)_b$ symmetry of $\mathbbm{C}^4/\mathbbm{Z}_k$. In appendix \ref{app:probeflat} (see also ref. \cite{Fujita:2009kw}) we find that the M9-branes preserve 6 real supercharges.

If we take $N_c \rightarrow \infty$, we can replace the M2-branes by their near-horizon geometry, which is $AdS_4 \times S^7/\mathbbm{Z}_k$. The M9-branes are extended along $AdS_3$ inside $AdS_4$ and wrap all of $S^7/\mathbbm{Z}_k$. In appendix \ref{app:probeads} we analyze the $\kappa$-symmetry condition for these branes and find that after the near-horizon limit the number of preserved supercharges has doubled to 12.

For large $k$, such that $k^5 \gg N_c$, the M9-branes reduce to D8-branes in type IIA that wrap $AdS_3 \times \cp$. These probe D8-branes were first studied in ref. \cite{Fujita:2009kw}.

To summarize: in the type IIB setup we can add D7-branes that produce fundamental matter for both gauge group factors. They have 8 ND directions (with respect to the D3-branes), so the flavor fields will be chiral. These D7-branes become M9-branes in M-theory on $\mathbbm{C}^4/\mathbbm{Z}_k$. These M9-branes preserve 6 real supercharges, so we expect a dual field theory with (in (1+1)-dimensional notation) $\N=(0,6)$ superconformal symmetry. The corresponding R-symmetry group must be $SU(4) \cong SO(6)$, which fits into the symmetry of the brane construction. We now proceed to construct the dual field theory and check that is has the right symmetries and amount of supersymmetry.

\subsection{The Field Theory}
\label{codonechiralfieldtheory}

Let us apply our recipe.

\bigskip

\textbf{Step 1: Construct the D3/D7 Theory}

Once again we consider a single stack of D3-branes alone in flat space (so again let $x^6$ be non-compact and remove the NS5-brane and $(1,k)$5-brane), and add a codimension-two D7-brane. Such a D3/D7 intersection has 8 ND directions and preserves 8 real supercharges. What is the field theory for such a D3/D7 intersection? This question was answered\footnote{Much of the analysis of refs.  \cite{Harvey:2007ab,Buchbinder:2007ar,Harvey:2008zz} focused on what we would call ``back-reaction,'' that is, effects that result from leaving the probe limit. Strictly speaking, all of our statements apply only in the probe limit.} in refs. \cite{Harvey:2007ab,Buchbinder:2007ar,Harvey:2008zz}. With 8 ND directions, the NS sector zero-point energy is 1/2, so the ground state is in the Ramond sector. What survives the GSO projection is a single Weyl spinor confined to the (1+1)-dimensional intersection, transforming in the $(N_c,\bar{N}_f)$. We thus obtain chiral flavor. Our Weyl fermion will be left-handed.

The immediate question is: with only fermions in the ground state, how can the theory be supersymmetric? The answer is that all of the preserved supercharges are \textit{right}-handed. The theory has (1+1)-dimensional $\N=(0,8)$ supersymmetry. The flavor fermions are completely inert under both supersymmetry and the R-symmetry.

The action is then remarkably simple. From the D3-branes we of course have the (3+1)-dimensional $\N=4$ $U(N_c)$ SYM theory action. For the defect flavor fields, the claim of refs. \cite{Harvey:2007ab,Buchbinder:2007ar,Harvey:2008zz} is that the \textit{only} marginal and gauge-invariant terms that respect all of the symmetries are
\be
S_{\text{fund}} = \int dx^+ dx^- \, \psi_q^{\dagger} \left( i \partial_- - A_- \right) \psi_q,
\ee
where we have used (1+1)-dimensional coordinates $x^{\pm} = x^0 \pm x^1$, $\psi_q$ is our left-handed Weyl fermion, and $A_-$ is the restriction of the ambient $U(N_c)$ gauge field to the defect. Of crucial importance is the fact that $A_-$ is inert under $\N=(0,8)$ supersymmetry transformations \cite{Harvey:2007ab,Buchbinder:2007ar,Harvey:2008zz}.

\bigskip

\textbf{Step 2: Add the NS5-branes}

We add the NS5-brane and NS5$^{\prime}$-brane along $(012345)$ and separated in $x^6$ (which for now is still non-compact), and let the D3-branes end on them in $x^6$. The NS5-brane boundary conditions set to zero the (2+1)-dimensional $\N=4$ hypermultiplet. The flavor fields only couple to the gauge field, however, so adding the NS5-branes does not alter the action in the flavor sector. The supersymmetry is reduced to $\N=(0,4)$, however.

\bigskip

\textbf{Step 3: Compactify $x^6$, form the $(1,k)$5-brane, and lift to M-theory}

We compactify $x^6$ and add another stack of D3-branes, so that the gauge group of the D3-branes' worldvolumes is $U(N_c)\times U(N_c)$, and another stack of $N_f$ D7-branes. We obtain two sets of flavor fields, which we will denote as $\psi_q^i$ with $i=1,2$. We obtain two copies of the action above, one for each $\psi_q^i$. The rest of the construction (forming the $(1,k)$5-brane, T-duality, etc.) also leaves the action in the flavor sector untouched. The Chern-Simons terms break the supersymmetry to $\N=(0,3)$.

\bigskip

\textbf{Step 4: Take the low-energy limit}

Lastly, we must take the low-energy limit, which means writing all terms consistent with the symmetries of the field theory. Let us review the symmetries of the theory at the end of Step 3. Our theory has $\N=(0,3)$ supersymmetry and the corresponding $SU(2)_R$ R-symmetry. (The $SU(2)_R$ is easy to see in the type IIB setup, being exactly the same $SU(2)_R$, which rotates (345) and (789) simultaneously, that appears in the theory without flavor.) The theory also has a baryon number symmetry that shifts the phase of $\psi_q^i$ and leaves all other fields invariant. Recall also that the theory of course has (1+1)-dimensional Lorentz invariance and gauge invariance. We will now argue that in fact these symmetries forbid any new (relevant or marginal) terms.

First let us do some dimension counting. The fields $\psi_q$ are (1+1)-dimensional fermions, hence they are dimension 1/2. (We will drop the $i$ index on $\psi_q^i$ for now.) We must also consider the restriction of the (2+1)-dimensional fields to (1+1) dimensions. We will use $\phi$ to denote a generic (2+1)-dimensional scalar restricted to the defect, and $\Psi$ to denote a (2+1)-dimensional fermion restricted to the defect. $\phi$ is dimension 1/2 and $\Psi$ is dimension 1.

Terms with an odd number of $\psi_q$ and $\psi_q^{\dagger}$, whether relevant or marginal, are forbidden by gauge invariance and by the $U(1)$ baryon number that shifts the phase of $\psi_q$. Terms with two $\psi_q$ that are relevant include couplings to scalars, of the form $\phi \, \psi_q^{\dagger} \psi_q$, which is dimension $3/2$. These are forbidden by Lorentz invariance. $\psi_q$ is a (1+1)-dimensional left-handed fermion. Its conjugate $\psi_q^{\dagger}$ is also left-handed, hence $\psi_q^{\dagger} \psi_q$ is not a Lorentz singlet. Marginal couplings of the form $\phi^2 \psi_q^{\dagger} \psi_q$ and $\Psi \psi_q^{\dagger} \psi_q$, and the marginal quartic term $(\psi_q^{\dagger} \psi_q)^2$, are forbidden for the same reason. (We can also eliminate many such terms, and/or linear combinations of them, using the R-symmetry and/or supersymmetry.) The only term involving derivatives and/or the gauge field that is allowed by the symmetries is the gauge-covariant kinetic term itself. The overall normalization of that term can change, but of course such an overall constant can be removed by a rescaling of $\psi_q$.

Our conclusion is that the form of the defect action does not change in Step 4. We can thus write the defect action easily. We have two Weyl fermions, $\psi_q^i$, where again $i=1,2$ labels the gauge group, that is, under $U(N_c)_k \times U(N_c)_{-k} \times U(N_f) \times U(N_f)$ the $\psi_q^{1}$ fermion transforms as $(N_c,\mathbf{1},\bar{N_f},\mathbf{1})$ and the $\psi_q^2$ fermion transforms as $(\mathbf{1},N_c,\mathbf{1},\bar{N_f})$. We add to the ABJM action the terms
\be
S_{\text{fund}} = \int dx^+ dx^- \, \psi_q^{i \dagger} \left( i \partial_- - A_-^i \right) \psi_q^i,
\label{eq:codonechiralaction}
\ee
where here again $A_-^i$ are the defect values of the bulk gauge fields, and summation over $i$ is implicit.

We show in appendix \ref{app:n06} that $A_-$ is invariant under $\N=(0,6)$ supersymmetry transformations, hence the flavor action preserves $\N=(0,6)$ supersymmetry. The action is also trivially invariant under the full $SU(4)_R \times U(1)_b$ symmetry. These symmetries perfectly match those of the brane construction.

\section{Codimension One $\N=(3,3)$ Supersymmetric Flavor}
\label{codonenonchiral}

In this section we study codimension-one non-chiral flavor, that is, non-chiral flavor fields propagating in a (1+1)-dimensional subspace of the (2+1)-dimensional ABJM theory. The flavor fields will have $\N=(4,4)$ supersymmetry broken to $\N=(3,3)$ supersymmetry when the Chern-Simons level $k\geq2$. The flavor brane in this case is a D3-brane in type IIB, a D4-brane in type IIA, and an M5-brane in M-theory. We perform our complete supergravity analysis for these branes, for example, we compute the supersymmetry that they preserve. We also comment on the structure of the field theory and explain in detail the complications that arise in applying our recipe for the field theory.

\subsection{Supergravity with M5/D4-brane probes}
\label{codoneM5}

We begin by adding D3-branes extended along (0137) in the type IIB setup. We are free to choose their position on the $x^6$ circle. We will add two stacks of D3-branes, each with $N_f$ coincident D3-branes, on opposite sides of the circle, away from the NS5-brane and $(1,k)$5-brane. The strings from the flavor D3-branes to the color D3-branes thus introduce massless flavor in both gauge groups. (In contrast, the authors of ref. \cite{Fujita:2009kw} considered matter fields that coupled only to a single gauge group.) Notice also that the flavor D3-branes and color D3-branes have 4 ND directions, hence the flavor fields will be non-chiral. As shown in table \ref{table:IIB} in section \ref{gravityanalysis}, these D3-branes preserve 3 real supercharges in the type IIB setup.

After T-duality in $x^6$ the D3-branes become D4-branes. The $2 N_f$ D4-branes are coincident, and have a $U(2N_f)$ symmetry broken to $U(N_f) \times U(N_f)$ by a $\mathbbm{Z}_2$-valued Wilson line. After uplift to M-theory and the ``near-horizon'' limit, the D4-branes become M5-branes in M-theory on $\mathbbm{C}^4/\mathbbm{Z}_k$. Using the results of appendix \ref{app:IIBtoM}, we find that the embedding of the M5-branes is described by the equations $z^1=z^2, z^3=z^4$. Here again we can use an $SU(4)$ transformation to produce new embedding equations that make the symmetries transparent.  Explicitly, the $SU(4)$ transformation is
\be
z^1_{new} = \frac{1}{\sqrt{2}}(z^1-z^2), \quad z^2_{new} = \frac{1}{\sqrt{2}}(-z^3+z^4), \quad z^3_{new} = \frac{1}{\sqrt{2}}(z^1+z^2), \quad z^4_{new} = \frac{1}{\sqrt{2}}(z^3+z^4).
\ee
The embedding equation is then $z_{new}^1=z_{new}^2=0$. The M5-branes are thus extended along (01) and $z_{new}^3$ and $z_{new}^4$.

For the embedding $z_{new}^1=z_{new}^2=0$ we can easily see that the M5-branes breaks the $SU(4)$ symmetry of $\mathbbm{C}^4/\mathbbm{Z}_k$ down to $SU(2) \times SU(2) \times U(1)$ where the first $SU(2)$ acts on $(z^1,z^2)$ and the second $SU(2)$ acts on $(z^3,z^4)$. The $U(1)$ acts as $(z^1, z^2,z^3,z^4) \rightarrow (e^{i \alpha} z^1, e^{i \alpha} z^2,e^{-i \alpha} z^3,e^{-i \alpha} z^4)$. We can also see that the embedding equations are invariant under the $U(1)_b$ symmetry, $z^i \rightarrow e^{i \alpha} z^i, \, \forall i$. We thus conclude that such branes preserve the symmetry group $SU(2) \times SU(2) \times U(1) \times U(1)_b$. In appendix \ref{app:probeflat}, we find that with the above embedding the M5-branes preserve 6 real supercharges (see also ref. \cite{Fujita:2009kw}).

If we take $N_c \rightarrow \infty$, we can replace the M2-branes by their near-horizon geometry, which is $AdS_4 \times S^7/\mathbbm{Z}_k$. The M5-branes are now extended along $AdS_3$ inside $AdS_4$ and wrap a three cycle in $S^7/\mathbbm{Z}_k$. In appendix \ref{app:probeads} we analyze the $\kappa$-symmetry condition for these M5-branes and find (as expected) that after the near-horizon limit the number of preserved supercharges has doubled to 12 (see also ref. \cite{Chandrasekhar:2009ey}).

As shown in ref. \cite{Fujita:2009kw} (see also ref. \cite{Chandrasekhar:2009ey}), when $k$ is large, such that $k^5 \gg N_c$, the M5-branes reduce to D4-branes in type IIA that wrap $AdS_3 \times \mathbbm{CP}^1$ in $AdS_4 \times \cp$, where $\mathbbm{CP}^1$ is the unique two-cycle in $\cp$.

To summarize: in the type IIB setup we can add D3-branes that produce flavor fields for both gauge group factors. These flavor fields will be non-chiral. The D3-branes become M5-branes in M-theory on $\mathbbm{C}^4/\mathbbm{Z}_k$ that preserve 6 real supercharges. We thus expect that the flavor fields will preserve (1+1)-dimensional $\N=(3,3)$ superconformal symmetry. The corresponding R-symmetry group must be $SO(3) \times SO(3)$, which fits into the symmetry group we found above. We now proceed to the dual field theory.

\subsection{Comments about the Field Theory}
\label{codonenonchiralfieldtheory}

Here we will again apply our recipe, although we will find some complications when we add the NS5-branes in Step 2.

\bigskip

\textbf{Step 1: Construct the D3/D3 Theory}

We start in type IIB in flat space (so $x^6$ non-compact and no NS5- or $(1,k)$5-brane) and introduce $N_c$ color D3-branes that intersect $N_f$ flavor D3-branes in (1+1) dimensions. The table below shows the intersection.

\begin{center}
        \begin{tabular}{|c|cccccccccc|}\hline
                &0&1&2&3&4&5&6&7&8&9\\ \hline
$N_c$ D3&$\bullet$&$\bullet$&$\bullet$&--&--&--&$\bullet$&--&--&--\\
$N_f$ D3&$\bullet$&$\bullet$&--&$\bullet$&--&--&--&$\bullet$&--&--\\ \hline
        \end{tabular}
\end{center}

The field theory for such a D3/D3 intersection was constructed in ref. \cite{Constable:2002xt}. The D3/D3 intersection has 4 ND directions, so the flavor fields are non-chiral. The full theory preserves 8 real supercharges. For this case, we have in the flavor sector two (1+1)-dimensional $\N=(2,2)$ chiral superfields $Q$ and $\tilde Q$, which together form an $\N=(4,4)$ hypermultiplet.

In the adjoint sector we must decompose the (3+1)-dimensional $\N=4$ multiplet into (1+1)-dimensional multiplets. The full decomposition appears in ref. \cite{Constable:2002xt}. We write the bosonic content, including auxiliary fields, in the table below. In our notation, the superscript on the vector field $A_{\mu}$ indicates which components are included in the multiplet, for example, $A_{\mu}^{0126}$ indicates that $A_0$, $A_1$, $A_2$ and $A_6$ are included. Scalars are denoted by the number of the corresponding direction in the type IIB construction. The subscript on auxiliary fields indicates the spacetime dimensionality: $D_4$ is the auxiliary field in a (3+1)-dimensional vector multiplet, while $D_2$ is the auxiliary field in a (1+1)-dimensional vector multiplet. The superscript, $a,$ $b$, or $c$, on the auxiliary fields $F$ is simply a label to distinguish them among each other (the superscript has no deeper meaning).

\begin{center}
\begin{tabular}{|c|c|c|}\hline
(3+1)d&$\N=4$ V $(A_{\mu}^{0126},345789,D_4,F_4^a,F_4^b,F_4^c)$&\\\hline
(1+1)d&$\N=(4,4)$ V $(A_{\mu}^{01},4589,D_2,F_2)$&$\N=(4,4)$ H $(A_2,A_6,37,F_2^a,F_2^b)$\\\hline
\end{tabular}
\end{center}

In (3+1) dimensions the $\N=4$ vector multiplet decomposes into an $\N=1$ vector multiplet and three $\N=1$ chiral multiplets. The bosonic content is the vector field (with components (0126)), the six scalars (345789) (transverse to the color D3-branes), the real auxiliary field $D_4$ in the $\N=1$ vector multiplet, and three complex auxiliary fields $F_4^a$, $F_4^b$, and $F_4^c$, from the three $\N=1$ chiral multiplets.

As shown ref. \cite{Constable:2002xt}, the (3+1)-dimensional $\N=4$ vector multiplet reduces to, in (1+1) dimensions, an $\N=(4,4)$ vector multiplet and an $\N=(4,4)$ hypermultiplet. The $\N=(4,4)$ vector multiplet includes the (01) components of the vector field, the four scalars (4589) (transverse to both the color and flavor D3-branes), the real auxiliary field $D_2$ and the complex auxiliary field $F_2$. Notice that the $\N=(4,4)$ vector multiplet can be decomposed into two $\N=(2,2)$ multiplets, a vector multiplet and a chiral multiplet. $D_2$ is the auxiliary field in the $\N=(2,2)$ vector multiplet, while $F_2$ is the auxiliary in the chiral multiplet. An important identification that we will use later is $D_2 = \frac{1}{\sqrt{2}} \left( D_4 + F_{26}\right)$ \cite{Constable:2002xt}, where $F_{26}$ is the field strength in the (26) directions (along the color D3-branes but transverse to the flavor D3-branes). The $\N=(4,4)$ hypermultiplet includes four scalars, namely the components $A_2$ and $A_6$ of the vector field and the scalars (37) (transverse to the color D3-branes but along the flavor D3-branes).

The key point is that the defect flavor fields couple only to the $\N=(4,4)$ vector multiplet. As described in ref. \cite{Constable:2002xt}, the easiest way to write the action is using $\N=(2,2)$ superspace. The kinetic term includes the usual coupling to the $\N=(2,2)$ vector superfield, and then a superpotential coupling to an $\N=(2,2)$ chiral superfield provides the completion to $\N=(4,4)$ supersymmetry. The action is written explicitly in appendix D of ref. \cite{Constable:2002xt}.

At this stage the R-symmetry is $SU(2) \times SU(2) \times U(1) \times U(1)$ \cite{Constable:2002xt}. The two $SU(2)$'s correspond to the $SO(4)$ isometry that acts in the overall transverse directions (4589). The first $U(1)$ corresponds to rotations along the color D3-branes but transverse to the flavor D3-branes, in the (26) plane. Similarly, the second $U(1)$ corresponds to rotations transverse to the color D3-branes but along the flavor D3-branes, in the (37) plane.

\bigskip

\textbf{Step 2: Add the NS5-branes}

We next add the NS5-brane and NS5$^{\prime}$-brane. The arrangement of branes is as follows (here we include explicitly only the NS5-brane):

\begin{center}
        \begin{tabular}{|c|cccccccccc|}\hline
                &0&1&2&3&4&5&6&7&8&9\\ \hline
$N_c$ D3&$\bullet$&$\bullet$&$\bullet$&--&--&--&$\bullet$&--&--&--\\
$N_f$ D3&$\bullet$&$\bullet$&--&$\bullet$&--&--&--&$\bullet$&--&--\\
NS5&$\bullet$&$\bullet$&$\bullet$&$\bullet$&$\bullet$&$\bullet$&--&--&--&--\\\hline
        \end{tabular}
\end{center}

We need to impose the NS5-brane boundary conditions on the $\N=(4,4)$ supersymmetric defect action written in appendix D of ref. \cite{Constable:2002xt}, that is, we need to know which fields of the (1+1)-dimensional $\N=(4,4)$ vector multiplet are killed by the NS5-brane boundary conditions, and hence what couplings are eliminated in the defect action. The relevant decomposition of fields from (3+1) dimensions to (2+1) and (1+1) dimensions is as follows (the first and last lines are simply repeated from the similar table above):

\begin{center}
\begin{tabular}{|c|c|c|}\hline
(3+1)d&$\N=4$ V $(A_{\mu}^{0126},345789,D_4,F_4^a,F_4^b,F_4^c)$&\\\hline
(2+1)d&$\N=4$ V $(A_{\mu}^{012},345,F_3,D_3)$&$\N=4$ H $(A_6,789,F_3^a,F_3^b)$\\\hline
(1+1)d&$\N=(4,4)$ V $(A_{\mu}^{01},4589,D_2,F_2)$&$\N=(4,4)$ H $(A_2,A_6,37,F_2^a,F_2^b)$\\\hline
\end{tabular}
\end{center}

As we have reviewed several times now, when the D3-branes end on the NS5-branes, the (3+1)-dimensional $\N=4$ fields decompose into (2+1)-dimensional $\N=4$ fields, namely a vector and a hypermultiplet. We have included that decomposition in the table above. The boundary conditions set to zero the fields in the (2+1)-dimensional $\N=4$ hypermultiplet. In particular, $A_6$ and the scalars (789) are all set to zero. That means that various parts of the (1+1)-dimensional multiplets are set to zero. Specifically, in the $\N=(4,4)$ vector multiplet, the scalars (89) are set to zero.

A cursory analysis suggests that the coupling of the flavor fields is described by an $\N=(2,2)$ kinetic term alone, that is, that the NS5-brane boundary conditions eliminate the superpotential of the D3/D3 theory and leave only the kinetic term. The argument goes as follows. As mentioned in section \ref{gravityanalysis} and appendix \ref{app:IIBprobesusy}, the brane intersection above preserves 4 real supercharges. The defect flavor fields obviously need a kinetic term, and are non-chiral, hence we expect that the flavor action is the $\N=(2,2)$ supersymmetric completion of the kinetic term. Indeed, the $\N=(2,2)$ vector multiplet includes the (01) components of the vector field, two real scalars, and the real auxiliary field $D_2$. The boundary conditions leave two scalars, (45), untouched, which is nicely consistent. Furthermore, the action seems to have the right symmetries. $\N=(2,2)$ supersymmetry comes with a $U(1) \times U(1)$ R-symmetry. In the brane description, with two NS5-branes, the flavor D3-brane obviously preserves two $U(1)$'s, namely the independent rotations in (45) and (89).

When we look more closely at the auxiliary fields, we find that the auxiliary field $D_2$ is set to zero by the NS5-brane boundary conditions. To see why, we examine how the auxiliary fields transform under R-symmetry. In (3+1) dimensions, $D_4$ is a singlet of the $SO(6)$ R-symmetry, while $F_4^a$, $F_4^b$, and $F_4^c$ form a \textbf{6} of $SO(6)$.  In (2+1) dimensions, the auxiliary fields in the $\N=4$ vector multiplet, the real $D_3$ and the complex $F_3$, form a \textbf{3} of the $SO(3)$ R-symmetry. The auxiliaries in the $\N=4$ hypermultiplet are more subtle. The auxiliary fields $F_3^a$ and $F_3^b$ have four real degrees of freedom. Of those, three real degrees of freedom form a $\textbf{3}$ of $SO(3)$.

In the decomposition from (3+1) dimensions to (2+1) dimensions, then, we must assign three real degrees of freedom from $(F_4^a,F_4^b,F_4^c)$ to the $\N=4$ vector multiplet, that is, to $(D_3,F_3)$. The remaining three real degrees of freedom, and $D_4$, must be assigned to the $\N=4$ hypermultiplet, that is, to $(F_3^a,F_3^b)$. Notice that such an identification makes sense: $F_3^a$ and $F_3^b$ describe four real degrees of freedom, yet only three of those form a $\textbf{3}$ of $SO(3)$. The ``extra'' degree of freedom is $D_4$, which is indeed a singlet of the R-symmetry. The boundary conditions then set $(F_3^a,F_3^b)$ to zero, so that $D_4$ is set to zero.

Now we recall the identification from ref. \cite{Constable:2002xt}, and mentioned above, $D_2 = \frac{1}{\sqrt{2}} \left ( D_4 + F_{26} \right )$. Notice that such an identification also makes sense, since $D_2$, $D_4$, and the gauge field are all singlets under R-symmetry. We can easily argue that in our case $F_{26} = \partial_2 A_6 - \partial_6 A_2 + i \left [ A_2, A_6\right]$ is zero. First, the boundary conditions set $A_6 = 0$. Second, we perform a dimensional reduction along $x^6$, hence none of the fields in the low-energy theory depend on $x^6$ (we keep only zero modes), so $\partial_6 A_2 = 0$. That leaves $D_2 = \frac{1}{\sqrt{2}} D_4$. As we argued in the last paragraph, however, $D_4$ ends up in the (2+1)-dimensional $\N=4$ hypermultiplet, and is thus set to zero by the boundary conditions. We thus have $D_2 = 0$.

Our conclusion is that the defect flavor fields do not couple to an $\N=(2,2)$ vector multiplet, since the appropriate coupling to the auxiliary field $D_2$ is absent. In other words, we (apparently) cannot write the defect action in $\N=(2,2)$ superspace, although we expect the theory to have $\N=(2,2)$ supersymmetry. The technical question is thus how to demonstrate that the theory has $\N=(2,2)$ supersymmetry. We will leave that question for the future, and turn now to other issues that arise in Steps 3 and 4.

\bigskip

\textbf{Step 3: Compactify $x^6$, form the $(1,k)$5-brane, and lift to M-theory}

When we use the NS5$^{\prime}$-brane to form the $(1,k)$5-brane, the supersymmetry analysis in appendix \ref{app:IIBprobesusy} shows that the system has 3 real supercharges. In field theory language, that na\"ively suggests that the Chern-Simons term breaks $\N=(2,2)$ down to $\N=(2,1)$ or $\N=(1,2)$. Notice that then the R-symmetry would be a single $U(1)$, which would also be consistent with the brane description, where the separate $SO(3)$'s, corresponding to independent rotations in (345) and (789), are broken to a single $SO(3)$, corresponding to simultaneous rotations in (345) and (789). The flavor D3-brane obviously only preserves a $U(1)$ subgroup of that, namely simultaneous rotations in (45) and (89).

At first glance, what is puzzling about $\N=(2,1)$ or $\N=(1,2)$ supersymmetry is how a (2+1)-dimensional term ``knows'' about (1+1)-dimensional chirality. In other words, why does the Chern-Simons term only break a single \textit{left}-handed supercharge (for example)? We can make sense of this very simply.\footnote{We thank Andreas Karch for the following observation.} Consider for the moment a \textit{single} stack of $N_f$ flavor D3-branes, so that we obtain flavor fields in only one gauge group. Such a configuration clearly breaks the parity symmetry of the type IIB setup, which involves an exchange of the two gauge groups. In the field theory with Chern-Simons terms, we can perform an integration by parts, producing a theta term on the defect, which breaks parity, so the idea that the Chern-Simons term may break $\N=(2,2)$ to $\N=(2,1)$ is not unnatural.

What is then curious is that if we add the second stack of flavor D3-branes the system seems to preserve parity again, so in that case how can we obtain $\N=(2,1)$ supersymmetry? Here we must be careful, and distinguish two $\mathbbm{Z}_2$ operations. The first $\mathbbm{Z}_2$ is normal parity, which reverses the sign of a spatial coordinate. In our case we are interested in the spatial coordinate along the (1+1)-dimensional defect. The second $\mathbbm{Z}_2$ involves an exchange of the two gauge groups. The simultaneous action of \textit{both} $\mathbbm{Z}_2$'s is the ``parity symmetry'' of the ABJM theory, as reviewed in section \ref{abjmtheory}. Now suppose we have only a single stack of flavor D3-branes, describing flavor fields in only one gauge group, and preserving $\N=(2,1)$ supersymmetry. If we perform the first $\mathbbm{Z}_2$, which exchanges left-movers and right-movers, then $\N=(2,1)$ becomes $\N=(1,2)$. If we then perform the second $\mathbbm{Z}_2$, we find flavor in the fundamental representation of the second gauge group. Clearly, then, when we add two stacks of flavor D3-branes, one stack will describe flavor fields in one gauge group, preserving $\N=(2,1)$ supersymmetry, while the other stack will describe flavor fields in the second gauge group, preserving $\N=(1,2)$ supersymmetry. The entire system can thus preserve 3 real supercharges and still be invariant under the simultaneous action of both $\mathbbm{Z}_2$'s. As an aside, notice that the conventional notation of supersymmetry is a source of confusion here. The notation $\N=(2,1)$ makes reference only to the first $\mathbbm{Z}_2$ operation. When we talk about supercharges in this theory, however, a better convention may be to let ``left-handed'' and ``right-handed'' refer to the transformation of the supercharges under the simultaneous action of both $\mathbbm{Z}_2$'s. We will leave a detailed investigation of these issues for future research.

\bigskip

\textbf{Step 4: Take the low-energy limit}

Our supersymmetry analysis of appendix \ref{Mtheorysusy} shows that the D3-brane, which becomes an M5-brane in M-theory, preserves 6 real supercharges on $\mathbbm{C}^4/\mathbbm{Z}_k$, which suggests that, in the field theory, when we integrate out the (2+1)-dimensional $\N=4$ vector multiplet we should find $\N=(3,3)$ supersymmetry. (Notice that when $k=1$, so that $\mathbbm{C}^4/\mathbbm{Z}_k$ becomes just $\mathbbm{C}^4$, the system has 8 real supercharges, so we expect $\N=(4,4)$ supersymmetry.) The procedure of integrating out fields will be much more complicated than in either the ABJM theory or in the codimension-zero flavor case above. In particular, the process of integrating out will probably not be possible at the level of superfields, but rather may have to be done using the components of the superfields.

\bigskip

Given the various complications in constructing the action and integrating out the (2+1)-dimensional $\N=4$ vector multiplet, we will try to ``work backwards'': we will use the symmetries of the gravity description to \textit{guess} the form of the low-energy theory. More precisely, we will write a scalar potential that has the symmetries we expect. The scalar potential will describe the coupling of the defect flavor scalars to the scalar components of the superfields $A_a$ and $B_a$, restricted to the defect. We will denote the defect flavor scalars as $q_i^n$, where $i=1,2$ labels the two gauge groups and $n=1,2$ labels the two complex scalars of an $\N=(4,4)$ hypermultiplet.

The scalar potential is of course constrained by gauge invariance and dimensional analysis. The scalars $A_a$ and $B_a$ are (2+1)-dimensional fields (though we will be interested in their restriction to the (1+1)-dimensional defect), so they have dimension $1/2$. The defect scalars $q_i^n$ are dimension zero. The potential on the (1+1)-dimensional defect must therefore involve four of the $A_a$ and $B_a$ fields.

The crucial question is what symmetries the potential must have. Here we turn to the gravity analysis. As mentioned above, the codimension-one M5-brane on $\mathbbm{C}^4/\mathbbm{Z}_k$ preserves $SU(2) \times SU(2) \times U(1) \times U(1)_b$. Which symmetries are these in the field theory?

Let us return for the moment to the theory without flavor, and review the symmetries. We start with Step 2, at which point the field theory is a (2+1)-dimensional $\N=4$ supersymmetric Yang-Mills theory (without Chern-Simons terms). The R-symmetry is an $SO(4) \sim SO(3) \times SO(3)$ where the two $SO(3)$'s correspond to independent rotations in (345) and (789). We will call these $SU(2)_1$ and $SU(2)_2$, where $SU(2)_1$ acts on $(A_1,B_1^*)$ as a doublet while $SU(2)_2$ acts on $(B_2^*,A_2)$ as a doublet. When we proceed to Step 3 and form the $(1,k)$5-brane, which introduces Chern-Simons terms in the field theory, the R-symmetry breaks to $SU(2)_R$, the diagonal part of $SU(2)_1 \times SU(2)_2$, which acts on $(A_1,B_1^*)$ and $(B_2^*,A_2)$ simultaneously as doublets, and which corresponds to simultaneous rotations in (345) and (789). At this stage the theory also has the $SU(2)_D$ symmetry we mentioned in section \ref{codzerofieldtheory} as well as the $U(1)_b$ symmetry. The low-energy limit of Step 4 then enhances the $SU(2)_D$ to $SU(2)_A \times SU(2)_B$. As we reviewed in section \ref{abjmtheory}, the key observation then is that $SU(2)_R$ and $SU(2)_A \times SU(2)_B$ do not commute, hence the full R-symmetry is $SU(4)$.

When we add our defect flavor, the symmetry in Step 2 breaks to $U(1)_1 \times U(1)_2$, corresponding to independent rotations in (45) and (89). That breaks in Step 3 to the diagonal $U(1)_R$, corresponding to simultaneous rotations in (45) and (89). The symmetry at that stage is then $U(1)_R$, as well as $SU(2)_D$ and $U(1)_b$. That must be the case because the defect flavor fields couple only to adjoint fields: they do not couple directly to the bifundamentals $A_a$ and $B_a$, whose couplings are of the form written in eq. (\ref{abjmsupo1}), which preserves $SU(2)_D \times U(1)_b$.

We then know from the gravity side that after Step 4 the symmetry becomes $SU(2) \times SU(2) \times U(1) \times U(1)_b$. Without knowing the couplings of the defect flavor, we cannot say exactly which two $SU(2)$'s these are. For example, the $U(1)_R$ may be enhanced back to $SU(2)_R$, in which case the other $SU(2)$ must be $SU(2)_D$. Another possibility is that the $U(1)_R$ is enhanced back to the full $SU(2)_1 \times SU(2)_2$, in which case the $U(1)$ must be $U(1)_D$, the $U(1)$ part of $SU(2)_D$, which commutes with all of $SU(2)_1 \times SU(2)_2$.

For concreteness, we will take the latter scenario as a working assumption, and write a scalar potential that respects $SU(2)_1 \times SU(2)_2 \times U(1)_D \times U(1)_b$. The scalar potential must thus be built from $SU(2)_1 \times SU(2)_2$ invariants. For convenience we define
\be
C =\left(\begin{array}{c} A_1 \\ B_1^* \end{array} \right), \qquad D = \left(\begin{array}{c} B_2^* \\ A_2 \end{array} \right) \, ,
\ee
where $C$ transforms as a doublet under $SU(2)_1$ and $D$ as a doublet under $SU(2)_2$. The two $U(1)$ symmetries act as follows. The first is $U(1)_D$ (the $U(1)$ part of $SU(2)_D$), which acts as
\be
A_1 \ra e^{i \a} A_1, \qquad A_2 \ra e^{-i\alpha} A_2, \qquad B_1^* \ra e^{i \a} B_1^*, \qquad B_2^* \ra e^{-i\alpha} B_2^*. \nonumber \\
\ee
The second is the ABJM baryon number $U(1)_b$, which acts as $A_a \ra e^{i\a} A_a$ and $B_a \ra e^{-i\a} B_a$.  Moreover we have also two $\N=(2,2)$ fundamental multiplets confined to the defect. Their \textit{scalar} parts and the corresponding gauge transformations are given by
\be
q_1^n \rightarrow e^{i \Lambda_1} q_1^n, \qquad q_2^n \rightarrow e^{i \Lambda_2} q_2^n, \qquad n=1,2. 
\ee
Note that these scalar fields are confined to the (1+1)-dimensional defect, and have dimension zero. $n=1,2$ labels the two complex scalars in an $\N= (4,4)$ hypermultiplet.

These scalars are inert under all four global symmetry groups. To see why, we return to the D3/D3 theory of ref. \cite{Constable:2002xt}. As we mentioned above, that theory has an $SU(2) \times SU(2) \times U(1) \times U(1)$ R-symmetry, where the $SU(2) \times SU(2)$ part corresponds to the $SO(4)$ that acts on the directions transverse to both the color and flavor D3-branes, (4589), one $U(1)$ corresponds to rotations in (26) (along the color D3-branes but transverse to the flavor D3-branes), and the other $U(1)$ corresponds to rotations in (37) (transverse to the color D3-branes but along the flavor D3-branes). In the D3/D3 theory, the scalars transform as $(0,0)_{(\frac{1}{2}, - \frac{1}{2})}$ under the $SU(2)\times SU(2)\times U(1) \times U(1)$ R-symmetry. In the current system, the $SU(2)_1$ symmetry rotates (345) and $SU(2)_2$ rotates (789). (Recall that the diagonal of $SU(2)_1 \times SU(2)_2$ is $SU(2)_R$, corresponding to simultaneous rotations in (345) and (789).) The key point is that the two $U(1)$'s from the D3/D3 theory are broken here: rotations in (26) are clearly broken by the NS5-branes (in Step 2), and rotations in (37) alone are not part of the product group (rotations in (345)) $\times$ (rotations in (789)). The scalars are clearly neutral under all of the symmetries here.

With the above ingredients, the possible contributions to a scalar defect potential preserving $SU(2)_1 \times SU(2)_2 \times U(1)_D \times U(1)_b$ are
\ba
S_{\mathrm{def}} & = & \sum\limits_{n=1,2} \, \int \! d^2x \,  \big[ \bar q_1^n \bar D \bar C CD q_1^n \, + \bar q_1^n \bar D D \bar D D q_1^n + \, \bar q_2^n \bar C \bar D DC q_2^n + \bar q_2^n \bar C C \bar C C  q_2^n \big] \nonumber \\ & & + \int \! d^2x \, \big[ \bar q_1^1 \bar D \bar C CD q_1^2 \, + \, \bar q_1^1 \bar D D \bar D D q_1^2 \,+\, \bar q_2^1 \bar C \bar D DC q_2^2 + \bar q_2^1 \bar C C \bar C C  q_2^2 + (c.c.) \big].
\label{eq:codonenonchiralscalarpot}
\ea
The coefficient of each of these terms, which we have suppressed for notational clarity, remains to be determined. Parity ensures that the first and third terms in each line must have the same coefficient, and similarly for the second and fourth terms in each line.

Note that similar terms with additional factors of higher powers of the zero-dimensional scalars, such as $(\bar q_1^n q_1^n)^l$ with some integer $l$, are compatible with the symmetries listed above and might in principle be present in the potential terms given. However for a conformal theory with six real supercharges we expect such terms to be absent due to supersymmetric non-renormalization theorems.

We emphasize that in writing eq. (\ref{eq:codonenonchiralscalarpot}) we assumed that the final symmetry is $SU(2)_1 \times SU(2)_2 \times U(1)_D \times U(1)_b$. We stress that this identification of the symmetry is an assumption at this stage. Indeed, in this section we have seen that many questions arise for the theory describing codimension-one $\N=(3,3)$ supersymmetric flavor. We plan to further investigate these questions in the future.

\section{Codimension Two $\N=4$ Supersymmetric Flavor}
\label{codtwo}

In this section we study codimension-two flavor, that is, flavor fields propagating in a (0+1)-dimensional subspace of the (2+1)-dimensional ABJM theory. The flavor fields will preserve (0+1)-dimensional $\N=4$ supersymmetry. The flavor brane in this case is a D3-brane in type IIB, a D2-brane in type IIA, and an M2-brane in M-theory. We perform a complete supergravity analysis for these branes and comment on the structure of the field theory.

\subsection{Supergravity with M2/D2-brane probes}
\label{codtwogravity}

We begin by adding $N_f$ coincident D3-branes extended along (0346) in the type IIB setup. The strings from the flavor D3-branes to the two stacks of color D3-branes introduce massless flavor in both gauge groups. As shown in table \ref{table:IIB} in section \ref{gravityanalysis}, these D3-branes preserve 2 real supercharges.

After T-duality in $x^6$ the D3-branes become D2-branes. After uplift to M-theory and the ``near-horizon'' limit, the D2-branes become M2-branes on $\mathbbm{C}^4/\mathbbm{Z}_k$. Using the results of appendix \ref{app:IIBtoM}, we find that the embedding of the M2-branes is described by the equations $z^1=z^2=0$, $z^3 = \bar{z}^4$ in $\mathbbm{C}^4/\mathbbm{Z}_k$. Such an embedding breaks the $SU(4) \times U(1)_b$ symmetry of $\mathbbm{C}^4/\mathbbm{Z}_k$ to $SU(2) \times U(1) \times U(1)$. The $SU(2)$ symmetry acts on $(z^1,z^2)$, the first $U(1)$ acts as $(z^1,z^2) \rightarrow (e^{i \alpha} z^1,e^{i \alpha} z^2)$ and the second $U(1)$ acts as $(z^3,z^4) \rightarrow (e^{i \alpha} z^3,e^{-i \alpha} z^4)$. Note that in this case the $U(1)_b$ symmetry is broken. In appendix \ref{app:probeflat} we find that the M2-branes preserve 4 real supercharges.

If we take $N_c \rightarrow \infty$, we can replace the color M2-branes by their near-horizon geometry, which is $AdS_4 \times S^7/\mathbbm{Z}_k$. The probe M2-branes are now extended along $AdS_2$ inside $AdS_4$ and wrap a trivial one-cycle in $S^7/\mathbbm{Z}_k$. In appendix \ref{app:probeads} we analyze the $\kappa$-symmetry condition for these branes and find (as expected) that after the near-horizon limit the number of preserved supercharges has doubled to 8.

We have embedded the M2-branes such that they do not wrap the M-theory circle that shrinks when we reduce to type IIA. Put another way, they break the $U(1)_b$ symmetry of the geometry because they are localized in that circle direction. We conclude that for large $k$, such that $k^5 \gg N_c$, the M2-branes reduce to D2-branes in type IIA that wrap $AdS_2$ and a trivial one-cycle in $\cp$.

To summarize: in the type IIB setup we can add D3-branes that produce massless flavor fields for both gauge group factors. These D3-branes become M2-branes in M-theory on $\mathbbm{C}^4/\mathbbm{Z}_k$ that preserve 4 real supercharges. We thus expect that the flavor fields will preserve (in (0+1)-dimensional notation) $\N=4$ superconformal symmetry.  The corresponding R-symmetry group should be $SU(2)$ \cite{Claus:1998us,BrittoPacumio:1999ax}, and indeed an $SU(2)$ appears in the symmetry of the brane contruction. In the next subsection we will comment on the coupling of the (0+1)-dimensional flavor fields to the fields of the ABJM theory.

\subsection{Comments about the Field Theory}
\label{codtwofieldtheory}

We will apply our recipe again, and discuss some complications that arise when we add the NS5-branes in Step 2.

\bigskip

\textbf{Step 1: Construct the D3/D3 Theory}

In the type IIB setup we introduced D3-branes along (0346). As in section section \ref{codonenonchiralfieldtheory}, the relevant theory is the D3/D3 theory written in ref. \cite{Constable:2002xt}. The full theory preserves 8 real supercharges. In the flavor sector we have two (1+1)-dimensional $\N=(2,2)$ chiral superfields $Q$ and $\tilde Q$, which together form an $\N=(4,4)$ hypermultiplet. The flavor fields couple to an $\N=(4,4)$ vector multiplet. The action is written explicitly in components in appendix D of ref. \cite{Constable:2002xt}.

\bigskip

\textbf{Step 2: Add the NS5-branes}

We once again add the NS5- and NS5$^{\prime}$-branes along (012345). The brane configuration appears in the table below (where we write explicitly only the NS5-brane), followed by a second table with the relevant arrangement of fields into (2+1)-dimensional and (1+1)-dimensional multiplets. In the latter table we use the same notation as in section \ref{codonenonchiralfieldtheory}.

\begin{center}
        \begin{tabular}{|c|cccccccccc|}\hline
                &0&1&2&3&4&5&6&7&8&9\\ \hline
$N_c$ D3&$\bullet$&$\bullet$&$\bullet$&--&--&--&$\bullet$&--&--&--\\
$N_f$ D3&$\bullet$&--&--&$\bullet$&$\bullet$&--&$\bullet$&--&--&--\\
NS5&$\bullet$&$\bullet$&$\bullet$&$\bullet$&$\bullet$&$\bullet$&--&--&--&--\\\hline
        \end{tabular}
\end{center}

\begin{center}
\begin{tabular}{|c|c|c|}\hline
(2+1)d &$\N=4$ V $(A_{\mu}^{012},345,F_3,D_3)$&$\N=4$ H $(A_6,789,F_3^a,F_3^b)$\\\hline
(1+1)d &$\N=(4,4)$ V $(A_{\mu}^{06},5789,D_2,F_2)$&$\N=(4,4)$ H $(A_1,A_2,34,F_2^a,F_2^b)$\\\hline
\end{tabular}
\end{center}

The (1+1)-dimensional defect flavors couple to the (1+1)-dimensional $\N=(4,4)$ vector multiplet, which includes the components of the vector field along both the color and flavor D3-branes, $A_0$ and $A_6$, as well as the scalars transverse to both the color and flavor D3-branes, (5789). Following ref. \cite{Constable:2002xt}, we identify $D_2 = \frac{1}{\sqrt{2}} \left( D_4 + F_{12} \right)$, where $D_4$ is the auxiliary field from the (3+1)-dimensional $\N=1$ vector multiplet of the $\N=4$ SYM theory, and $F_{12}$ is the gauge field strength in the (12) directions. The components of the vector field along the color D3-branes but transverse to the flavor D3-branes, $A_1$ and $A_2$, and the scalars transverse to the color D3-branes but along the flavor D3-branes, (34), appear in an $\N=(4,4)$ hypermultiplet that does not couple to the defect flavors.

We next need to dimensionally reduce in the $x^6$ direction. The defect flavor action will then become (0+1)-dimensional, giving us our codimension-two flavor.  As mentioned in section \ref{gravityanalysis} and appendix \ref{app:IIBprobesusy}, the brane intersection above preserves 4 real supercharges, hence we expect (0+1)-dimensional $\N=4$ supersymmetry. Clearly the gauge field component $A_6$ will become a scalar in (0+1)-dimensions, hence the codimension-two defect flavor will couple to five scalars, $A_6$ and (5789). The defect flavors of course also couple to the gauge field component $A_0$.

After dimensional reduction we then impose the NS5-brane boundary conditions. These set to zero $A_6$ and (789), (they are in the (2+1)-dimensional $\N=4$ hypermultiplet, as shown above), so the defect flavors couple only to the single scalar 5. We may identify 5 as the single real scalar, called $\sigma$ in section \ref{abjmtheory}, in the (2+1)-dimensional $\N=2$ vector multiplet that is part of the $\N=4$ vector multiplet written above. Notice also that after dimensionally reducing and imposing the NS5-brane boundary conditions, a coupling to $D_2$ seems to survive, where now $D_2 = \frac{1}{\sqrt{2}} F_{12}$, since the NS5-brane boundary conditions set $D_4=0$, as explained in section \ref{codonenonchiralfieldtheory}.

\bigskip

\textbf{Step 3: Compactify $x^6$, form the $(1,k)$5-brane, and lift to M-theory}

As mentioned above, when we replace one NS5-brane with the $(1,k)$5-brane, the D3-brane along (0346) preserves 2 real supercharges. In the field theory we thus expect the Chern-Simons terms to break the supersymmetry to (0+1)-dimensional $\N=2$ supersymmetry. The R-symmetry should then be $U(1)$. That is consistent with the symmetry in the brane picture, where the two $U(1)$'s, corresponding to independent rotations in (34) and (78), are broken to a single $U(1)$, corresponding to simultaneous rotations in (34) and (78).

\bigskip

\textbf{Step 4: Take the low-energy limit}

As mentioned above, the flavor D3-brane becomes a codimension-two M2-brane in M-theory on $\mathbbm{C}^4/\mathbbm{Z}_k$, preserving an $SU(2) \times U(1) \times U(1)$ subgroup of the isometries, and 4 real supercharges. In the field theory, we thus expect an enhancement back to (0+1)-dimensional $\N=4$ supersymmetry. We suspect that the $SU(2)$ symmetry is the R-symmetry group (which would be consistent with the results of refs. \cite{Claus:1998us,BrittoPacumio:1999ax}). Notice also the interesting feature that the flavor fields should break the ABJM $U(1)_b$ symmetry. This is our only example in which that occurs.

\section{$SU(4)$ Equivalence of Probe Flavor}
\label{su4}

Although the ABJM construction starts with a fairly complicated brane setup in type IIB, we have seen in section \ref{reviewabjm} that after the ``near-horizon'' limit we end up with M2-branes probing $\mathbbm{C}^4/\mathbbm{Z}_k$. The ``near-horizon'' limit, that is, zooming in on the $\mathbbm{C}^4/\mathbbm{Z}_k$ singularity of the space $X_8$ mentioned in section \ref{typeIIAandM}, discards much of the complicated information of the type IIB setup. After taking $N_c \rightarrow \infty$ we reach M-theory on $AdS_4 \times S^7/\mathbbm{Z}_k$.

As mentioned in the introduction, for $k=1$ the addition of flavor branes in M-theory, namely codimension-two M2-branes and codimension-one M5-branes, was studied in ref. \cite{Myers:2006qr}. There the authors had to consider only one embedding for each probe brane since the $SO(8)$ isometry group of $\mathbbm{C}^4$ or $S^7$ can map any two supersymmetric embeddings into each other. If two brane embeddings are related by such an $SO(8)$ symmetry transformation, then they are physically equivalent. In other words, when $k=1$ all supersymmetric codimension-two M2-branes are physically equivalent, and similarly for supersymmetric codimension-one M5-branes.

For general $k$, the $\mathbbm{Z}_k$ orbifold of $\mathbbm{C}^4$ breaks the $SO(8)$ isometry group to $SU(4) \times U(1)_b$. Two supersymmetric brane embeddings may be related by an $SO(8)$ element that is not contained in $SU(4) \times U(1)_b$. In that case, we have two physically distinct ways of adding flavor. An interesting question is whether we can fully classify the supersymmetric embeddings of flavor branes in the ABJM theory, but that is beyond the scope of this paper. Here, we will discuss how to use the unbroken $SU(4) \times U(1)_b$ symmetry to show that certain probe branes are physically equivalent although they look very different in the type IIB setup. When that occurs, we will call the two type IIB D-branes ``$SU(4)$-equivalent.''

On the gravity side, we will present three examples of $SU(4)$-equivalent pairs. Two of these examples appeared above, in section \ref{codzerogravity}, for the codimension-zero KK monopole, and section \ref{codoneM5}, for the codimension-one M5-brane. Here we will present one more example, for a codimension-two M2-brane, and we will explore the field theory side more. In the field theory, $SU(4)$ equivalence occurs when two different theories flow to the same low-energy fixed point. In the language of our recipe, the two different theories are the theories we obtain at the end of Step 3, which flow to the same theory at low energy in Step 4. We will present one explicit example of such flow in what follows, for the codimension-zero case.

Two necessary conditions for two D-branes to be $SU(4)$-equivalent are 1.) they become the same object in M-theory and 2.) they have the same codimension. More precisely, as we do a T-duality along $x^6$ to go from type IIB to type IIA, two $SU(4)$-equivalent D-branes must have the same codimension in the directions (012). Furthermore, if both D-branes wrap $x^6$ or both do not wrap $x^6$, then they have to be both Dp-branes. Another possibility is that a type IIB D(p+1)-brane is equivalent to a type IIB D(p-1)-brane, if the D(p+1)-brane wraps $x^6$ and the D(p-1)-brane does not. Notice also that, in M-theory on $\mathbbm{C}^4/\mathbbm{Z}_k$, the orientation of the object does not affect the symmetries it preserves. (That is obvious in the $k=1$ case.) That means that, in addition to an $SU(4)\times U(1)_b$ transformation, we can also reverse the orientation of an object, so that, in type IIB, Dp-branes and anti-Dp-branes may be $SU(4)$ equivalent. Finally, an especially important point is that, due to the ``near-horizon'' limit in which the R-symmetry $SO(3)_R$ is enhanced to $SU(4)_R$, two branes in type IIB that preserve different amounts of supersymmetry and different subgroups of the $SO(3)_R$ may still be $SU(4) \times U(1)_b$ equivalent.

\subsection{Codimension-zero KK monopoles}

We start with the codimension-zero D5-branes along (012789) from section \ref{codzero}. As shown in ref. \cite{Hikida:2009tp}, these D5-branes become KK-monopoles on $\mathbbm{C}^4/\mathbbm{Z}_k$ with the embedding equations $z^1 = \bar{z}^3$, $z^2 =\bar{z}^4$. As we argued above, we can perform an $SU(4)$ transformation from the old coordinates $z^i$ to new coordinates $z^i_{new}$, such that the embedding becomes Im$(z^i_{new}) =0$, $\forall i$. Explicitly, the $SU(4)$ transformation is
\be
z^1_{new} = \frac{1}{\sqrt{2}}(z^1+z^3), \quad z^2_{new} = \frac{-i}{\sqrt{2}}(z^1-z^3), \quad z^3_{new} = \frac{1}{\sqrt{2}}(z^2+z^4), \quad z^4_{new} = \frac{-i}{\sqrt{2}}(z^2-z^4).
\ee
What happens if we start with a KK monopole described by Im$(z^i)=0$, $\forall i$ and return to type IIB (using the results of appendix \ref{app:IIBtoM})? Up to an $SU(4) \times U(1)_b$ transformation that only changes the constant value of $x^6$, we find D7-branes along (01235679) (If we reverse the orientation of the KK monopole, we can obtain anti-D7-branes, as explained above). These D7-branes are $SO(3)_R$ equivalent to the D7-branes along (01234678) listed in table \ref{table:IIB}. We summarize the $SU(4)$ equivalence in the following table.

\begin{center}
\begin{tabular}{|c|cc|cc|}\hline
Type IIB & D5 & (012789) & D7 & (01235679)\\\hline
M-theory & KK & $z^1 = \bar{z}^3, \,z^2 =\bar{z}^4$ & KK & $Im(z^i)=0$\\\hline
\end{tabular}
\end{center}

We have found two different types of D-branes in type IIB that lead to the same configuration in M-theory, and are therefore physically identical in M-theory. That might be surprising since the D-branes preserve different amounts of supersymmetry and different subgroups of the $SO(3)_R$ symmetry in the type IIB setup. We will, therefore, now show on the field theory side that both D-branes lead to the same theory upon taking the low-energy limit.

For the flavor D5-branes we reviewed the field theory in section \ref{codzerofieldtheory}, following refs. \cite{Hohenegger:2009as,Gaiotto:2009tk,Hikida:2009tp}. The action of the $\N=3$ supersymmetric (2+1)-dimensional flavor appears in eqs. \ref{eq:D7kineticterms} and \ref{eq:D7superpotential}. Here we will begin instead with anti-D7-branes along (01234678). We will apply our recipe once again.

\bigskip

\textbf{Step 1: Construct the D3/D7 Theory}

We begin with D3-branes along (0126) and anti-D7-branes along (01234678). Such an intersection preserves 8 real supercharges. The intersection has 4 ND directions, hence we obtain non-chiral flavor propagating in (3+1) dimensions (along (0126)). The field theory of the 4 ND D3/D7 intersection is well known: it is (3+1)-dimensional $SU(N_c)$ $\N=4$ supersymmetric Yang-Mills theory coupled to $\N=2$ supersymmetric hypermultiplets in the fundamental representation of $SU(N_c)$. The action is usually written in $\N=1$ superspace, and includes the usual kinetic terms for the flavor fields as well as a superpotential coupling for the flavor fields whose form is dictated by $\N=2$ supersymmetry. If we decompose the $\N=4$ vector multiplet into an $\N=1$ vector multiplet and three $\N=1$ chiral multiplets, then the superpotential includes a coupling of the flavor fields to the $\N=1$ chiral multiplet whose scalars represent fluctuations of the branes in the overall transverse directions, which here are (59). For more details about the D3/D7 theory, see ref. \cite{Erdmenger:2007cm} and references therein.

\bigskip

\textbf{Step 2: Add the NS5-branes}

We now add the NS5-brane and NS5$^{\prime}$-brane along (012345), and let the D3-branes end on them. As mentioned in appendix \ref{app:IIBprobesusy}, the system then preserves 4 real supercharges. Technically, we should perform a dimensional reduction from (3+1) dimensions to (2+1) dimensions (since the anti-D7-branes are extended along $x^6$) and then impose the NS5-brane boundary conditions. We know what the result has to be, however. The NS5-brane boundary conditions set to zero the scalars (789). That means that after imposing those boundary conditions, the flavor fields couple only to the single scalar 5. We also know that the theory has 4 real supercharges, or in (2+1)-dimensional language, $\N=2$ supersymmetry. The flavor fields must of course have kinetic terms, with the usual coupling to the $\N=2$ vector multiplet. The key observation is that the $\N=2$ vector multiplet includes a single real scalar (which, recalling the type IIB construction of the ABJM theory, must indeed be 5). We can conclude that the (2+1)-dimensional flavors have no superpotential: any superpotential coupling must preserve $\N=2$ supersymmetry, and hence must be a coupling to an $\N=2$ chiral superfield, but that would introduce couplings to additional scalars that are obviously absent here. In short, the flavor fields only couple to enough scalars for an $\N=2$ vector multiplet! The (2+1)-dimensional action in the flavor sector is then simply the $\N=2$ kinetic term, whose explicit form appears in eq. (\ref{eq:D7kineticterms}).

\bigskip

\textbf{Step 3: Compactify $x^6$, form the $(1,k)$5-brane, and lift to M-theory}

As usual, these steps leave the form of the flavor action untouched. Notice also that in this case the supersymmetry remains $\N=2$ throughout. For example, table \ref{table:IIB} in section \ref{gravityanalysis} shows that after we form the $(1,k)$5-brane the system still preserves 4 real supercharges. Notice also that the symmetries of the field theory and the brane construction agree. $\N=2$ supersymmetry has a $U(1)$ R-symmetry, and the anti-D7-brane along (01234678) clearly preserves the $U(1)$ subgroup of $SO(3)_R$ that rotates (34) and (78) simultaneously.

\bigskip

\textbf{Step 4: Take the low-energy limit}

Now we come to the crucial step. We must write all terms consistent with $\N=2$ supersymmetry, the $U(1)$ R-symmetry, and the $U(1)_b$ symmetry. Here we will borrow some arguments from ref. \cite{Gaiotto:2007qi}. Only one such term exists, a coupling to the $\N=2$ chiral fields $\Phi_i$, of the form written in eq. (\ref{eq:D7superpotentialbeforeintegratingout}). We must therefore add such a term to the superpotential, with some coefficient. Arguments similar to those in ref. \cite{Gaiotto:2007qi}, based on the sign of the two-loop beta function, then suggest that the coefficient flows to precisely the right value to produce the enhancement to $\N=3$ supersymmetry. The coupling is then \textit{identical} to the term in eq. (\ref{eq:D7superpotentialbeforeintegratingout}), and we thus recover exactly the same theory as in section \ref{codzero}.

\bigskip

We have thus seen how two different field theories flow to the same low-energy fixed point, and hence how $SU(4)$-equivalence appears on the field theory side. Notice that these two theories preserved different symmetries: the D5-brane along (012789) preserved the whole $SO(3)_R$ while the anti-D7-brane along (01234678) preserved only a $U(1)$ subgroup. We now turn to other examples, of higher codimension.

\subsection{Codimension-one M5-branes}

We return to the codimension-one D3-brane along (0137) from section \ref{codonenonchiral}. These become D4-branes after T-duality in $x^6$. Uplifting to M-theory and taking the ``near-horizon'' limit gives M5-branes in M-theory on $\mathbbm{C}^4/\mathbbm{Z}_k$. Using the results of appendix \ref{app:IIBtoM}, we find that the embedding of the M5-branes is described by the equations $z^1=z^2, z^3=z^4$. In section \ref{codonenonchiral} we used an $SU(4)$ transformation to produce new embedding equations. The transformation was
\be
z^1_{new} = \frac{1}{\sqrt{2}}(z^1-z^2), \quad z^2_{new} = \frac{1}{\sqrt{2}}(-z^3+z^4), \quad z^3_{new} = \frac{1}{\sqrt{2}}(z^1+z^2), \quad z^4_{new} = \frac{1}{\sqrt{2}}(z^3+z^4),
\ee
so that the embedding equation becomes $z_{new}^1=z_{new}^2=0$. The M5-brane is thus extended along (01) and $z_{new}^3$ and $z_{new}^4$. Going back to type IIB this embedding corresponds to the D5-branes along (013456). The D3-brane along (0137) and the D5-brane along (013456) are thus $SU(4)$ equivalent. We summarize the $SU(4)$ equivalence in a table:
\begin{center}
\begin{tabular}{|c|cc|cc|}\hline
Type IIB & D3 & (0137) & D5 & (013456)\\\hline
M-theory & M5 & $z^1=z^2, z^3=z^4$ & M5 & $z^1=z^2=0$\\\hline
\end{tabular}
\end{center}

Here we will briefly comment on the action that we obtain from the D5-brane along (013456).

\bigskip

\textbf{Step 1: Construct the D3/D5 Theory}

We begin with the action describing the (2+1)-dimensional defect fields in the D3/D5 intersection, which is the same action we mentioned in section \ref{codzero}, originally constructed in refs. \cite{DeWolfe:2001pq,Erdmenger:2002ex}. Once again, the (3+1)-dimensional $\N=4$ vector multiplet decomposes into two (2+1)-dimensional multiplets, an $\N=4$ vector multiplet and an $\N=4$ hypermultiplet. The flavor fields couple only to the $\N=4$ vector multiplet.

\bigskip

\textbf{Step 2: Add the NS5-branes}

We now add the NS5- and NS5$^{\prime}$-branes along (012345). The brane configuration appears in the table below.  Notice that we must dimensionally reduce the defect action along $x^6$ because the flavor D5-brane is now along $x^6$. We also of course have a new decomposition of fields from (2+1) dimensions to (1+1) dimensions.

\begin{center}
        \begin{tabular}{|c|cccccccccc|}\hline
                &0&1&2&3&4&5&6&7&8&9\\ \hline
$N_c$ D3&$\bullet$&$\bullet$&$\bullet$&--&--&--&$\bullet$&--&--&--\\
$N_f$ D5&$\bullet$&$\bullet$&--&$\bullet$&$\bullet$&$\bullet$&$\bullet$&--&--&--\\
NS5&$\bullet$&$\bullet$&$\bullet$&$\bullet$&$\bullet$&$\bullet$&--&--&--&--\\\hline
        \end{tabular}
\end{center}

We write the reduction of the fields in the table below. The first line is the arrangement of fields that is relevant for the D3/D5 intersection, that is, the flavor fields in that intersection couple to the $\N=4$ vector multiplet listed in the first line. The second line is then the arrangement of fields relevant for ABJM. The last line is the dimensional reduction of the first line to (1+1) dimensions, which is relevant for the D3/D5 intersection when $x^6$ is compact. In particular, notice that if we rewrite the action for the (2+1)-dimensional defect fields in the D3/D5 system in terms of (1+1)-dimensional fields, the flavors couple to the $\N=(4,4)$ vector multiplet whose four scalars are (789) (transverse to both color D3-branes and flavor D5-branes) and the vector field component $A_6$ (along the reduced direction).

\begin{center}
\begin{tabular}{|c|c|c|}\hline
(2+1)d D5&$\N=4$ V $(A_{\mu}^{016},789,F_{D5},D_{D5})$&$\N=4$ H $(A_2,345,F_{D5}^a,F_{D5}^b)$\\\hline
(2+1)d NS5&$\N=4$ V $(A_{\mu}^{012},345,F_{NS5},D_{NS5})$&$\N=4$ H $(A_6,789,F_{NS5}^a,F_{NS5}^b)$\\\hline
(1+1)d &$\N=(4,4)$ V $(A_{\mu}^{01},A_6,789,D_2,F_2)$&$\N=(4,4)$ H $(A_2,345,F_2^a,F_2^b)$\\\hline
\end{tabular}
\end{center}

At this stage, the system preserves 4 real supercharges, as mentioned in section \ref{gravityanalysis} and appendix \ref{app:IIBprobesusy}. The flavor fields will necessarily be non-chiral, since they come from the dimensional reduction of a (2+1)-dimensional theory. We thus expect an $\N=(2,2)$ supersymmetric action, which will include the kinetic term and possibly a superpotential.

The theory we obtain in this fashion is very different from what we obtain from the D3-brane along (0137). Here, the boundary conditions due to the NS5-branes set to zero all four scalars in the $\N=(4,4)$ vector multiplet, namely $A_6$ and (789). That suggests that the flavor fields do not couple to any $\N=(2,2)$ multiplet, simply because the defect flavor fields couple to no adjoint scalars \footnote{As shown in refs. \cite{DeWolfe:2001pq,Erdmenger:2002ex}, the defect flavor fields in the D3/D5 intersection do couple to normal derivatives of some adjoint scalars. More precisely, terms involving the derivative $\partial_2$ of the scalars (345) appear in the defect action as $F$-terms, that is, they appear in combination with the auxiliary field we called $F_{D5}$. Whether these scalars can solve this problem is not clear to us.}. Furthermore, $\N=(2,2)$ supersymmetry has the wrong R-symmetry to describe a D5-brane along (013456). $\N=(2,2)$ supersymmetry has a $U(1) \times U(1)$ R-symmetry, while the D5-brane clearly preserves the full $SO(3) \times SO(3)$ symmetry acting on (345) and (789).

\bigskip

\textbf{Step 3: Compactify $x^6$, form the $(1,k)$5-brane, and lift to M-theory}

Forming the $(1,k)$5-brane breaks one real supercharge. The theory we obtain at the end of Step 3 thus has 3 real supercharges and describes defect flavor fields that couple to no adjoint scalars, although of course they still couple to the gauge field components $A_0$ and $A_1$. Whatever this theory is, we predict that in Step 4 it flows at low energies to the theory we discussed in section \ref{codonenonchiral}. We leave the details of that analysis for future research.

\subsection{Codimension-two M2-branes}
Here we return to the D3-branes along (0346) from section \ref{codtwo}. As mentioned in section \ref{codtwogravity}, these D3-branes along (0346) become M2-branes on $\mathbbm{C}^4/\mathbbm{Z}_k$ with the embedding given by $z^1 = z^2=0$, $z^3 = r e^{i (y_0+ \frac{x^6_0+\phi}{2})}$, $z^4 = r e^{i (y_0 + \frac{x^6_0 -\phi}{2})}$, where $y_0$ and $x^6_0$ are constants. We now perform an $SU(4)$ transformation,
\ba
z^1_{new} = \frac{1}{\sqrt{2}}\lp z^1 + e^{-i x^6_0} z^4 \rp, & \quad & z^2_{new}  = \frac{1}{\sqrt{2}} \lp z^2 + e^{-i x^6_0} z^3\rp, \nn \\
z^3_{new} = \frac{1}{\sqrt{2}}\lp -e^{i x^6_0} z^2 + z^3\rp, & \quad & z^4_{new} = \frac{1}{\sqrt{2}}\lp -e^{i x^6_0} z^1 + z^4\rp, \nn
\ea
such that the embedding becomes
\ba
z^1_{new} = \frac{r}{\sqrt{2}} e^{i (y_0 + \frac{-x^6_0 -\phi}{2})}, & \quad & z^2_{new} = \frac{r}{\sqrt{2}} e^{i (y_0 + \frac{-x^6_0 +\phi}{2})}, \nn \\
z^3_{new} = \frac{r}{\sqrt{2}} e^{i (y_0 + \frac{x^6_0 +\phi}{2})}, & \qquad & z^4_{new} = \frac{r}{\sqrt{2}} e^{i (y_0 + \frac{x^6_0 -\phi}{2})}. \nn
\ea
This embedding describes, in type IIB, a D3-brane along (0678). We have thus shown that the D3-branes along (0346) and (0678) are $SU(4)$-equivalent. We summarize the $SU(4)$ equivalence in a table:
\begin{center}
\begin{tabular}{|c|cc|cc|}\hline
Type IIB & D3 & (0346) & D3 & (0678)\\\hline
M-theory & M5 & $z^1 = z^2=0, \, z^3 = e^{i(2y_0+ x^6_0)} \bar{z}^4$ & M5 & $z^1= e^{i(2y_0- x^6_0)} \bar{z}^2= e^{2 i y_0}\bar{z}^3= e^{-i x^6_0} z^4$\\\hline
\end{tabular}
\end{center}

Here again we will just comment briefly on the field theory.

\bigskip

\textbf{Step 1: Construct the D3/D3 Theory}

We begin again with the D3/D3 theory of ref. \cite{Constable:2002xt}, which we mentioned already in sections \ref{codonenonchiral} and \ref{codtwo}. In this case, the flavor fields are confined to propagate along the (1+1)-dimensional defect in the (06) directions. The defect flavors couple to an $\N=(4,4)$ supersymmetric vector multiplet that includes the scalars (3459), transverse to both the color and flavor D3-branes.

\bigskip

\textbf{Step 2: Add the NS5-branes}

We add the NS5- and NS5$^{\prime}$-branes along (012345). At this stage the system preserves 4 real supercharges, as mentioned in section \ref{gravityanalysis} and appendix \ref{app:IIBprobesusy}. We must also dimensionally reduce the defect action along $x^6$, since the flavor D3-branes are extended in $x^6$. We thus expect a (0+1)-dimensional defect action with $\N=4$ supersymmetry. The theory should preserve the same symmetries as the theory we obtain from the D3-brane along (0346): clearly in the brane description both flavor D3-branes preserve the same $SO(2)$ subgroup, rotating (34) and (78) simultaneously, of the $SO(3)_R$, that rotates (345) and (789) simultaneously. The couplings of the two theories are very different, however. For the theory from the D3-brane along (0346), the defect flavors couple to the scalar 5. For the theory from the D3-brane along (0678), the defect flavors will couple to the three scalars 345, since the scalar 9 is set to zero by the NS5-brane boundary conditions.

\bigskip

\textbf{Step 3: Compactify $x^6$, form the $(1,k)$5-brane, and lift to M-theory}

Forming the $(1,k)$5-brane breaks two real supercharges, so the theory then has (0+1)-dimensional $\N=2$ supersymmetry. The prediction of $SU(4)$ equivalence is then that, in Step 4, this theory flows at low energy to the same $\N=4$ supersymmetric theory we obtain from the D3-brane along (0346). We leave the details of that analysis to future research.

\section{Conclusion}
\label{conclusion}

We have studied a large class of supersymmetric flavor branes in the brane construction of the ABJM theory, and provided a general method to derive the corresponding field theories. We applied our method to four different examples. We first studied codimension-zero $\N=3$ supersymmetric flavor, which appeared in the supergravity description as D5-branes/D6-branes/KK-monopoles. We then studied codimension-one chiral $\N=(0,6)$ supersymmetric flavor, which appeared in supergravity as a D7/D8/M9-brane. Next we studied codimension-one non-chiral $\N=(3,3)$ supersymmetric flavor, which appeared in supergravity as a D3/D4/M5-brane. Finally we studied codimension-two $\N=4$ supersymmetric flavor, which appeared in supergravity as a D3/D2/M2-brane. In all cases we discussed both the supergravity description and the field theory description. On the supergravity side we studied the embedding of the brane/monopole in supergravity and the symmetries it preserved, including the number of supersymmetries. For the first two cases, on the field theory side we wrote the kinetic terms and couplings of the flavor fields explicitly and matched the symmetries to the supergravity description. In the last two cases we took steps toward constructing the Lagrangians, commenting in particular on the symmetries. Finally, we argued how in general different probe branes in type IIB can become physically equivalent in M-theory and therefore give rise to the same field theory. 

Our work is only the tip of the iceberg. Many extensions and generalizations are possible. We did not explore the matching of supergravity fields to field theory operators. Many deformations are also possible, for example, we can give the flavor fields a (supersymmetric or non-supersymmetric) mass in various ways. A nonzero mass (which breaks scale invariance) would allow us to compute meson spectra along the lines of ref. \cite{Kruczenski:2003be}. We can also deform the background, for example by replacing $\mathbbm{C}^4$ with a cone over some non-trivial seven-dimensional manifold, such that the theory to which we add flavor would have reduced supersymmetry. We can also ask what role flavor fields play in various dualities, such as mirror symmetry \cite{Jensen:2009xh}. More general types of probe branes are also possible \cite{Koerber:2007jb}, for example the author of ref. \cite{Koerber:2009he} studied the addition of co-isotropic codimension zero probe D8-branes on the gravity side, and the authors of ref. \cite{Haack:2009jg} studied more generally the addition of codimension-one domain walls. Many applications are also possible, especially in the context of hydrodynamics and/or condensed matter physics, for example along the lines of refs. \cite{Myers:2008me,Alanen:2009cn,Kachru:2009xf} (as just a small sample).

Another particularly interesting extension of our work would be to study back-reaction effects. The effect of the KK monopoles on the metric of eleven-dimensional supergravity has already been studied in refs.  \cite{Hohenegger:2009as,Gaiotto:2009tk,Fujita:2009xz}. As stated in ref. \cite{Fujita:2009kw}, when we include back-reaction we should see that the D4/M5-branes change the rank of the gauge group(s). The back-reaction of the M5-branes can in principle be studied using the methods of ref. \cite{D'Hoker:2009gg}. As stated in refs. \cite{Fujita:2009kw,Gaiotto:2009mv,Gaiotto:2009yz}, when we include back-reaction the D8-brane will source the Romans mass (Ramond-Ramond zero-form field strength), which means that the sum of the Chern-Simons levels of the two gauge groups will no longer be zero. 

\nobreak

We plan to study these and various related issues in the future.

\section*{Acknowledgements}
We thank Oren Bergman, Ralph Blumenhagen, Davide Cassani, John Estes, Amihay Hanany, Stefan Hohenegger, Daniel Jafferis, Dieter L\"ust, Juan Maldacena, Andrew Royston, Oliver DeWolfe, and Marco Zagermann for useful discussions. We are especially grateful to Andreas Karch, Ingo Kirsch, and Paul Koerber for reading and commenting on the manuscript. This work was supported in part by the Cluster of Excellence ``Origin and Structure of the Universe.'' M. A. would also like to thank the Studienstiftung des deutschen Volkes for financial support. The work of T.W. is also supported by the German Research Foundation (DFG) within the Emmy-Noether Programme (Grant number ZA 279/1-2). J.E. thanks the Kavli Institute for Theoretical Physics in Santa Barbara, R. M. thanks the Erwin Schroedinger International Institute for Mathematical Physics in Vienna, and A. O'B. thanks the Aspen Center for Physics for hospitality while this work was in progress.

\vspace{10mm}

\noindent \textbf{\Large Appendix}

\appendix

\section{Supersymmetry of Type IIB Probes}
\label{app:IIBprobesusy}

In this appendix we will add supersymmetry-preserving D-branes to the type IIB setup of the ABJM theory. Our starting point is the list of probes in ref. \cite{Skenderis:2002vf}, since these are known to be mutually supersymmetric with respect to the D3-branes, but now we have two new ingredients: the NS5-branes and the (1,k)5-brane\footnote{Another new ingredient is that the $x^6$ direction is compact. While that is crucially important for deriving the low-energy (2+1)-dimensional worldvolume field theory, it will not affect the counting of supersymmetries.}. These new ingredients break the ``usual'' $SO(6)$ symmetry, which rotates the $(345789)$ directions into one another, down to the $SO(3)$ that rotates $(345)$ and $(789)$ simultaneously.

As mentioned in section \ref{gravityanalysis}, we limit our search for supersymmetric probes by imposing four constraints. First, we consider only D1-, D3-, D5- and D7-brane probes (the list from ref. \cite{Skenderis:2002vf}). Second, we do not separate any probes from the D3-branes in overall transverse directions. Third, when we consider multiple probes, \textit{i.e.} $N_f > 1$, we do not separate them from each other, so that they retain a $U(N_f)$ symmetry. Fourth, we consider only probes aligned along the coordinate axes. More generally the probe brane could be at an angle with respect to these axes. We studied a few special cases of probes at angles (see below) and found that such probes never appear to exhibit enhanced supersymmetry.

Our results are summarized in table \ref{table:IIB} in section \ref{gravityanalysis}, reproduced here as table \ref{table:IIB2}. The details of the notation (such as the column headings) appear in section \ref{gravityanalysis}.

\begin{table}[h!]
\begin{center}
\begin{tabular}{|c|c|c|c|c|c|c|}
  \hline
  Type IIB & Type IIA & M theory & codim & wrapping & SUSY & SUSY (anti)\\ \hline \hline
  D1 & D2 & M2 & 2 & 0(7) & 2 & 2  \\\hline
  D3 & D2 & M2 & 0 & 0126 & 6 & 0 \\\hline
  D3 & D4 & M5 & 1 & 01(37) & 3 & 3 \\\hline
  D3 & D4 & M5 & 1 & 01(38) & 2 & 2\\\hline
  D3 & D2 & M2 & 2 & 0(34)6 & 2 & 2\\\hline
  D3 & D2 & M2 & 2 & 06(78) & 2 & 2\\\hline
  D5 & D6 & KK & 0 & 012(347) & 2 & 2 \\\hline
  D5 & D6 & KK & 0 & 012(349) & 4 & 2\\\hline
  D5 & D6 & KK & 0 & 012789 & 6 & 0\\\hline
  D5 & D4 & M5 & 1 & 013456 & 3 & 3\\\hline
  D5 & D4 & M5 & 1 & 01(378)6 & 2 & 2\\\hline
  D5 & D4 & M5 & 1 & 01(389)6 & 3 & 3\\\hline
  D5 & D6 & KK & 2 & 0(34)789 & 2 & 2\\\hline
  D7 & D6 & KK & 0 & 0126(3478) & 2 & 4\\\hline
  D7 & D6 & KK & 0 & 0126(3479) & 2 & 2\\\hline
  D7 & D8 & M9 & 1 & 01345789 & 3 & 3\\\hline
\end{tabular}
\end{center}
\caption{List of D-branes (extended along the coordinate axes) that we can add to the type IIB construction while still preserving some supersymmetry.}\label{table:IIB2}
\end{table}

To count unbroken supercharges we follow ref. \cite{Kitao:1998mf} very closely. In particular, we perform a different T-duality from the one that leads to the ABJM theory: we T-dualize in $x^2$, not $x^6$, and then lift to M-theory. The type IIB construction reviewed in section \ref{typeIIBconstruction} then has a very simple interpretation in terms of M-branes alone (rather than M-branes in some nontrivial geometry). The D3-branes become M2-branes along $(016)$, the NS5-brane becomes an M5-brane along $(012345)$, and the $(1,k)5$-brane becomes an M5-brane tilted at an angle $\theta$ in the $(37)$, $(48)$, and $(59)$ directions and an angle $-\theta$ in the $(2\#)$ directions, relative to the other M5-brane, where $\tan \theta = k$.

Let $\e$ denote the 32-component Majorana spinor and $\G_A$ the $32\times32$ $\G$-matrices  of 11-dimensional supergravity. The $\G$ matrices obey the flat space Clifford algebra $\{\G_A ,\G_B\} = 2 \eta_{A B}$, where we use a mostly-plus metric. Let $\G_{ABC\ldots}$ denote the totally antisymmetric product of $\G$-matrices, equivalent to the usual product due to the Clifford algebra. The product $\G_{0123456789\#}=\mathbbm{1}_{32}$, where $\mathbbm{1}_{32}$ is the $32\times32$ identity matrix.

The M2- and M5-branes give rise to projection conditions on $\e$,
\be
\G_{016} \e = \e, \qquad \G_{012345} \e = \e, \qquad R \, \G_{012345} \, R^{-1} \e = \e,
\ee
where the last condition, for the rotated M5-brane, involves the rotation matrix
\be
R(\theta) = \exp \left(- \frac{\theta}{2} \G_{2\#} + \frac{\theta}{2} \G_{37} + \frac{\theta}{2} \G_{48} + \frac{\theta}{2} \G_{59} \right).
\ee
Notice that $R^{-1}(\theta) = R(-\theta)$. Making use of the fact that all of the $\G$-matrices in $R$ anti-commute with $\G_{012345}$, we can simplify the condition for the rotated M5-brane,
\be
R \, \G_{012345} \, R^{-1} \e = R^2 \G_{012345} \e = R^2 \e = \e,
\ee
where we used the projection condition for the un-rotated M5-brane. We can then write the rotated M5-brane's projection condition as $(R^2 - \mathbbm{1}_{32}) \e = 0$. At this point we need to write the matrices appearing in the projection conditions explicitly, in order to count the number of preserved supercharges. Following ref. \cite{Kitao:1998mf}, we use a basis in which the following set of mutually-commuting matrices are diagonal:
\ba
\G_{012345} & = & (\mathbbm{1}_{16},-\mathbbm{1}_{16}) \nonumber \\
\G_{016} & = & (\mathbbm{1}_2,-\mathbbm{1}_2,-\mathbbm{1}_2,\mathbbm{1}_2,-\mathbbm{1}_2,\mathbbm{1}_2,\mathbbm{1}_2,-\mathbbm{1}_2,\ldots) \nonumber \\
\G_{2\#37} & = & (\mathbbm{1}_8,-\mathbbm{1}_8,\ldots)\nonumber \\
\G_{2\#48} & = & (\mathbbm{1}_4,-\mathbbm{1}_4,\mathbbm{1}_4,-\mathbbm{1}_4,\ldots)\nonumber\\
\G_{2\#59} & = & (\mathbbm{1}_2,-\mathbbm{1}_2,\mathbbm{1}_2,-\mathbbm{1}_2,\mathbbm{1}_2,-\mathbbm{1}_2,\mathbbm{1}_2,-\mathbbm{1}_2,\ldots) \nonumber.
\ea
In this basis, the matrix in the projection condition for the rotated M5-brane becomes
\be
R^2 - \mathbbm{1}_{32}= 2 R \G_{2\#} \left( \sin (- 2 \theta) \mathbbm{1}_2, \sin (-\theta) \mathbbm{1}_2, \sin(-\theta) \mathbbm{1}_2, 0_2, \sin(-\theta) \mathbbm{1}_2, 0_2, 0_2, \sin(\theta) \mathbbm{1}_2,\ldots\right).
\ee
The $0_2$'s in this equation that overlap with the $\mathbbm{1}_2$'s in $\G_{016}$ indicate which components of $\e$ will be preserved, hence the full system of M2-brane, M5-brane, and rotated M5-brane preserves 6 real supercharges.

To study probe branes we first need to translate all the type IIB D-branes to the M-theory description, which produces various M2- and M5-branes, as well as KK monopoles. We will not present every case in detail, rather, we will just show a few representative examples, with decreasing amounts of supersymmetry.

First we note that when $k=0$, such that the rotation matrix $R$ is simply the identity (and in the type IIB description we have just NS5-branes), all of the objects we study preserve 4 real supercharges, with two exceptions: the D3-branes along (0126) and the D5-branes along (012789). These two D-branes preserve 8 real supercharges when $k=0$.

For nonzero $k$, the easiest example is in fact the D5-brane along (012789) (see also refs. \cite{Hanany:1996ie,Kitao:1998mf}), which upon T-duality in $x^2$ and lift to M-theory becomes an M5-brane along $(01789\#)$. The projection condition is $\G_{01789\#} \e = \e$. We can use $\G_{0123456789\#}=\mathbbm{1}_{32}$ to write $\G_{01789\#} = \G_{016} \G_{012345}$, hence this M5-brane does not impose any additional constraint on $\e$, and preserves 6 real supercharges.

An example preserving 4 real supercharges is a D5-brane along $(012349)$, which upon T-duality in $x^2$ and uplift to M-theory becomes an M5-brane along $(01349\#)$. In this case we use $\mathbbm{1}_{32} = - \G_{25} \G_{25}$ to write $\G_{01349\#} = - \G_{012345} \G_{2\#59}$. In the upper-left $16 \times 16$ block, $\G_{012345}$ is simply the identity, so in this subspace $\G_{01349\#} = - \G_{2\#59}$. Using the $\G$-matrices written explicitly above, we can count that this probe preserves 4 real supercharges. The same steps obviously apply for cases related to this one by $SO(3)$ rotations. An anti-D5-brane will have $-\G_{01349\#} = \G_{2\#59}$, and hence will preserve 2 real supercharges.

An example preserving 3 real supercharges is a D3-brane along $(0137)$. Here we first rotate the D3-brane so that it is extended along (0237), which then becomes an M2-brane along $(037)$. We insert $\mathbbm{1}_{32} = - \G_{2\#} \G_{2\#}$ to write $\G_{037} = - \G_{02\#} \G_{2\#37}$. We need to know the additional $\G$ matrix,
\be
\G_{02\#} = \left(\s_1,\s_1,-\s_1,-\s_1,\s_1,\s_1,-\s_1,-\s_1,\ldots \right),
\ee
where $\s_1$ is the first Pauli matrix, $\s_1 = ((0,1),(1,0))$. We thus have
\be
\G_{037} \e = - \G_{02\#} \G_{2\#37} \e  = \left(-\s_1,-\s_1,\s_1,\s_1,\s_1,\s_1,-\s_1,-\s_1,\ldots \right) \e = \e.
\ee
Each $\s_1$ imposes an additional constraint on the two components of $\e$ in that $2 \times 2$ block, hence this brane ``kills'' half of the supercharges, \textit{i.e} it preserves 3 real supercharges.

The cases preserving 2 real supercharges are slightly more involved. For example, consider the D1-brane along $(07)$, which becomes an M2-brane along $(027)$. Inserting $\mathbbm{1}_{32} = \G_3 \G_3$ we have $\G_{027} \e = - \G_{23} \G_{037} \e = \e$. We know that $\G_{037}$ is $2 \times 2$ block-diagonal in this basis, but $\G_{23}$ is not (it can be written as $\G_{23} = - i \, \mathbbm{1}_2 \otimes \mathbbm{1}_2 \otimes \s_1 \otimes \s_1 \otimes \mathbbm{1}_2$). Nevertheless, these cases are straightforward to check explicitly, although we will not present the details.

As mentioned above, we could also consider probe branes rotated with respect to the coordinate axes. In principle, such branes could have enhanced supersymmetry. We have not analyzed all possible rotations. For many of the branes in our table we considered the special case in which the brane is rotated by one angle in the $(37)$, $(48)$ and $(59)$ planes simultaneously and by an independent angle in the $(2\#)$ plane. In all cases the rotated brane never exhibits enhanced supersymmetry, and in most cases preserves fewer supersymmetries.

\section{Type IIB to M-theory}
\label{app:IIBtoM}
When we add flavor branes to the brane construction of the ABJM theory, many aspects of the field theory are best understood from the type IIB description, while the symmetries and the amount of supersymmetry preserved become manifest after the ``near-horizon'' limit in M-theory. To determine where the probe D-branes of the type IIB setup end up in M-theory on $\mathbbm{R}^{2,1} \times \mathbbm{C}^4/\mathbbm{Z}_k$, we need to find the coordinate transformations from the type IIB coordinates, $x^m$ with $m=0,\ldots,9$, to the M-theory coordinates $z^i, \, i=1,2,3,4$ on $\mathbbm{C}^4$. Our objective in this appendix is to write the $z^i$ in terms of the $x^m$ (and vice versa) explicitly, so that we can translate directly between the two coordinate systems.

As reviewed in section \ref{typeIIAandM}, when we T-dualize along $x^6$ and then lift to M-theory, the NS5-brane and $(1,k)$5-brane both become KK monopoles in M-theory extended along $(012)$, so that we only need to consider the eight other directions, which we denote by
\be
\varphi_1 = x^6, \quad \varphi_2 = x^\sh, \quad \vec{x}^1 = \left(
                                  \begin{array}{c}
                                    x^7 \\
                                    x^8 \\
                                    x^9 \\
                                  \end{array}
                                \right), \quad \vec{x}^2 = \left(
                                  \begin{array}{c}
                                    x^3 \\
                                    x^4 \\
                                    x^5 \\
                                  \end{array}
                                \right).
\ee
Both $x^6$ and $x^\sh$ have $2\pi$ periodicity. The metric describing the intersection of the two KK monopoles is \cite{Gauntlett:1997pk}
\ba\label{eq:intersectingKKmetric}
ds^2 &=& U_{ij} d\vec{x}^i \cdot d\vec{x}^j + U^{ij}(d \varphi_i+A_i)(d\varphi_j + A_j),\\
A_i &=& d\vec{x}^j \cdot \vec{\om}_{ji} = d x^j_a \om^a_{ij}, \quad \p_{x_a^j} \om^b_{ki} - \p_{x_b^k} \om^a_{ji} = \e^{abc} \p_{x^j_c} U_{ki},
\ea
where $U^{ij}$ is the transposed inverse of $U_{ij}$. Notice in particular that the metric is uniquely determined by the matrix $U$ and that the equations are linear in $U$ so that we can obtain the configuration for two monopoles simply by linear superposition.

The NS5-brane becomes a KK monopole associated with the circle $\varphi_1 = x^6$ and transverse directions $\vec{x}^1$. The corresponding $U$ matrix reads
\be
U = \left(
      \begin{array}{cc}
        1 & 0 \\
        0 & 1 \\
      \end{array}
    \right) + \left(
                \begin{array}{cc}
                  h_1 & 0 \\
                  0 & 0 \\
                \end{array}
              \right), \quad h_1 = \frac{1}{2 |\vec{x}^1|}.
\ee
The identity matrix in $U$ indicates that asymptotically the space is $\mathbb{R}^6 \times T^2$. The $(1,k)$5-brane is rotated in the ($\vec{x}^1,\vec{x}^2$)- and ($\varphi_1, \varphi_2$)-plane. The $U$ matrix of such a KK monopole is
\be
U = \left(
      \begin{array}{cc}
        1 & 0 \\
        0 & 1 \\
      \end{array}
    \right) + \left(
                \begin{array}{cc}
                  h_2 & k h_2 \\
                  k h_2 & k^2 h_2 \\
                \end{array}
              \right), \quad h_2 = \frac{1}{2 |\vec{x}^1+k \vec{x}^2|}.
\ee
The metric describing the two intersecting KK monopoles is then
\be\label{eq:U}
U = \left(
      \begin{array}{cc}
        1 & 0 \\
        0 & 1 \\
      \end{array}
    \right) + \left(
                \begin{array}{cc}
                  h_1 & 0 \\
                  0 & 0 \\
                \end{array}
              \right) + \left(
                \begin{array}{cc}
                  h_2 & k h_2 \\
                  k h_2 & k^2 h_2 \\
                \end{array}
              \right).
\ee
The metric in eq. (\ref{eq:intersectingKKmetric}), with the $U$ matrix in eq. (\ref{eq:U}), is the metric of the space $X_8$ mentioned in section \ref{typeIIAandM}.

To relate the type IIB coordinates $\varphi_1$, $\varphi_2$, $\vec{x}^1$ and $\vec{x}^2$ to the $\mathbb{C}^4$ coordinates $z^i$, we proceed in five steps. The first step is to take the ``near-horizon'' limit \cite{Aharony:2008ug} that we described in section \ref{typeIIAandM}. The four subsequent steps are simply changes of coordinates.

The ``near-horizon'' limit consists of taking $\vec{x}^1 \sim \vec{x}^2 \sim 0$, which in simple terms means we drop the identity matrix from the $U$ in eq. \eqref{eq:U}.

Now we change coordinates four times. The first change of coordinates will diagonalize the new $U$, producing the ``near-horizon'' metric of strictly perpendicular KK monopoles \cite{Aharony:2008ug}:
\ba
&\vec{x}^{'1} = \vec{x}^1, &\vec{x}^{'2} = \vec{x}^1 + k \vec{x}^2,\\
&\varphi_1' = x^6 - \frac1k x^\sh, \quad &\varphi_2'= \frac1k x^\sh,
\ea
with new $U$ matrix $U'$,
\ba
U' = \left(
      \begin{array}{cc}
        \frac{1}{2 |\vec{x}^{'1}|} & 0 \\
        0 & \frac{1}{2 |\vec{x}^{'2}|} \\
      \end{array}
    \right).
\ea
The new circle coordinates, $\varphi'_i, \, i=1,2$, are $2 \pi$ periodic, but the $2 \pi$ periodicity of $x^\sh$ leads to an extra identification
\be\label{eq:Zkaction}
(\varphi_1',\varphi_2') \sim (\varphi_1',\varphi_2') +\frac{2\pi}{k} (-1,1),
\ee
that is, if we shift $\varphi_1'$ by a multiple of $\frac{2\pi}{k}$ and simultaneously shift $\varphi_2'$ by the opposite amount, we end up at the same point. The above periodicity leads to the $\mathbb{Z}_k$ action on $\mathbbm{C}^4$, as we will see below. In the new coordinates we have two perpendicular KK monopoles so we will treat them simultaneously. The Taub-NUT metric of a single KK monopole in the ``near-horizon'' limit is
\be
ds^2_i = \frac{1}{2 |\vec{x}^{'i}|} d(\vec{x}^{'i})^2 + 2|\vec{x}^{'i}| \lp d\varphi'_i + A_i \rp^2,
\ee
where $i=1,2$ labels the KK monopoles.

Now we do the second change of coordinates, which is simply a change from Euclidean to spherical coordinates, $(d\vec{x}^{'i})^2 = dr_i^2 + r_i^2(d\th_i^2 +\sin{^2\th_i} d\phi_i^2)$, so that we obtain
\be
ds^2_i = \frac{1}{2 r_i} dr_i^2 + \frac{r_i}{2} \lp d\th_i^2 +\sin{^2\th_i} d\phi_i^2 \rp + 2 r_i \lp d\varphi'_i + \frac12 \cos{\th_i} d\phi_i \rp^2,
\ee
where we have used $A_i = \frac12 \cos{\th_i} d\phi_i$.

In the third change of coordinates we define a new radial coordinate $r_i =\r_i^2/2$ and new angles $\varphi'_i = \psi_i/2$. The metric then becomes that of flat space, with an extra $\mathbbm{Z}_k$ identification,
\be
ds^2_{i} = d\r_{i}^2 + \frac{\r_{i}^2}{4} (d\th_{i}^2 + d\phi_{i}^2 + d\psi_{i}^2 + 2 \cos{\th_{i}} d\phi_{i} d\psi_{i}).
\ee
The three angles have ranges $0 \leq \th_i < \pi$, $0 \leq \phi_i < 2 \pi$ and $0 \leq \psi_i < 4 \pi$, and the $\psi_i$ have the extra identification
\be\label{psiZkaction}
(\psi_1,\psi_2) \sim (\psi_1,\psi_2) +\frac{4\pi}{k} (-1,1),
\ee
following from eq. \eqref{eq:Zkaction}.\\

In the fourth and final change of coordinates, we introduce complex coordinates for the first KK monopole,
\be
z^1 = \r_1 \cos{\lp \frac{\th_1}{2} \rp} e^{-i (\psi_1+\phi_1)/2}, \quad z^2 = \r_1 \sin{\lp \frac{\th_1}{2} \rp} e^{-i (\psi_1-\phi_1)/2},
\ee
while for the second KK monopole we choose something similar, but with $i \rightarrow -i$,
\be
z^3 = \r_2 \cos{\lp \frac{\th_2}{2} \rp} e^{i (\psi_2+\phi_2)/2}, \quad z^4 = \r_2 \sin{\lp \frac{\th_2}{2}\rp} e^{i (\psi_2-\phi_2)/2}.
\ee
In these coordinates, the Taub-NUT metrics in the ``near-horizon'' limit have become
\be
ds^2_1 = dz^1 d\bar{z}^1 + dz^2 d\bar{z}^2, \quad ds^2_2 = dz^3 d\bar{z}^3 + dz^4 d\bar{z}^4,
\ee
and the $\mathbbm{Z}_k$ transformation of eq. \eqref{psiZkaction} acts as $z^i \rightarrow e^{\frac{2 \pi i}{k}} z^i$.

Tracing back through our coordinate transformations, we can write the original type IIB coordinates (plus $x^{\sh}$) in terms of the $z^i$:
\ba
x^6 &=& \frac{1}{2} \arg{(\bar{z}^1 \bar{z}^2 z^3 z^4)},\\
x^\sh &=& \frac{k}{2} \arg{(z^3 z^4)},\\
\left(
  \begin{array}{c}
    x^7 \\
    x^8 \\
    x^9 \\
  \end{array}\right) &=& \left(
                \begin{array}{c}
                  \text{Re}(z^1 \bar{z}^2) \\
                  -\text{Im}(z^1 \bar{z}^2) \\
                  \frac12 (|z^1|^2 - |z^2|^2) \\
                \end{array}
              \right),\\
\left(
  \begin{array}{c}
    x^3 \\
    x^4 \\
    x^5 \\
  \end{array}\right) &=& \frac{1}{k} \lp \left(
                \begin{array}{c}
                  \text{Re}(z^3 \bar{z}^4) \\
                  \text{Im}(z^3 \bar{z}^4) \\
                  \frac12 (|z^3|^2 - |z^4|^2) \\
                \end{array}
              \right)- \left(
                \begin{array}{c}
                  \text{Re}(z^1 \bar{z}^2) \\
                  -\text{Im}(z^1 \bar{z}^2) \\
                  \frac12 (|z^1|^2 - |z^2|^2) \\
                \end{array}
              \right) \rp.
\ea
Inverting the above expressions, we can write the $z^i$ in terms of the type IIB coordinates (plus $x^{\sh}$):
\ba\label{eq:zcoord}
|z^1|^2 &=& x^9 +\sqrt{(x^7)^2+(x^8)^2+(x^9)^2},\nn\\
|z^2|^2 &=& -x^9 +\sqrt{(x^7)^2+(x^8)^2+(x^9)^2},\nn\\
|z^3|^2 &=& (x^9+k x^5) +\sqrt{(x^7+k x^3)^2+(x^8+k x^4)^2+(x^9+ k x^5)^2},\nn\\
|z^4|^2 &=& -(x^9+k x^5) +\sqrt{(x^7+k x^3)^2+(x^8+k x^4)^2+(x^9+ k x^5)^2},\nn\\
\arg{z^1} &=& \frac{x^\sh}{k} - x^6 - \frac12 \arctan{\frac{x^8}{x^7}},\\
\arg{z^2} &=& \frac{x^\sh}{k} - x^6 + \frac12 \arctan{\frac{x^8}{x^7}},\nn\\
\arg{z^3} &=& \frac{x^\sh}{k} + \frac12 \arctan{\frac{x^8+k x^4}{x^7+k x^3}},\nn\\
\arg{z^4} &=& \frac{x^\sh}{k} - \frac12 \arctan{\frac{x^8+k x^4}{x^7+k x^3}}.\nn
\ea

Recall that the $z^i$ transform as a \textbf{4} of $SU(4)$, which clearly acts nontrivially on the $x^m$. The $U(1)_b$ is just a common phase shift $z^i \rightarrow e^{i \a} z^i$. The only coordinate that changes under the $U(1)_b$ is $x^{\sh}$, which shifts as $x^{\sh} \rightarrow x^{\sh} + k \alpha$.

As explained in section \ref{dualgravity}, we can take a large-$N_c$ limit in M-theory, so that the geometry becomes $AdS_4 \times S^7/\mathbbm{Z}_k$, and then take also large $k$, so that a circle in M-theory shrinks and we obtain type IIA in $AdS_4 \times \cp$. Where in the geometry is the circle that shrinks when $k \rightarrow \infty$? To answer this question, notice that the circle direction
\be
x^6 = \frac{1}{2} \arg{(\bar{z}^1 \bar{z}^2 z^3 z^4)}
\ee
is invariant under the $\mathbbm{Z}_k$ action, hence the circle that shrinks when $k \rightarrow \infty$ must be part of $x^\sh$. To show this explicitly, we return to our third change of coordinates, which involved the angles $\psi_1$ and $\psi_2$, with $\mathbbm{Z}_k$ acting as in eq. (\ref{psiZkaction}). Tracing back through the changes of coordinates, we find $x^6 = \frac{1}{2} \left( \psi_1 + \psi_2 \right)$, which is of course invariant, and $x^\sh = \frac{k}{2} \psi_2$, on which the $\mathbbm{Z}_k$ acts as a $2\pi$ shift. We can then define a coordinate $y$,
\be
y = \frac{1}{4} \left( \psi_2 - \psi_1 \right) = - \frac{1}{2} x^6 + \frac{1}{k} x^\sh.
\ee
such that the $\mathbb{Z}_k$ acts as $y \sim y + \frac{2 \pi}{k}$ but leaves all other coordinates invariant. The direction $y$ is thus the circle that shrinks when $k \rightarrow \infty$. From a type IIB perspective we are ``decomposing'' $x^\sh$ as $x^\sh = k y + \frac{k x^6}{2}$. In terms of the $z^i$,
\be
\arg{z^1} = y-\frac{x^6}{2}-\frac{\phi_1}{2}, \qquad \arg{z^2} = y-\frac{x^6}{2}+\frac{\phi_1}{2}, \nn
\ee
\be
\arg{z^3} =  y+\frac{x^6}{2}+ \frac{\phi_2}{2}, \qquad \arg{z^4} =  y+\frac{x^6}{2}- \frac{\phi_2}{2}\nn,
\ee
which shows that $y$ is simply the sum of the phases of the $z^i$.

A crucial question is whether a D-brane in type IIB remains a D-brane in type IIA on $AdS_4 \times \cp$. After T-duality from type IIB, when we first lift to M-theory, the circle $x^\sh$ opens up. We then take the ``near-horizon'' limit and $N_c \ra \infty$ to obtain M-theory on $AdS_4 \times S^7/\mathbbm{Z}_k$, and then we take large $k$, so that the $y$ circle shrinks, and the theory reduces to type IIA on $AdS_4 \times \cp$. In short, $x^\sh$ opens up but $y$ shrinks. A D4-brane in type IIA will become an M5-brane when $x^\sh$ opens up, but what happens when $y$ closes? Does the M5-brane reduce to a D4-brane again, or an NS5-brane with D4-brane flux?

The easiest way to see what happens is to return to our first change of coordinates, and in particular to consider the torus spanned by the coordinates $\varphi_1'=x^6 - \frac{1}{k} x^\sh$ and $\varphi_2' = \frac{1}{k} x^\sh$. These two coordinates are orthogonal (as opposed to, say, $x^6$ and $x^\sh$). The $\mathbbm{Z}_k$ action on these coordinates appears in eq. (\ref{eq:Zkaction}). We draw the fundamental domain of the $(\varphi_1',\varphi_2')$ torus in the figure. We also indicate the $y$ direction in the figure, where in these coordinates $y = \frac{1}{2} \left( \varphi_2' - \varphi_1'\right)$. The generators of homology are the $y$ and $\varphi_1'$ axes as drawn, \textit{i.e.} these form a basis of one-cycles. When $k \ra \infty$, the upper bound of the fundamental domain moves down, so that the parallelogram collapses (in the $y$ direction) onto the $\varphi_1'$ axis. The cycles that shrink in this process are all the ones that have net winding around $y$ and zero winding around $\varphi_1'$. The shortest cycles that shrink are parallel to the $y$-axis, so here again we identify $y$ as the M-theory circle (when descending to type IIA on $AdS_4 \times \cp$).

\begin{FIGURE}[t]
{
\centering
\includegraphics[width=0.8\textwidth]{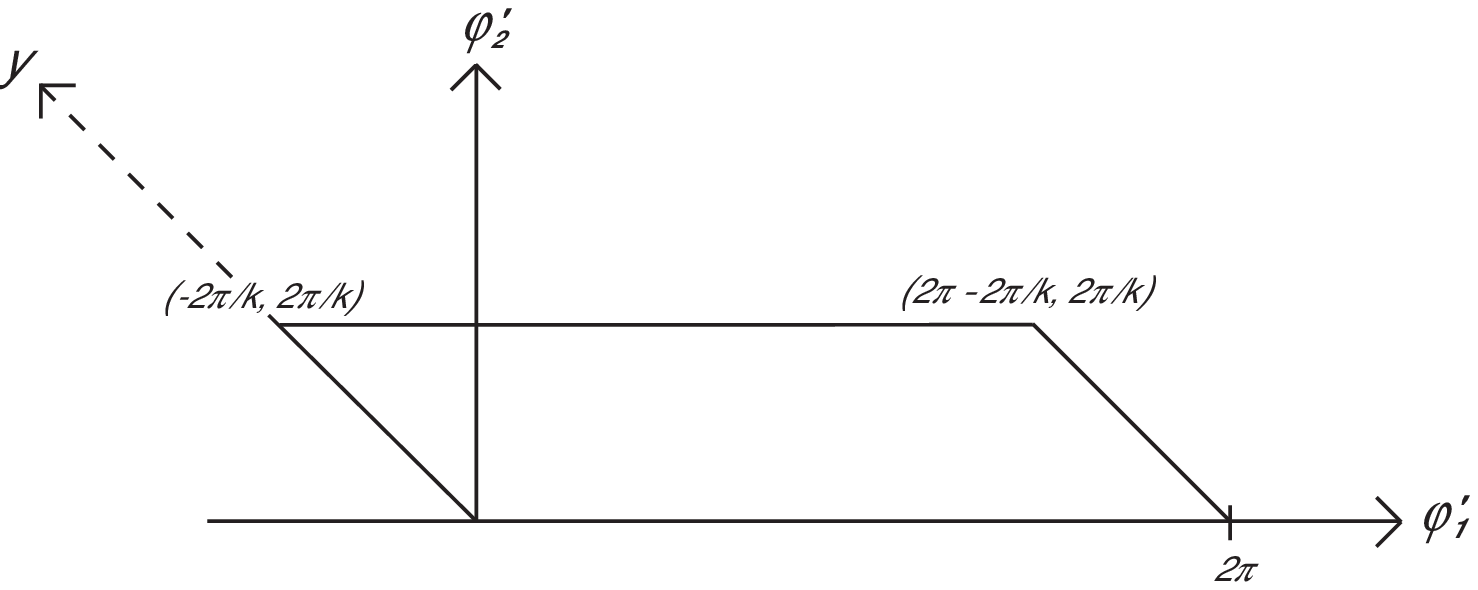}
\caption{The torus spanned by $(\varphi_1',\varphi_2')$. We have indicated the fundamental domain (the parallelogram) and the $y$ direction (the dotted line). A basis of one cycles is a curve in the $y$ direction and a curve in the $\varphi_1'$ direction. When $k \ra \infty$, the $y$ circle shrinks, and the parallelogram collapses onto the $\varphi_1'$ axis.}
}
\end{FIGURE}

Let's consider what happens to our flavor branes when $x^\sh$ opens up and then $y$ collapses. We have four options: a brane can wrap $x^\sh$ but not $x^6$, $x^6$ but not $x^\sh$, or a brane can wrap both, or a brane can wrap neither.

Consider a brane that wraps $x^\sh$ and sits at fixed $x^6$. The key point is that, from the definition of $\varphi_1'$ and $\varphi_2'$, we immediately see that such a brane will be parallel to the $y$-axis, so such a brane will return to whatever it was in IIA. For example, a D4-brane localized in $x^6$, which lifts to an M5-brane wrapping $x^\sh$, would descend back to a D4-brane localized in $x^6$ when $y$ closes.

Now consider a brane that wraps $x^6$ and sits at fixed $x^\sh$. Such a brane will extend along $\varphi_1'$ at fixed $\varphi_2'$ (a horizontal line in the figure). Since $\varphi_1'$ is the direction that remains when $k \rightarrow \infty$ we see that such a brane again returns to whatever it was (now in IIA on $AdS_4 \times \mathbbm{CP}^3$).

The last two cases are basically trivial. A brane that wraps both directions or neither will return to whatever it was. For example, a D2-brane localized in $x^6$ will lift to an M2-brane that wraps neither $x^6$ nor $x^\sh$, and will return to a D2-brane localized in $x^6$ when $y$ collapses. A D4-brane wrapping $x^6$ will lift to an M5-brane wrapping both $x^6$ and $x^\sh$, and hence will wrap the entire $(\varphi_1',\varphi_2')$ torus, and will become a D4-brane wrapping $x^6$ when $y$ collapses.

To summarize: $x^\sh$ opens up and $y$ shrinks, and all D-branes remain the same D-branes when we go to IIA on $AdS_4 \times \mathbbm{CP}^3$.

\section{Supersymmetry of M-theory Objects}
\label{Mtheorysusy}

In this appendix we will calculate the supersymmetry preserved by probes added to the $N_c$ M2-branes along $\mathbbm{R}^{2,1}$ sitting at the origin of $\mathbbm{C}^4/\mathbbm{Z}_k$ and also for probes added to the near-horizon geometry of $N_c \ra \infty$ M2-branes, $AdS_4 \times S^7/\mathbbm{Z}_k$ (see \cite{Nishioka:2008ib,Drukker:2008zx} for similar calculations).

The number of supersymmetries preserved by our probes is the number of solutions of the $\k$-symmetry condition
\be
\label{eq:kappa}
\G_\k \e = \e,  \quad \G_\k =\frac{1}{n!}\frac{1}{\sqrt{-g}} \, \e^{12\ldots n} \, \g_{12 \ldots n},
\ee
where $\e$ is the 32-component spinor of the background, $n$ is the dimensionality of the object (KK monopole or M-brane), and $g_{mn} = \p_m X^I \p_n X^J g_{IJ}$ and $\g_m = \p_m X^I e_{~I}^A \G_A$ are the pullbacks of the background metric and the $\G$-matrices to the worldvolume of the object. Here $X^I$ represent the scalars on the worldvolume of the object, $e_{~I}^A$ are vielbeins, and $A,B\ldots$ are tangent space indices. The $\G_A$ satisfy the tangent space Clifford algebra $\{ \G_A, \G_B \} = 2 \eta_{AB}$, where we use a mostly-plus metric. We calculate the spinor $\e$ by demanding that the supersymmetry transformation of the gravitino, $\Psi$, vanishes
\be
\delta \Psi_I = \left( \partial_I + \frac14 \omega^{AB}_I \, \G_{AB} \right) \e -\frac{1}{288} \left( \G^{ABCD}_I F^{(4)}_{ABCD} - 8 \G^{BCD} F^{(4)}_{I BCD} \right) \e = 0.
\ee
Here $\omega^{AB}_I$ is the spin connection and $F^{(4)}$ is the four-form field strength of M-theory, while $I$ is a general coordinate index ($A,B,C,D$ are still tangent space indices).

\subsection{Objects in $\mathbbm{R}^{2,1} \times \mathbbm{C}^4/\mathbbm{Z}_k$}
\label{app:probeflat}

In this subsection we will consider M-theory on $\mathbbm{R}^{2,1} \times \mathbbm{C}^4/\mathbbm{Z}_k$, without flux ($F^{(4)}=0$). We add M2-branes along $\mathbbm{R}^{2,1}$ and a variety of KK monopoles and M-branes. We will use polar coordinates on $\mathbbm{C}^4$ such that $z^i = r_i \, e^{i \varphi_i}$. The metric is
\be
ds^2 = -(dx^0)^2 + (dx^1)^2 + (dx^2)^2 + \sum_{i=1}^4 (dr_i^2 + r_i^2 d \varphi_i^2).
\ee
In these coordinates the spinor on $\mathbbm{R}^{2,1} \times \mathbbm{C}^4$ is
\be
\e = e^{i \frac{\varphi_1}{2} \G_{r_1 \varphi_1}} \, e^{i \frac{\varphi_2}{2} \G_{r_2 \varphi_2}}\, e^{i \frac{\varphi_3}{2} \G_{r_3 \varphi_3}} \, e^{i \frac{\varphi_4}{2} \G_{r_4 \varphi_4}} \, \e_0 \equiv M \e_0,
\ee
where $\e_0$ is a constant 32-component spinor. The $\mathbbm{Z}_k$ acts as $\varphi_i \rightarrow \varphi_i + \frac{2 \pi}{k}, \, \forall i$. We write $\epsilon_0$ as a sum of eigenspinors $\e_{s_1s_2s_3s_4}$ that satisfy $\G_{r_i \varphi_i} \e_{s_1s_2s_3s_4} = i s_i \e_{s_1s_2s_3s_4}$ for $i = 1, \ldots, 4$, where $s_i = \pm 1$. For the spinor to be invariant we demand that $\sum_i s_i =0$ for $k>2$. This means that of the 16 combinations of $(s_1,s_2,s_3,s_4)=(\pm1,\pm1,\pm1,\pm1)$, we project out 10 combinations and preserve 6. A spinor in $\mathbbm{R}^{2,1}$ has two real components so that the 6 preserved combinations correspond to a total of 12 real preserved supercharges. From $\sum_i s_i =0$ we find that $\prod_i s_i=1$ and therefore that $\e = \G_{01\ldots r_4 \varphi_4} \e = \G_{012} (i)^4 s_1 s_2 s_3 s_4 \e = \G_{012} \e$, so the projection condition for the color M2-branes is automatically satisfied, \ie the M2-branes do not break any additional supersymmetry.

Now we can calculate $\G_\kappa$ for any given embedding using eq. \eqref{eq:kappa}, check how many supercharges are preserved by the condition\footnote{The cautious reader might worry whether this procedure is applicable to KK monopoles and the mysterious M9-branes. We will not give a direct proof, rather we will think of this one condition as a combination of the two projection conditions for the left- and right-handed spinors for D6- and D8-branes in type IIA.} $\G_\kappa \e = \e \Leftrightarrow M^{-1} \G_\kappa M \e_0 = \e_0$, where $\e_0$ is the 12-component spinor from above. The calculation is fairly easy. We summarize our results in table \ref{table:flat}, and work out explicit examples in the more complicated background of $AdS_4 \times S^7/\mathbbm{Z}_k$ in the next subsection. In the table we restrict ourselves to objects that sit at the origin of $\mathbbm{C}^4/\mathbbm{Z}_k$. We can use the $SU(4) \times U(1)$ symmetry to set any constant phase factor to zero so that configurations that differ from those in the table by constant shifts in any of the $\varphi_i$ preserve the same amount of supersymmetry. The table contains the four examples studied in this paper and also some other easy configurations. The second column indicates what the resulting object is in type IIA for $k \rightarrow \infty$ and the third column gives the codimension of the probe in $\mathbbm{R}^{2,1}$.

\begin{table}[h!]
\begin{center}
\begin{tabular}{|c|c|c|c|c|c|c|c|}
\hline
M-theory & Type IIA & codim & real supercharges & worldvolume coordinates\\ \hline \hline
M2 & D2 & 0 & 12 & $x^0, x^1, x^2$\\ \hline \hline
KK & D6 & 0 & 6 & $x^0, x^1, x^2, r_1, r_2, r_3, r_4$\\ \hline
KK & KK & 0 & 8 & $x^0, x^1, x^2, z^1=z^2,z^3=z^4$\\ \hline
M5 & D4 & 1 & 6 & $x^0, x^1, z^1=z^2,z^3=z^4 $\\ \hline
M5 & NS5 & 1 & 6 & $x^0, x^1, r_1, r_2, r_3, r_4$\\ \hline
M9 & D8 & 1 & 6 & $x^0, x^1, r_i, \varphi_i, i=1,2,3,4$\\ \hline
M2 & D2 & 2 & 4 & $x^0, r_3 = r_4, \varphi_3 = -\varphi_4$\\ \hline
M2 & F1 & 2 & 6 & $x^0, z^1=z^2=z^3=z^4$\\ \hline
\end{tabular}
\end{center}
\caption{List of supersymmetry-preserving objects of given codimension and given worldvolume directions in $\mathbbm{C}^4/\mathbbm{Z}_k$. The second column indicates what the probes become in type IIA (large $k$). For details see the accompanying paragraph.}\label{table:flat}
\end{table}

\subsection{Objects in $AdS_4 \times S^7/\mathbbm{Z}_k$}
\label{app:probeads}

In this section we study objects in the geometry obtained as the near-horizon limit ($N_c \rightarrow \infty$) of the $N_c$ M2-branes. First we introduce new coordinates
\be
z^1 = r \cos\a \sin \b \, e^{\z_1}, \quad z^2 = r \cos \a \cos \b \, e^{\z_2} \quad z^3 = r \sin \a \sin \g \, e^{\z_3}, \quad z^4 = r \sin \a \cos \g \, e^{\z_4}\nn
\ee
With these coordinates, the metric of $AdS_4 \times S^7$ becomes
\ba
ds^2 &=& \frac{R^2}{4} \left( dr^2+e^{2r} (-dt^2+dx_1^2 +dx_2^2) \right)+  R^2 \left( d\a^2+ \cos{^2\a}\, d\b^2 + \sin{^2\a} \, d\g^2 \right.\\ \nn
&& \left. + \cos{^2 \a} \sin{^2 \b} \, d\z_1^2 + \cos{^2 \a} \cos{^2 \b} \, d\z_2^2 + \sin{^2 \a} \sin{^2 \g} \, d\z_3^2 + \sin{^2 \a} \cos{^2 \g} \, d\z_4^2\right),
\ea
where $0 \leq \a,\b,\g \leq \frac{\pi}{2}$ and $0 \leq \z_i < 2 \pi$. We also have the flux $F^{(4)} = \frac38 R^3 \Omega_{AdS_4}$ with $\Omega_{AdS_4}$ being the volume form of $AdS_4$.

From the supersymmetry variation of the gravitino we find that the spinor preserved by this background is
\ba
\e &=& e^{-\frac{r}{2} \G_r \hat{\G}}\lp \mathbbm{1}_{32} +\frac12 x^\m \G_\m \hat{\G} (\mathbbm{1}_{32}+ \G_r \hat{\G}) \rp e^{\frac{\a}{2} \G_{\a} \hat{\G}} e^{\frac{\b}{2} \G_{\b} \hat{\G}} e^{\frac{\g}{2} \G_{\a\g}} e^{\frac{\z_1}{2} \G_{\b\z_1}} e^{\frac{\z_2}{2} \G_{\z_2} \hat{\G}} e^{\frac{\z_3}{2} \G_{\g \z_3}} e^{\frac{\z_4}{2} \G_{\a \z_4}} \e_0 \nn \\
&\equiv& M_{AdS} M_{\a\b\g} M_\z \e_0 \equiv M \e_0,
\ea
where $\e_0$ is a constant 32-component spinor, $\hat{\G} = \G^{012r} = -\G_{012r}$, $M_{AdS}$ is the part of $M$ that depends on $AdS_4$ coordinates, and similarly for $M_{\a\b\g}$ and $M_{\z}$. The $\mathbb{Z}_k$ quotient acts as $\z_i \rightarrow \z_i +\frac{2 \pi}{k}$. We write the spinor $\e_0$ as a sum of eigenspinors of $(\G_{\b \z_1},\G_{\z_2} \hat{\G},\G_{\g \z_3}, \G_{\a \z_4})$, that is, $\G_{\b \z_1} \e_{s_1s_2s_3s_4} = i s_1\e_{s_1s_2s_3s_4}$, etc.; since only the eigenspinors that satisfy $\sum_{i=1}^4 s_i = 0$ are invariant under $\mathbb{Z}_k$ for $k>2$, the background preserves 24 real supercharges for $k>2$. This can be seen from $\mathbbm{1}_{32} = \G_{01\ldots\z_3 \z_4} = s_1 s_2 s_3 s_4 \mathbbm{1}_{32}$ which implies an even number of positive and negative $s_i$. The condition $\sum_{i=1}^4 s_i = 0$ therefore projects out the two cases where all $s_i$ are the same, and is satisfied for the other six cases, so $\frac68$ of the 32 supercharges (hence 24) are preserved.

Now we will explicitly solve the $\kappa$-symmetry equation for the four objects discussed in sections \ref{codzero}, \ref{codonechiral}, \ref{codonenonchiral}, and \ref{codtwo}. As mentioned in the previous subsection, we think of this one condition in M-theory as a combination of the two projection conditions for the left- and right-handed spinors for D-branes in type IIA.

We start with the codimension-zero KK monopole of section \ref{codzero}. Although we can use the $SU(4) \times U(1)$ symmetry of the background to set constant phases to zero, we will keep them explicitly in our calculation. This is useful if we want to consider multiple stacks of probes that sit at different constant phases. Instead of choosing the embedding such that Im$(z^i)=0, \, \forall i$, we will be more general and take the worldvolume coordinates to be $x^0,x^1,x^2, r, \a, \b, \g$ and set the phase to constant values $\z_i^0$. For this embedding we find $\G_\k= \G_{012r\a\b\g}$ and
\be
M^{-1} \G_\k M = M^{-1}_\z \G_\k M_\z = \G_\k M_\z^2 = \G_{012r\a\b\g} e^{\z_1^0 \G_{\b\z_1}} e^{\z_2^0 \G_{\z_2} \hat{\G}} e^{\z_3^0 \G_{\g\z_3}} e^{\z_4^0 \G_{\a\z_4}}.
\ee
Solving $\G_{012r\a\b\g} e^{\z_1^0 \G_{\b\z_1}} e^{\z_2^0 \G_{\z_2} \hat{\G}} e^{\z_3^0 \G_{\g\z_3}} e^{\z_4^0 \G_{\a\z_4}} \e_0 = \e_0$ we find that 12 real components are preserved. Comparing with the second row in table \ref{table:flat}, we see that after the near-horizon limit the amount of supersymmetry has doubled, as expected.

Now we look at the codimension-one case of section \ref{codonechiral}, that is, the lift of D8-branes to M-theory. We choose the ten worldvolume coordinates to be $(x^0,x^1,r,\a,\b,\g,\z_1,\z_2,\z_3,\z_4)$, and find that $\G_\k = \G_{01r\a\b\g\z_1\z_2\z_3\z_4} = \G_2$, where we have used $\G_{012r\a\b\g\z_1\z_2\z_3\z_4}=\mathbbm{1}_{32}$. $\G_\k$ commutes with $M_{\a\b\g}$ and $M_\z$, so we find
\be
M^{-1} \G_\k M = M_{AdS}^{-1} \G_\k M_{AdS} = (\mathbbm{1}_{32} - x^2 \G_2 \hat{\G} (\mathbbm{1}_{32} + \G_r \hat{\G})) \G_\k = (\mathbbm{1}_{32} - x^2 \G_2 \hat{\G} (\mathbbm{1}_{32} + \G_r \hat{\G})) \G_2.
\ee
Demanding that $M^{-1} \G_\k M \e_0 = \e_0$ again reduces the components of $\e_0$ by a factor of $\frac12$, so that we find 12 preserved supercharges.

The next codimension-one case, from section \ref{codonenonchiral}, are M5-branes extended along $AdS_3$ inside $AdS_4$ and embedded such that $z^1=z^2=0$. Recall that this embedding is $SU(4)$-equivalent to the one used in table \ref{table:flat} (see section \ref{codoneM5}). The worldvolume coordinates are $(x^0,x^1,r,\g,\z_3,\z_4)$ and we have to set $\a=\frac{\pi}{2}$, which leads to $\G_\k = \G_{01r\g\z_3\z_4}$. Since $M_{\a\b\g}^{-1} \G_\k M_{\a\b\g} = - \hat{\G} \G_\a \G_\k$ and $M_\z$ commutes with that, we find
\ba
M^{-1} \G_\k M = - M_{AdS}^{-1} \hat{\G} \G_\a \G_\k M_{AdS} &=& - (\mathbbm{1}_{32} - x^2 \G_2 \hat{\G} (\mathbbm{1}_{32} + \G_r \hat{\G})) \hat{\G} \G_\a \G_\k \nn \\
&=& -(\mathbbm{1}_{32} - x^2 \G_2 \hat{\G} (\mathbbm{1}_{32} + \G_r \hat{\G})) \G_{2 \a \g \z_3\z_4}.
\ea
We then find that 12 real supercharges are preserved. Again we see that the probes reduce the amount of supersymmetry of the background by a factor of $\frac12$, and that the near-horizon limit leads to a doubling of the preserved supercharges (cf. table \ref{table:flat}).

Finally we look at the example from section \ref{codtwo}, codimension-two M2-branes embedded such that $z^1=z^2=0$, $z^3=\bar{z}^4$. Again we will be slightly more general and allow for constant phases. We take the worldvolume coordinates to be $x^0, r, \z_3$ and set $\a=\frac{\pi}{2}, \g = \frac{\pi}{4}, \z_1 = \z_1^0, \z_2 =\z_2^0, \z_4 = -\z_3 +\z^0$, where $\z_1^0, \z_2^0, \z^0$ are constants. We then find
\be
\Gamma_\k = \frac{1}{\sqrt{2}} \Gamma_{0r} (\Gamma_{\z_3} - \Gamma_{\z_4}).
\ee
Next we write $M = M_{AdS} M_{\a \b \g} M_\z$ and note that the AdS part commutes with the rest, so we first calculate
\be
M^{-1}_{AdS} \G_\k M_{AdS} = (\mathbbm{1}_{32}- (x^1 \G_1 +x^2 \G_2) \hat{\G}(\mathbbm{1}_{32}+\G_r \hat{\G})) \G_\k.
\ee
We then use $\alpha = \frac{\pi}{2}$ to find
\ba
\label{eq:m2brane}
&& M^{-1}_{\a \b \g} M^{-1}_{AdS} \G_\k M_{AdS} M_{\a \b \g} \nn \\
&=& (\mathbbm{1}_{32}- (x^1 \G_1 +x^2 \G_2) \hat{\G}(\mathbbm{1}_{32}+\G_r \hat{\G})) e^{-\frac{\g}{2} \G_{\a\g}}  e^{-\frac{\b}{2} \G_{\b} \hat{\G}} e^{-\frac{\a}{2} \G_{\a} \hat{\G}} \G_\k e^{\frac{\a}{2} \G_{\a} \hat{\G}} e^{\frac{\b}{2} \G_{\b} \hat{\G}} e^{\frac{\g}{2} \G_{\a\g}} \nn \\
&=& (\mathbbm{1}_{32}- (x^1 \G_1 +x^2 \G_2) \hat{\G}(\mathbbm{1}_{32}+\G_r \hat{\G})) e^{-\frac{\g}{2} \G_{\a\g}}  e^{-\frac{\b}{2} \G_{\b} \hat{\G}} \G_\k e^{\a \G_{\a} \hat{\G}} e^{\frac{\b}{2} \G_{\b} \hat{\G}} e^{\frac{\g}{2} \G_{\a\g}} \nn \\
&=& (\mathbbm{1}_{32}- (x^1 \G_1 +x^2 \G_2) \hat{\G}(\mathbbm{1}_{32}+\G_r \hat{\G})) e^{-\frac{\g}{2} \G_{\a\g}} \G_\k \G_{\a} \hat{\G} e^{\frac{\g}{2} \G_{\a\g}} \\
&=& (\mathbbm{1}_{32}- (x^1 \G_1 +x^2 \G_2) \hat{\G}(\mathbbm{1}_{32}+\G_r \hat{\G})) \G_\k \frac{1}{\sqrt{2}} (\G_{\a} +\G_{\g}) \hat{\G}\nn\\
&=& (\hat{\G}+ (x^1 \G_1 +x^2 \G_2) (\mathbbm{1}_{32} - \G_r \hat{\G})) \G_\k \frac{1}{\sqrt{2}} (\G_{\a} +\G_{\g}) \nn \\
&=& \frac12 (\hat{\G}+ (x^1 \G_1 +x^2 \G_2) (\mathbbm{1}_{32} - \G_r \hat{\G})) \G_{0r} (-\G_{\a\z_3} + \G_{\a\z_4} -\G_{\g\z_3} + \G_{\g\z_4}).\nn
\ea
Since eq. \eqref{eq:m2brane} commutes with $\Gamma_{\b\z_1}$ and $\Gamma_{\z_2} \hat{\G}$ we have
\ba
M^{-1} \G_\k M &=& e^{-\frac{\z_4}{2} \G_{\a\z_4}} e^{-\frac{\z_3}{2} \G_{\g\z_3}} \frac12 (\hat{\G}+ (x^1 \G_1 +x^2 \G_2) (\mathbbm{1}_{32} - \G_r \hat{\G})) \G_{0r} \nn \\
&& \qquad \qquad \times \, (-\G_{\a\z_3} + \G_{\a\z_4} -\G_{\g\z_3} + \G_{\g\z_4}) e^{\frac{\z_3}{2} \G_{\g\z_3}} e^{\frac{\z_4}{2} \G_{\a\z_4}} \\
&=& \frac12 (\hat{\G}+ (x^1 \G_1 +x^2 \G_2) (\mathbbm{1}_{32} - \G_r \hat{\G})) \G_{0r} \nn \\
&& \qquad \qquad \times \, (\G_{\a\z_4} -\G_{\g\z_3} + \cos{\z^0}(\G_{\g\z_4} - \G_{\a\z_3})+ \sin{\z^0} (\G_{\a\g}+\G_{\z_3\z_4})).\nn
\ea
Acting with this on the 24-component constant spinor $\e_0$, we find that such branes preserve 8 real supercharges. Note that although the projector depends on $\z^0$, we find that the preserved supercharges depend only on the position in $x^1, x^2$ and not on the constant phases $\z_1^0, \z_2^0, \z^0$.

\section{$\N=(0,6)$ Supersymmetry Transformations}
\label{app:n06}

In this appendix we discuss the supersymmetry transformations of the codimension-one flavor field theories of sections \ref{codonechiral} and \ref{codonenonchiral}. In particular, we show that the gauge field somponent $A_-,$ appearing in the Lagrangian of the codimension-one chiral field theory of section \ref{codonechiral}, eq. (\ref{eq:codonechiralaction}), is invariant under $\mathcal{N}=(0,6)$ supersymmetry. 

The supersymmetry algebra of the ABJM theory is
\begin{equation}
\{ \mathcal{Q}^{(I)}_\alpha, \mathcal{Q}^{(J)}_\beta \} = - 2 \delta^{IJ} \, ( \gamma^\mu )_{\alpha\beta} \, P_\mu,
\end{equation}
where $(\gamma^\mu)_{\alpha\beta}$ is given by $\gamma^\mu = ( - \mathbbm{1}, - \sigma^3, \sigma^1 )$ and $\mu = 0, 1, 2.$ The index $\alpha = 1,2$ labels the components of the real two-component spinor $\mathcal{Q}.$ (Notice that we are using different conventions from those in section \ref{abjmtheory}.)

Let us place the defect at $x^2 = 0.$ Since the translational invariance in $x_2$ direction and therefore the momentum $P_2$ is broken, some of the supersymmetry charges are also broken. Let us discuss the broken supersymmetry generators for the $\mathcal{N}=(0,6)$ and $\mathcal{N}=(3,3)$ supersymmetric flavor theories.

The broken supersymmetry generators for the $\mathcal{N}=(0,6)$ supersymmetric flavor theory are $\mathcal{Q}^{(I)}_1.$ An explicit check shows that, upon setting $\mathcal{Q}^{(I)}_1=0$, the algebra reduces to a supersymmetry algebra for a (1+1)-dimensional theory, \textit{i.e.} $P_2$ drops out.

The broken supersymmetry generators for the $\mathcal{N}=(3,3)$ supersymmetric flavor theory are more complicated since the obvious guess, setting half of the supersymmetry generators $\mathcal{Q}^{(I)}$ to zero, is wrong. For simplicity let us consider the algebra just for two supersymmetry generators, say $I=1$ and $I=2$,
\begin{equation}
\{ \mathcal{Q}^{(1)}_\alpha, \mathcal{Q}^{(1)}_\beta \} =\{ \mathcal{Q}^{(2)}_\alpha, \mathcal{Q}^{(2)}_\beta \}= - 2 \, ( \gamma^\mu )_{\alpha\beta} \, P_\mu, \qquad \{ \mathcal{Q}^{(1)}_\alpha, \mathcal{Q}^{(2)}_\beta \} = 0 \, .
\end{equation}
Since we want to eliminate $P_2$ and $\sigma_2$ is off-diagonal, we have to define the new supersymmetry charges,
\begin{equation}
\tilde{\mathcal{Q}}_1 \equiv \mathcal{Q}_{1}^{(1)}, \quad \bar{\tilde{\mathcal{Q}}}_2 \equiv \mathcal{Q}_2^{(2)},
\quad \tilde{\slashed{\mathcal{Q}}}_1 \equiv \mathcal{Q}_1^{(2)}, \quad \tilde{\slashed{\mathcal{Q}}}_2 \equiv \mathcal{Q}_2^{(1)},
\end{equation}
and set $\tilde{\slashed{\mathcal{Q}}}_1 = \tilde{\slashed{\mathcal{Q}}}_2 = 0.$ The remaining supersymmetry generators $\tilde{\mathcal{Q}}_\alpha, \ \ \alpha=1,2,$ satisfy the (1+1)-dimensional supersymmetry algebra. This procedure can be straightforward generalized to six supersymmetry generators. The unbroken supercharges generate a $\mathcal{N}=(3,3)$ supersymmetric algebra in (1+1) dimensions.

In order to determine the supersymmetry transformation for $A_-$ in the $\mathcal{N}=(0,6)$ supersymmetric theory, we use the conventions and supersymmetry transformations of ref. \cite{Bandres:2008ry}. Let us quote the supersymmetry transformation of the gauge fields $A_\mu$ and $\hat A_\mu$ (the gauge fields of the two gauge groups),
\begin{eqnarray}
\delta A_\mu &=& \Gamma^I_{AB} \, \bar{\epsilon}^I \gamma_\mu \Psi^A X^B - \tilde\Gamma^{IAB} X_B \bar\Psi_A \gamma_\mu \epsilon^I \, , \\
\delta \hat A_\mu &=& \Gamma^I_{AB} \, X^B \bar{\epsilon}^I \gamma_\mu \Psi^A - \tilde\Gamma^{IAB} \bar\Psi_A \gamma_\mu \epsilon^I X_B \, ,
\end{eqnarray}
where $X_A, \, A=1, \dots, 4$ are the four complex scalars and $\Psi_A$ are the spinor fields of the ABJM theory (in the notation of ref. \cite{Bandres:2008ry}). The spinor field $\Psi_{A \, \alpha}$ has a lower spinor index, whereas the conjugated field $\bar{\Psi}_A^{\alpha}$ carries an upper spinor index. Note that $\left( \gamma^\mu \right)_\alpha^{ \ \beta} = ( i \sigma^2, \sigma^1, \sigma^3)$ for $\mu = 0,1,2.$ The conjugate fields $\Psi^A$ and $X^B$ are denoted by upper indices. Moreover, $\epsilon^I$ are real two-component spinors for $I=1, \dots, 6$, and the $4 \times 4$ matrices $\Gamma^I, \, I=1, \dots, 6$ satisfy the commutation relation
\begin{equation}
\Gamma^I \tilde \Gamma^J + \Gamma^J \tilde \Gamma^I = 2 \, \delta^{IJ},
\end{equation}
where $\tilde \Gamma^I = \left( \Gamma^I \right)^\dagger.$ Let us decompose $\epsilon^I$ into 
\begin{equation}
\epsilon^I = \left( \begin{array}{c} \epsilon^I_R \\ \epsilon^I_L \end{array} \right),
\end{equation}
and set $\epsilon^I_L$ to zero since the $\tilde{\mathcal{Q}}^{(I)}_2$ are broken in the $\mathcal{N}=(0,6)$ algebra. Finally, introducing lightcone coordinates $x^{\pm} = x^0 \pm x^1$, the unbroken right-handed supersymmetry transformations $\delta_{R,I}$ with respect to $\epsilon_R^I$ of the gauge field components read
\begin{eqnarray}
\delta_{R,I} \, A_+ &=&  \Gamma^I_{AB} \epsilon_R^I \Psi_R^A X^B - \tilde\Gamma^{IAB} X_B \Psi_{R \, A} \epsilon_R^I \, , \\
\delta_{R,I} \, A_- &=& 0 \, , \\
\delta_{R,I} \, \hat A_+ &=& \Gamma^I_{AB} X^B \epsilon_R^I \Psi_R^A - \tilde\Gamma^{IAB} \Psi_{R \, A} \epsilon_R^I X_B \, , \\
\delta_{R,I} \, \hat A_- &=& 0 \, . 
\end{eqnarray}
In particular we see that $A_-$ and $\hat A_-$ do not transform under $\mathcal{N}=(0,6)$ supersymmetry.

\bibliographystyle{JHEP}
\bibliography{abjmflavor}

\end{document}